\documentclass[runningheads]{svmult}

\usepackage{graphicx}  
\usepackage{subeqnar}  
\usepackage{multicol}  
\usepackage{physprbb}  

\def\Ref#1{(\ref{#1})}
\def\BA{\begin{eqnarray}}
\def\BE{\begin{equation}}
\def\BF{\begin{figure}[htb]}
\def\BT{\begin{table}[t]}
\def\EA{\end{eqnarray}}
\def\EE{\end{equation}}
\def\EF{\end{figure}}
\def\ET{\end{table}}
\def\vr{\vec R\,}  \def\rT{r} \def\brT{{\vec r}}
\def\vp{\vec p\,}  \def\pT{p_T} \def\bpT{{\vec p_T}}

\def\Jpsi{J\!/\!\psi}
\def\sqq{{\sigma_{q\bar q}}}

\newcommand{\beq}{\begin{equation}}
\newcommand{\eeq}{\end{equation}}
\newcommand{\beqn}{\begin{eqnarray}}
\newcommand{\eeqn}{\end{eqnarray}}
 
\newcommand\la{\langle}
\newcommand\ra{\rangle}
\newcommand\eps\varepsilon
\newcommand\euler{{\rm e}}
\newcommand\kzero{{\rm K}_0}
\newcommand\kone{{\rm K}_1}
\newcommand\unint{{\cal F}}
\newcommand\as{{\alpha_{\rm s}}}  
\newcommand\aem{{\alpha_{\rm em}}}

\def\sig{\sigma_{q\bar q}}

\def\mb{\,\mbox{mb}}
\def\fm{\,\mbox{fm}}
\def\GeV{\,\mbox{GeV}}
\def\TeV{\,\mbox{TeV}}
\def\lsim{\mathrel{\rlap{\lower4pt\hbox{\hskip1pt$\sim$}}
    \raise1pt\hbox{$<$}}}         
\def\gsim{\mathrel{\rlap{\lower4pt\hbox{\hskip1pt$\sim$}}
    \raise1pt\hbox{$>$}}}         

\begin{document}

\title*{Heavy flavor production off protons and
\protect\newline in a nuclear environment}

\toctitle{Heavy flavor production off protons and in a nuclear 
environment }

\titlerunning{Heavy flavor production off protons and in a nuclear 
environment}

\author{B.~Z.~Kopeliovich \inst{1,2,3} 
\and J.~Raufeisen \inst{4}}

\authorrunning{B.~Z.~Kopeliovich and J.~Raufeisen }

\institute{Max-Planck Institut f\"ur Kernphysik, Postfach 103980,
69029 Heidelberg, Germany
\and
Institut f\"ur Theoretische Physik der Universit\"at, 93040
Regensburg, Germany
\and
Joint Institute for Nuclear Research, Dubna, 141980 Moscow
Region, Russia
\and
Los Alamos National Laboratory, MS H846, Los Alamos,
		  NM 87545, USA} 

 \maketitle

\begin{abstract} These lectures present an overview of the current status
of the QCD based phenomenology for open and hidden heavy
flavor production at high energies. A unified description based on the
light-cone color-dipole approach is employed in all cases. A good
agreement with available data is achieved without fitting to the data to
be explained, and nontrivial predictions for future experiments are made.
The key phenomena under discussion are: (i) formation of the wave
function of a heavy quarkonium; (ii) quantum interference and coherence
length effects; (iii) Landau-Pomeranchuk suppression of gluon radiation
leading to gluon shadowing and nuclear suppression of heavy flavors; (iv)
higher twist shadowing related to the finite size of heavy quark dipoles;
(v) higher twist corrections to the leading twist gluon shadowing making
it process dependent.
\end{abstract}

\section{Introduction}
\label{sec:intr}

Reactions in which heavy flavors are produced 
involve a hard scale that allows 
one to
employ perturbative QCD (pQCD). In particular, production off
nuclei has always been an important topic in heavy quark physics. On the
one hand, interest in this field is stimulated by demand to provide a
proper interpretation for available and forthcoming data, especially from
heavy ion collisions. On the other hand, nuclei have been traditionally
employed as an analyzer for the dynamics and time scales of hadronic
interactions.

In these lectures, we shall review the color dipole formulation
of heavy flavor production, which was developed in \cite{HIKT,kth,ikth,kt}.
The dipole approach  to heavy flavor production
expresses the cross section for production of open or hidden heavy flavor
in terms of the cross section for scattering a color neutral quark-antiquark
($q\bar q$) pair off a nucleon.
This approach is
motivated by the need for a theoretical framework for the description
of nuclear effects.
Its main distinction from the conventional parton model 
is the possibility to actually calculate nuclear
effects in the dipole formulation, rather than absorbing them
into the initial conditions. In addition, the dipole approach is
not restricted to the leading twist approximation. This is especially
important for nuclear effects in $J/\psi$ 
production, which are dominated by higher twists.
Another advantage of the dipole approach is that it correctly describes the
absolute normalization of cross sections in different 
processes, without introducing an arbitrary
overall normalization factor (``$K$-factor'') \cite{pp,rpn}.
However, the dipole approach is applicable only in the
kinematical domain, where the heavy quark mass $m_Q$ is much smaller than
the center of mass ({\em cm.}) energy $\sqrt{s}$. The latter condition is
fulfilled for charm production at the BNL Relativistic Heavy Ion Collider
(RHIC) and for both, charm and bottom 
production, at the CERN Large Hadron Collider (LHC). 

An alternative approach to heavy quark production that is designed especially
for high energies and which
is able to describe nuclear effects is desirable for a
variety of reasons.
At low $x$, the heavy quark pair is produced over large
longitudinal distances, which can exceed the radius of a large nucleus
by orders of magnitude. 
Indeed, even though the matrix element of a hard 
process is dominated by short distances, of the order of the
inverse of the hard scale, the cross section of that process also depends
on the phase space element. 
Due to gluon radiation, the latter becomes very large at high energies,
and it is still a challenge how to resum the corresponding low-$x$ logarithms.
The dipole formulation allows for a 
simple phenomenological recipe to include these low-$x$ logs.
The large length scale in the problem 
leads to pronounced nuclear effects, giving one the possibility to
use the nuclear medium as microscopic detectors to study the space-time 
evolution of heavy flavor production. 

In addition, heavy quark 
production is of particular interest, because this
process directly probes the gluon distributions of the colliding particles.
Note that at the tremendous center of mass energies of RHIC and LHC, 
charm (and at LHC also bottom) decays 
will dominate the dilepton continuum \cite{gmrv}. Thus,
a measurement of the heavy quark production cross section at RHIC and LHC
will be relatively easy to accomplish
and can yield invaluable information 
about the (nuclear) gluon density \cite{ev}.
It is expected that at very low $x$, 
the growth of the gluon density will be slowed 
down by nonlinear terms  
in the QCD evolution equations \cite{glr}. The onset of this
non-linear regime is controlled by the so-called saturation scale $Q_s(x,A)$,
which is already of 
order of the charm quark mass at RHIC and LHC energies.
Moreover, $Q_s(x,A)\propto A^{1/3}$ ($A$ is the atomic mass of the nucleus), 
so that 
one can expect sizable higher twist corrections in $AA$ collisions \cite{kt}. 
Note that saturation will lead to a breakdown of the
twist expansion, since 
one cannot conclude any more that terms suppressed by powers of the 
heavy quark mass $m_Q$ are small, $Q_s^n(x,A)/m_Q^n\in O(1)$ for any $n$
\cite{kr}.
Saturation effects are most naturally described in the dipole 
picture.

The most prominent motivation for a theoretical investigation of nuclear 
effects in heavy flavor production is perhaps the 
experimentally observed suppression
of $J/\psi$ mesons in nuclear collisions, see Ref.~\cite{reviews} for 
a review. Suppression of $J/\psi$ production in 
nucleus-nucleus ($AA$) collisions has been proposed as signal of quark-gluon
plasma (QGP) formation \cite{Satz}. There are however several ``mundane''
nuclear effects 
that also lead to $J/\psi$ suppression, such as gluon shadowing, final state
absorption and breakup due to interactions with comovers. 
This issue has been widely and controversially discussed in the literature,
and
one clearly needs a reliable theoretical description of all
these effects, before definite conclusions about any non-standard dynamics
in heavy ion collisions can be drawn.
Since the creation and study of the QGP is the main physics motivation
for RHIC, the theoretical investigation of nuclear 
effects in $J/\psi$ production is at the heart of the RHIC
heavy ion program. Note that the production mechanism for heavy
quarkonia itself
is poorly understood, even in proton-proton ($pp$) collisions.
However,
since heavy flavors can be produced in many different reactions, and one may
hope that a detailed experimental and theoretical study 
of quarkonium production in different environments and over a wide
kinematical range will eventually 
clarify the underlying production mechanism.

\section{The foundations of the color dipole approach to high 
energy scattering}
\label{sec:found}

It was first realized in \cite{ZKL} that at high energies color dipoles
with a well defined transverse separation are the eigenstates of interaction
at, i.e. can experience only diagonal transitions when interacting
diffractively with a target. The eigenvalues of the amplitude operator
are related to the cross section $\sig(r)$ of interaction of a 
$q\bar q$ dipole with a nucleon. This dipole cross section
is a universal, flavor independent
quantity which depends only on transverse ${q\bar q}$ separation. Then, the
total hadron (meson) nucleon cross section can be presented in a
factorized form,
 \beq
\sigma^{hN}_{tot} = \int d^2r\,|\Psi^h_{{q\bar q}}(\vec r)|^2\,
\sig(r)\ ,
\label{bzk1}
 \eeq
 where $\Psi^h_{{q\bar q}}(\vec r)$ is the light-cone wave function
of the ${q\bar q}$ component of the hadron. It was assumed in \cite{ZKL}
that $|\Psi^h_{{q\bar q}}(\vec r)|^2$ is integrated over longitudinal
coordinate.

One of the advantages of this approach is simplicity of calculation
of the effects of multiple interactions which have the eikonal form 
(exact for eigenstates),
 \beq
\sigma^{hA}_{tot} = 2\int d^2b
\int d^2r\,|\Psi^h_{{q\bar q}}(\vec r)|^2
\left\{1-\exp\left[-{1\over2}\sig(r)T_A(b)\right]\right\}\ ,
\label{bzk2}
 \eeq
 where $T_A(b)$ is the nuclear thickness function which depends on impact 
parameter $\vec b$. 

Later the energy dependence of the dipole cross section was taken into
account \cite{BBFS}, and was found that one can apply the same
color-dipole formalism to variety of QCD processes, including radiation
\cite{hir}. Here we review this approach applied to production of heavy
flavors.

The simplest example for heavy flavor production is open heavy quark
production in deep inelastic scattering (DIS) off a proton, which can be
measured at the DESY $ep$ collider HERA. We use this example to explain
the basic ideas underlying the color dipole approach to high energy
scattering.

\begin{figure}[t]
  \centerline{
  \scalebox{1}{\includegraphics{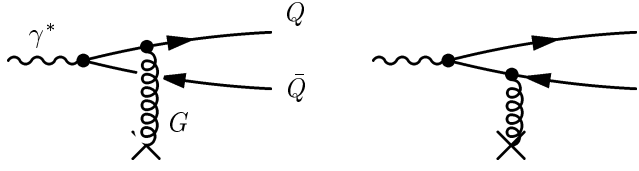}}}
\center{
{\caption{\em
  \label{fig:2graphs} Perturbative QCD graphs for heavy quark 
($Q\bar Q$) production
in DIS. The virtual photon ($\gamma^*$) fluctuates into a (virtual)
$Q\bar Q$-pair
far before the target. The interaction with the target, which is denoted by 
gluon ($G$) exchange, can put this fluctuation on mass shell.
  }  
    }  }
\end{figure}

The dipole approach is formulated in the target rest frame, where DIS
looks like pair creation in the gluon field of the target, 
see Fig.~\ref{fig:2graphs}. For further simplification, we consider only the
case of a longitudinally polarized $\gamma^*$ 
with virtuality $Q^2$ in this section, since the
transverse polarization does not contain any qualitatively new
physics. A straightforward calculation of the two Feynman diagrams
in Fig.~\ref{fig:2graphs} yields for the transverse momentum 
($\vec\kappa_\perp$)
distribution of the heavy quark \cite{catani},
\beqn\nonumber\label{eq:kt}
\frac{d\sigma\left(\gamma^*_Lp\to\{Q\bar Q\}X\right)}{d^2\kappa_\perp}&=&
\frac{4\aem\as e_Q^2Q^2}{\pi}\int d\alpha\, \alpha^2\left(1-\alpha\right)^2\\
&\times&
\int\frac{d^2k_T}{k_T^2}\left(\frac{1}{\kappa_\perp^2+\eps^2}-
\frac{1}{\left(\vec\kappa_\perp-\vec k_T\right)^2+\eps^2}\right)^2
\unint(x,k_T^2),
\eeqn
where $\alpha$ is the light-cone momentum fraction of the heavy quark
and $\eps^2=\alpha\left(1-\alpha\right)Q^2+m_Q^2$. The quark electric charge 
is denoted by $e_Q$; $\aem=1/137$ and $\as$ are the electromagnetic and 
the strong coupling constants, respectively. The $Q\bar Q$-pair exchanges a
gluon with transverse momentum $\vec k_T$ with the target. The latter is
characterized by the unintegrated gluon density $\unint(x,k_T^2)$.
Note that Eq.~(\ref{eq:kt}) is also valid for light flavors.

In the dipole approach, a mixed representation is employed, that
treats the longitudinal ($\gamma^*$) direction in momentum space,
but the two transverse directions are described in coordinate 
({\em i.e.} impact parameter) space. With help of the relation
\beq
\frac{1}{\kappa_\perp^2+\eps^2}=
\int\frac{d^2 r}{2\pi}\kzero(\eps r)\,\euler^{-\imag \vec\kappa_\perp\cdot
\vec r},
\eeq
where $\kzero$ is the zeroth order MacDonald function \cite{abramowitz},
one can Fourier transform Eq.~(\ref{eq:kt}) into impact parameter space,
\beqn\label{eq:ptdis}\nonumber
\frac{d\sigma\left(\gamma^*_Lp\to\{Q\bar Q\}X\right)}{d^2\kappa_\perp}&=&
\frac{1}{(2\pi)^2}\int d\alpha \int d^2{r}_1 d^2{r}_2 
\euler^{\imag\vec \kappa_\perp\cdot(\vec{r}_1-\vec{r}_2)}\\
\nonumber&\times&
\Psi_{\gamma^*\to Q\bar Q}(\alpha,\vec{r}_1)
\Psi^*_{\gamma^*\to Q\bar Q}(\alpha,\vec{r}_2)\\
&\times&
\frac{1}{2}\Bigl\{
\sigma_{q\bar q}({r}_1,x)
+\sigma_{q\bar q}({r}_2,x)
-\sigma_{q\bar q}(\vec{r}_1-\vec{r}_2,x).
\Bigr\}
\eeqn
Since the $\vec r_i$ are conjugate variables to $\vec\kappa_\perp$,
one can interpret $\vec r_1$ as the transverse size of the $Q\bar Q$-pair
in the amplitude and $\vec r_2$ as the size of the pair in the complex
conjugate amplitude. An expression similar to Eq.~(\ref{eq:ptdis})
was also obtained in Ref.~\cite{Mueller99}.

The light-cone (LC) wavefunctions for longitudinal ($L$)
and for transverse ($T$) photons
are given by
\beqn\label{eq:lcwfdis}
\Psi^L_{\gamma^*\to Q\bar Q}(\alpha,\vec{r}_1)
\Psi^{*L}_{\gamma^*\to Q\bar Q}(\alpha,\vec{r}_2)&=&
\frac{6\aem e_Q^2}{(2\pi)^2}4Q^2\alpha^2\left(1-\alpha\right)^2
\kzero(\eps{r}_1)\kzero(\eps{r}_2)\\
\nonumber
\Psi^T_{\gamma^*\to Q\bar Q}(\alpha,\vec{r}_1)
\Psi^{*T}_{\gamma^*\to Q\bar Q}(\alpha,\vec{r}_2)&=&
\frac{6\aem e_Q^2}{(2\pi)^2}\Biggl\{m_Q^2\kzero(\eps{r}_1)\kzero(\eps{r}_2)
\Biggr. \\
&+&\left.
\left[\alpha^2+\left(1-\alpha\right)^2\right]\eps^2
\frac{\vec{r}_1\cdot\vec{r}_2}{{r}_1{r}_2}
\kone(\eps{r}_1)\kone(\eps{r}_2)
\right\}.
 \eeqn
 The concept of LC wavefunction of a photon was first introduced in
Refs.~\cite{ks,bks}. These wavefunctions are simply the $\gamma^*\to
Q\bar Q$ vertex times the Feynman propagator for the quark line in
Fig.~\ref{fig:2graphs}, and can therefore be calculated in perturbation
theory.

The flavor independent dipole cross section $\sig$ in Eq.~(\ref{eq:ptdis})
carries all the information about the target.
It is related to the unintegrated gluon density by \cite{gBFKL}
\beq\label{eq:dipsec2}
\sig(r,x)=\frac{4\pi}{3}\int\frac{d^2k_T}{k_T^2}\as \unint(x,k_T)
\left\{1-\euler^{-\imag\vec k_T\cdot\vec r}\right\}.
\eeq
The color screening factor in the curly brackets in Eq.~(\ref{eq:dipsec2})
ensures that $\sig(r,x)$ vanishes $\propto r^2$ (modulo logs) at small
separations. This seminal property of the dipole cross section is 
known as {\em color transparency} \cite{ZKL,ct,bm}. The dipole cross
section cannot be calculated from first principles, but has to be determined
from experimental data, see sect.~\ref{sec-cross}. In principle,
the energy, {\em i.e.} $x$, dependence of $\sig$ could be calculated in
perturbative QCD. This has been attempted
in the generalized BFKL 
approach of Nikolaev and Zakharov, see {\em e.g.}
\cite{gBFKL}, by resumming higher orders in perturbation theory. 
However, the widely discussed next-to-leading order (NLO) 
corrections to the BFKL equation \cite{BFKL,nlo} has left the
theory of low-$x$ resummation in an unclear state. We shall account for
higher order effects by using a phenomenological parameterization
of the dipole cross section.

In the high energy limit, one can neglect the dependence of the gluon
momentum fraction $x$ on $\vec\kappa_\perp$ and integrate Eq.~(\ref{eq:ptdis})
over $\vec\kappa_\perp$ from $0$ to $\infty$. 
One then obtains a particularly simple formula
for the total cross section,
\beq\label{eq:totaldis}
\sigma_{\rm tot}\left(\gamma^*p\to\{Q\bar Q\}X\right)=
\sum_{T,L}\int d\alpha \int d^2{r}\left|
\Psi^{T,L}_{\gamma^*\to Q\bar Q}(\alpha,\vec{r}_1)\right|^2
\sigma_{q\bar q}({r},x).
\eeq
It was argued in Ref.~\cite{bialas} that the dipole formulation
is valid only in the leading $\log(x)$ approximation where 
Eq.~(\ref{eq:totaldis}) holds. Note however that Eq.~(\ref{eq:ptdis}) does
not rely on any high energy approximation and is exactly equivalent
the the $k_T$-factorized expression Eq.~(\ref{eq:kt}).

Eq.~(\ref{eq:totaldis}) has an illustrative interpretation, which is the 
key to calculating nuclear effects in the dipole approach:
The total cross section can be written in factorized form in impact
parameter space, because partonic configurations with fixed transverse 
separations are eigenstates of the interaction \cite{ZKL,mp}, {\em i.e.}
of the $T$ matrix restricted to diffractive processes.
Intuitively, the transverse size is frozen during the entire interaction 
because of time dilation.
In the dipole approach, the projectile is expanded in these eigenstates,
\beq
|\gamma^*\ra=\sum_kc_k^{\gamma^*}|\psi_k\ra,
\eeq
with
\beq
-\imag T|\psi_k\ra=\sigma_k|\psi_k\ra.
\eeq
Each eigenstate scatters independently off the target.
According to the optical theorem, the total cross section is then given by
\beq\label{eq:st}
\sigma_{\rm tot}={\rm Im}\la\gamma^* |T|\gamma^*\ra
=\sum_k|c_k^{\gamma^*}|^2 \sigma_k.
\eeq
Comparing this expression, Eq.~(\ref{eq:st}), with Eq.~(\ref{eq:kt}),
one can identify $\sig(r,x)$ as an eigenvalue of the $T$-matrix and
the coefficients $c_k^{\gamma^*}$ as LC wavefunctions. The 
summation over the index $k$ is
replaced by the integrals over $\alpha$ and $r$. 

Knowing the eigenstates of the interaction is a great advantage in
calculating multiple scattering effects in nuclear collisions. Note 
however that
the quark-antiquark pair is only the lowest Fock component of the virtual
photon. There are also higher Fock states containing gluons, which are
not taken into account by these simple considerations. These gluons 
cause the $x$-dependence of the dipole cross section and will
be included in a phenomenological way, see sect.~\ref{sec-cross}. In
addition, at lower energies color dipoles are no longer exact
eigenstates. A dipole of size $r_1$ may evolve into a dipole of a
different size $r_2$. This can be calculated from Eq.~(\ref{eq:ptdis}).

\section{The phenomenological dipole cross section}
\label{sec-cross}

The total cross sections for all hadrons and (virtual) photons are known
to rise with energy. It is obvious that 
the energy dependence cannot originate
from the hadronic wave functions, but
only from the dipole cross section. In the approximation of two-gluon
exchange used in \cite{ZKL} the dipole cross section is constant, the
energy dependence originates from higher order corrections related to
gluon radiation. 
Since no reliable way is known so far 
to sum up higher order corrections, especially in the semihard regime,
we resort to
phenomenology and employ a parameterization of $\sig(r,x)$. 

Few such parameterizations
are available in the literature, we choose two of them which are simple,  
but quite successful in describing data and denote them by the initials of
the authors as ``GBW'' \cite{GBW} and ``KST'' \cite{KST}.

We have
\BA
  \label{GBW}
  \mbox{``GBW'':}~~~~~~~~
  \sigma_{q\bar q}(\rT,x)&=&23.03\left[1-e^{-\rT^2/r_0^2(x)}\right]\mb\ ,\\
 r_0(x) &=& 0.4 \left(\frac{x}{x_0}\right)^{\!\!0.144} \fm\ ,
  \nonumber
\EA
where $x_0=3.04\cdot10^{-4}$. The proton structure function calculated with
this parameterization fits very well all available data at small $x$ and for
a wide range of $Q^2$ \cite{GBW}. However, it obviously fails describing the
hadronic total cross sections, since it never exceeds the value $23.03\mb$.
The $x$-dependence guarantees Bjorken scaling for DIS at high $Q^2$, however,
Bjorken $x$ is not a well defined quantity in the soft limit.

This problem as well as the difficulty with the definition of $x$ have been
fixed in \cite{KST}, where 
the dipole cross section is treated as a function
of the {\em c.m.} energy $\sqrt{s}$, rather than $x$, since $\sqrt{s}$ is more
appropriate for hadronic processes. A similarly simple form for the dipole
cross section is used
\BA
  \label{KST}
  \mbox{``KST'':}~~~~~~~~~
  \sigma_{{q\bar q}}(\rT,s) &=& \sigma_0(s) \left[1 - e^{-\rT^2/r_0^2(s)}
  \right]\ .~~~~~~
\EA
The values and energy dependence of hadronic cross sections is guaranteed   
by the choice of
\BA
  \sigma_0(s) &=& 23.6 \left(\frac{s}{s_0}\right)^{\!\!0.08}
  \left(1+\frac38 \frac{r_0^2(s)}{\left<r^2_{ch}\right>}\right)\mb\ ,\\   
  r_0(s)      &=& 0.88 \left(\frac{s}{s_0}\right)^{\!\!-0.14}  \fm\ .
\EA
The energy dependent radius $r_0(s)$ is fitted to data for the proton
structure function $F^p_2(x,Q^2)$, $s_0 = 1000\GeV^2$ and the mean square of
the pion charge radius $\left<r^2_{ch}\right>=0.44\fm^2$. The improvement at
large separations leads to a somewhat worse description of the proton structure
function at large $Q^2$. Apparently, the cross section dependent on energy,
rather than $x$, cannot provide Bjorken scaling. Indeed, the
parameterization
Eq.~(\ref{KST}) is successful only up to $Q^2\approx 10\GeV^2$.
   
In fact, the cases we are interested in, charmonium production and
interaction, are just in between the regions where either of these
parameterization is successful. Therefore, we suppose that the difference   
between predictions using Eqs.~\Ref{GBW} and \Ref{KST} is a measure of the  
theoretical uncertainty which fortunately turns out to be rather small.

%
\begin{figure}[t]
  \centerline{\scalebox{0.5}{\includegraphics{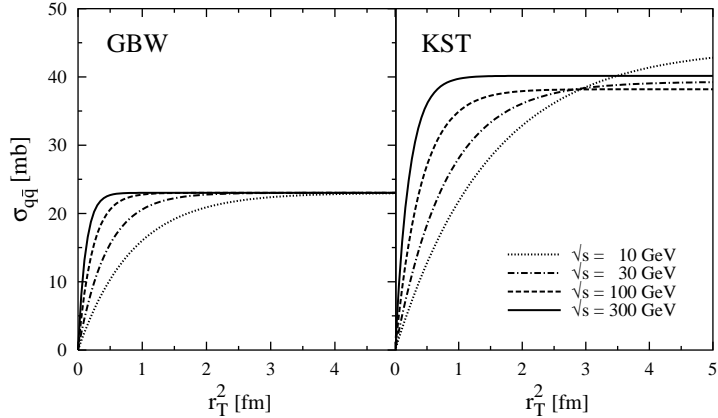}}}
{\caption{\em
      \label{Fig-dipole}
The dipole cross section as function of $r_T^2$ at energies
$\sqrt{s}=10,\ 30,\ 100$ and $300\ \GeV$ for GBW (left) and KST (right)
parameterizations.  In the left panel, we used
the prescription of \cite{Ryskin}, $x=(M^2_\psi+Q^2)/s$, where $M_\psi$ is
the charmonium mass. 
    }} 
\end{figure}

We demonstrate in Fig.~\ref{Fig-dipole}
a few examples of $r^2$-dependence of the dipole cross section
at different energies for both parameterization.
Both, GBW and KST cross section, 
vanish $\propto \rT^2$ at small $\rT$, but deviate  
considerably from this simple behavior at large separations.

Quite often, the simplest parameterization ($\propto\rT^2$) is used for
the dipole cross section. For the coefficient in front of $\rT^2$
we employ the expression obtained by the first term of the
Taylor expansion of
Eq.~\Ref{KST}:
\BA
  \label{rT2}
  \mbox{``$\rT^2$'':}~~~~~~~~~~~~~
  \sigma_{{q\bar q}}(\rT,s) &=& \frac{\sigma_0(s)}{r_0^2(s)} \cdot \rT^2 \ .
  ~~~~~~~~~~~~~~~~~~~
\EA
We shall refer to this form of the dipole cross section as 
$r^2$-approximation.

\section{Diffractive photoproduction of charmonia off protons}
\label{sec:cp}

The dynamics of production and interaction of charmonia has drawn attention
since their discovery back in 1974 \cite{jpsi}. 
As these heavy mesons have a small size
it has been expected that hadronic cross sections may be calculated relying
on perturbative QCD. The study of charmonium production became even more
intense after charmonium suppression had been suggested as a probe for the
creation and interaction of quark-gluon plasma in relativistic heavy ion
collisions \cite{Satz}.

Since we will never have direct experimental information on charmonium-nucleon
total cross sections one has to extract it from other data for example from
elastic photoproduction of charmonia $\gamma p \to \Jpsi(\psi')\ p\,$. The
widespread believe that one can rely on the vector dominance model (VDM)
is based on previous experience with photoproduction of $\rho$ mesons.
However, even a dispersion approach shows that this is quite a risky way,
because the $\Jpsi$ pole in the complex $Q^2$ plane is nearly 20 times
farther away from the physical region than the $\rho$ pole. The multichannel
analysis performed in \cite{HK} demonstrates that the corrections are huge,
$\sigma^{\Jpsi\,p}_{tot}$ turns out to be more that three times larger than
the VDM prediction. Unfortunately, more exact predictions of the multichannel
approach, especially for $\psi'$, need knowledge of many diagonal and 
off-diagonal amplitudes which are easily summed only if one uses the 
oversimplified oscillator wave functions and a $q\bar q$-proton cross 
section of the form $\sigma_{q\bar q}(\rT)\propto\rT^2$, where $\rT$ is
the transverse $q\bar q$ separation.

Instead, one may switch to the quark basis, which should be equivalent to
the hadronic basis because of completeness. In this representation the
procedure of extracting $\sigma^{\Jpsi\,p}_{tot}$ from photoproduction data
cannot be realized directly, but has to be replaced by a different strategy.
Namely, as soon as one has expressions for the wave functions of charmonia and
the universal dipole cross section $\sigma_{q\bar q}(\rT,s)$, one can predict
both, the experimentally known charmonium photoproduction cross sections and
the unknown $\sigma^{\Jpsi(\psi')\,p}_{tot}$. If the photoproduction data
are well described one may have some confidence in the predictions for the
$\sigma^{\Jpsi(\psi') p}_{tot}$. Of course this procedure will be model
dependent, but we believe that this is the best use of photoproduction
data one can presently make. 
This program was performed for the first time in \cite{KZ}. 
We do not propose a conceptually new scheme here,
but calculate within a given approach as accurately as possible and
without any free parameters. Wherever there is room for arbitrariness, like
forms for the color dipole cross section and those for for charmonium wave
functions, we use and compare other author's proposals, which have been tested
on data different from those used here.

\begin{figure}[t]
  \centerline{
  \scalebox{0.33}{\includegraphics{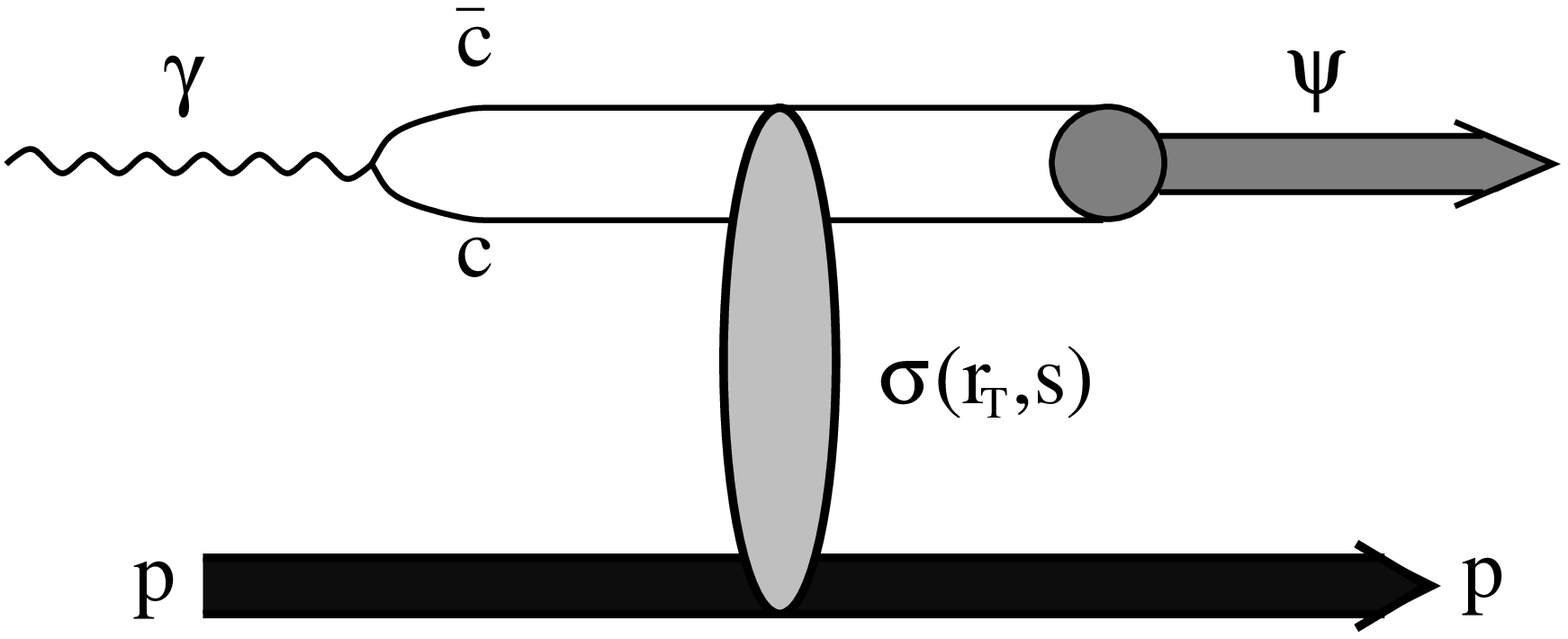}}~~
  \scalebox{0.33}{\includegraphics{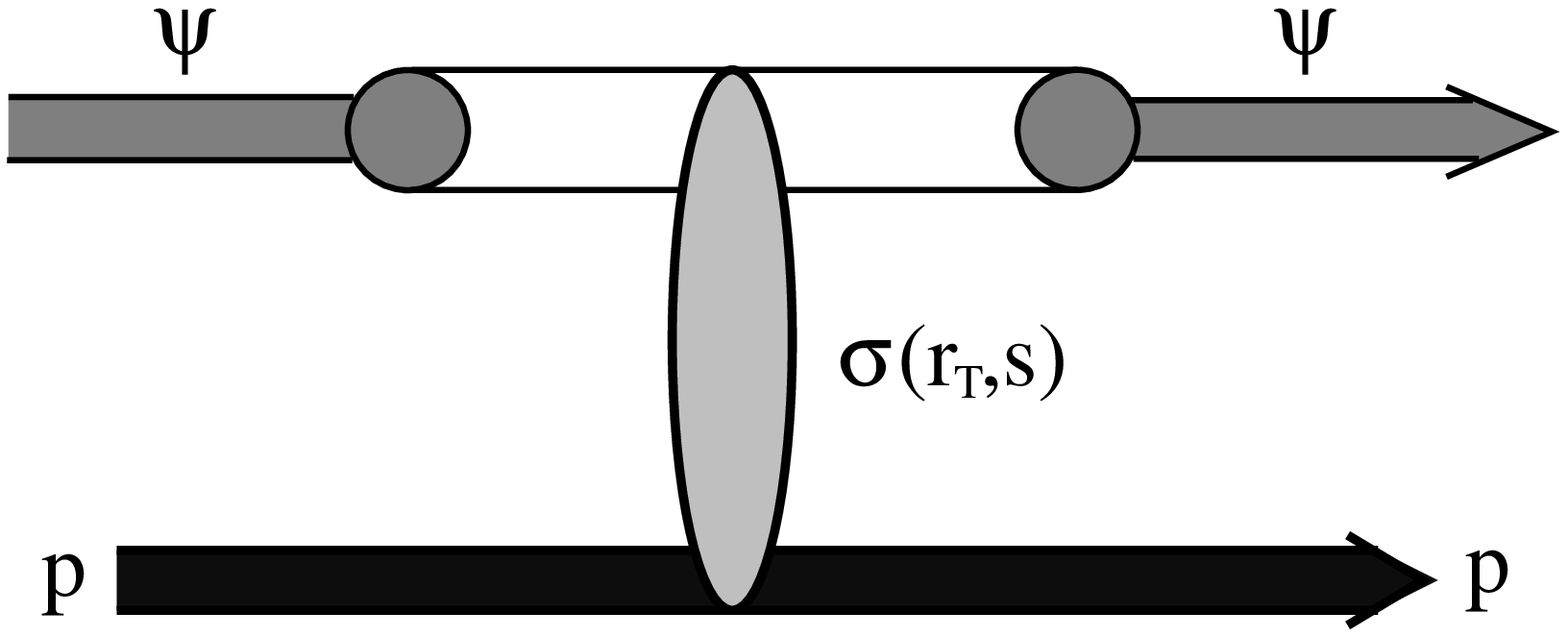}}}
\center{
{\caption{\em
  \label{Fig-D}
  Schematic representation of the amplitudes for the reactions $\gamma^*p%
  \to \psi p$ (left) and $\psi\,p$ elastic scattering (right) in the rest
  frame of the proton. The $c\bar c$ fluctuation of the photon and the $\psi$
  with transverse separation $\rT$ and c.m. energy $\sqrt{s}$ interact with
  the target proton via the cross section $\sigma(\rT,s)$ and produce
  a $\Jpsi$ or $\psi'$. }  
    }  }
\end{figure}

In the light-cone 
dipole approach the two processes, photoproduction and charmonium-nucleon
elastic scattering look as shown in Fig.~\ref{Fig-D} \cite{KZ}.
The corresponding expressions for the forward amplitudes read 
\BA
  \label{Mgam}\label{sigma-gp}
  {\cal M}_{\gamma^* p}(s,Q^2) &=& \sum_{\mu,\bar\mu}
   \,\int\limits_0^1 \!\!d\alpha \int d^2\brT
   \,\Phi^{\!*(\mu,\bar\mu)}_{\psi}(\alpha,\brT) \,\sigma_{q\bar q}(\rT,s)
   \,\Phi^{(\mu,\bar\mu)}_{\gamma^*}(\alpha,\brT,Q^2) \ ,\\
  \label{Mpsi}
  {\cal M}_{\psi\,p}(s)    &=& \sum_{\mu,\bar\mu}
    \,\int\limits_0^1 \!\!d\alpha \int d^2\brT 
    \,\Phi^{\!*(\mu,\bar\mu)}_{\psi}(\alpha,\brT) \,\sigma_{q\bar q}(\rT,s)
    \,\Phi^{(\mu,\bar\mu)}_{\psi}(\alpha,\brT) \ .
\EA
Here the summation runs over spin indexes $\mu$, $\bar\mu$ of the $c$ and
$\bar c$ quarks, $Q^2$ is the photon virtuality, $\Phi_{\gamma^*}(\alpha,%
\rT,Q^2)$ is the light-cone distribution function of the photon for a
$c\bar c$ fluctuation of separation $\rT$ and relative fraction $\alpha$ of
the photon light-cone momentum carried by $c$ or $\bar c$. Correspondingly,
$\Phi_{\psi}(\alpha,\brT)$ is the light-cone wave function of $\Jpsi$,
$\psi'$ and $\chi$ (only in Eq.~\ref{Mpsi}). The dipole cross section
$\sigma_{q\bar q}(\rT,s)$ mediates the transition ({\it cf\/}
Fig.~\ref{Fig-D}).

The light cone variable describing longitudinal motion which is invariant
to Lorentz boosts is the fraction $\alpha=p_c^+/p_{\gamma^*}^+$ of the
photon light-cone momentum $p_{\gamma^*}^+ = E_{\gamma^*}+p_{\gamma^*}$
carried by the quark or antiquark. In the nonrelativistic approximation
(assuming no relative motion of $c$ and $\bar c$) $\alpha=1/2$ (e.g.
\cite{KZ}), otherwise one should integrate over $\alpha$ (see Eq.~\Ref{Mgam}).
For transversely ($T$) and longitudinally ($L$) polarized photons  
the perturbative photon-quark distribution function in Eq.~\Ref{Mgam} 
reads \cite{ks,bks},
\BE
  \label{psi-g}
  \Phi_{T,L}^{(\mu,\bar\mu)}(\alpha,\brT,Q^2) =
    \frac{\sqrt{N_c\,\alpha_{em}}}{2\,\pi}\,Z_c
    \,\chi_c^{\mu\dagger}\,\widehat O_{T,L}
    \,\widetilde\chi_{\bar c}^{\bar\mu}\,\kzero(\eps\rT) \ ,
\EE
where 
\BE
  \label{tildechi}
  \widetilde\chi_{\bar c} = i\,\sigma_y\,\chi^*_{\bar c}\ ;
\EE
$\chi$ and $\bar\chi$ are the spinors of the $c$-quark and antiquark
respectively; $Z_c=2/3$. $K_0(\epsilon\rT)$ is the modified Bessel 
function with
\BE
  \label{eps-Q}
  \eps^2 = \alpha(1-\alpha)Q^2 + m_c^2\ .
\EE
The operators $\widehat O_{T,L}$ have the form:
\BA
  \widehat O_{T} &=& m_c \, \vec\sigma\cdot\vec e_\gamma
    + i(1-2\alpha)
      \,(\vec\sigma \cdot \vec n)
      \,(\vec e_\gamma \cdot \vec\nabla_{\rT})
    + (\vec n \times \vec e_\gamma)\cdot\vec\nabla_{\rT}\ ,\\
  \widehat O_{L} &=& 2\,Q\,\alpha(1-\alpha)\,\vec\sigma\cdot\vec n\ ,
\EA
where $\vec n=\vec p/p$ is a unit vector parallel to the photon momentum
and $\vec e$ is the polarization vector of the photon. Effects of the
non-perturbative interaction within the $q\bar q$ fluctuation are 
negligible for the heavy charmed quarks.

\subsection{Charmonium wave functions}
\label{sec-wave}

The spatial part of the $c\bar c$ pair wave function satisfying the 
Schr\"odinger equation
\BE
  \label{Schroed}
  \left(-\,\frac{\Delta}{m_c}+V(R)\right)
   \,\Psi_{nlm}(\vr)=E_{nl}
   \,\Psi_{nlm}(\vr)
\EE
is represented in the form
\BE
  \label{wf}
  \Psi(\vr) = \Psi_{nl}(R) \cdot Y_{lm}(\theta,\varphi) \ ,
\EE
where $\vr$ is 3-dimensional $c\bar c$ separation
(not to be confused with the 2-dimen\-sional argument $\vec r$ of the
dipole cross section), $\Psi_{nl}(R)$ and
$Y_{lm}(\theta,\varphi)$ are the radial and orbital parts of the wave
function. The equation for radial $\Psi(R)$ is solved with the help
of the program \cite{Lucha}. The following four potentials $V(R)$ 
have been used:

\begin{itemize}
\item ``COR'': Cornell potential \cite{COR},
  \BE
    \label{COR}
    V(r) = -\frac{k}{R} + \frac{R}{a^2}
  \EE
  with $k=0.52$, $a=2.34\GeV^{-1}$ and $m_c=1.84\GeV$.
\item ``BT'': Potential suggested by Buchm\"uller and Tye \cite{BT}
  with $m_c=1.48\GeV$. It has a similar structure as the Cornell
  potential: linear string potential at large separations and 
  Coulomb shape at short distances with some refinements, however.
\item ``LOG'': Logarithmic potential \cite{LOG}
  \BE
    \label{LOG}
    V(R) = -0.6635\GeV + (0.733\GeV) \log(R \cdot 1\GeV)
  \EE
  with $m_c=1.5\GeV$.
\item ``POW'': Power-law potential \cite{POW}
  \BE
    \label{POW}
    V(R) = -8.064\GeV + (6.898\GeV) (R \cdot 1\GeV)^{0.1}
  \EE
  with $m_c=1.8\GeV$.
\end{itemize}

\begin{figure}[t]
  \centerline{
  \scalebox{0.33}{\includegraphics{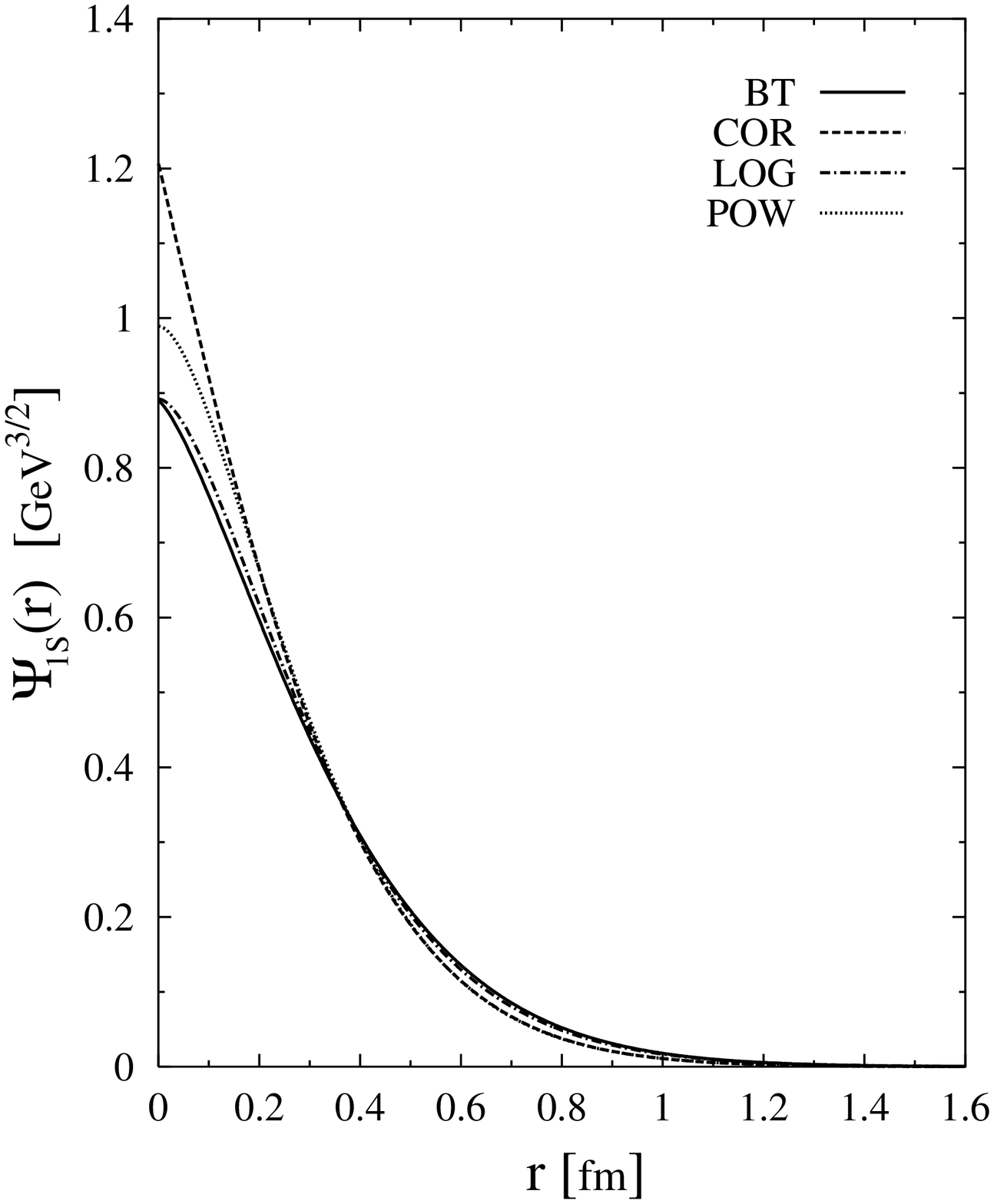}}~~
  \scalebox{0.33}{\includegraphics{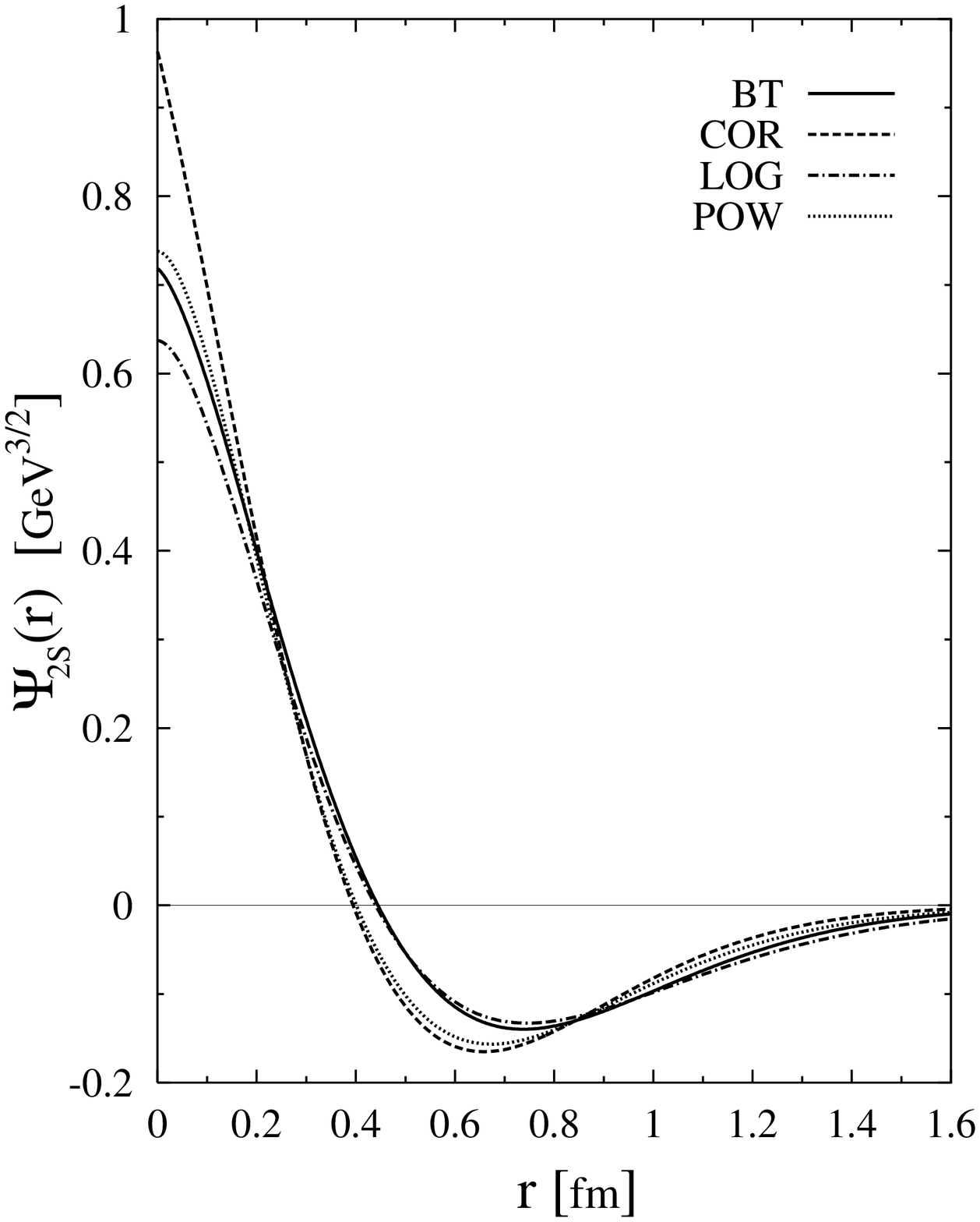}}}
\center{
{\caption{\em\label{Fig-y}
  The radial part of the wave function $\Psi_{nl}(r)$ for the $1S$ and
  $2S$ states calculated with four different potentials (see text).
In this figure, $r=R$. }  
    }  }
\end{figure}

The results of calculations for the radial part $\Psi_{nl}(R)$ of the $1S$
and $2S$ states are depicted in Fig.~\ref{Fig-y}. For the ground state all
the potentials provide a very similar behavior for $R>0.3\fm$, while for 
small $R$ the predictions are differ by up to $30\%$. The peculiar property
of the $2S$ state wave function is the node at $R\approx 0.4\fm$ which causes
strong cancelations in the matrix elements Eq.~\Ref{Mgam} and as a result,
a suppression of photoproduction of $\psi'$ relative to $\Jpsi$
\cite{KZ,Benhar}.

Note that the lowest Fock component $|c\bar c\ra$ in the
infinite momentum frame is not related by simple Lorentz boost to the
wave function of charmonium in the rest frame. This makes the problem
of building the light-cone wave function for the lowest $|c\bar c\ra$
component difficult, no unambiguous solution is yet known. There are
only recipes in the literature, a simple one widely used \cite{Terent'ev},
is the following. One applies a Fourier transformation from coordinate to
momentum space to the known spatial part of the non-relativistic wave
function \Ref{wf}, $\Psi(\vr)\Rightarrow\Psi(\vp)$, which can be written
as a function of the effective mass of the $c\bar c$, $M^2=4(p^2+m_c^2)$,
expressed in terms of light-cone variables
\BE
  M^2(\alpha,\pT) = \frac{\pT^2+m_c^2}{\alpha(1-\alpha)}\ .
\EE 
In order to change the integration variable $p_L$ to the light-cone variable
$\alpha$ one relates them through $M$, namely $p_L=(\alpha-1/2)M(p_T,\alpha)$.
In this way the $c\bar c$ wave function acquires a kinematical factor
\BE
  \label{lc-wf-p}
  \Psi(\vp) \Rightarrow
  \sqrt{2}\,\frac{(p^2+m_c^2)^{3/4}}{(\pT^2+m_c^2)^{1/2}}
  \cdot \Psi(\alpha,\bpT)
  \equiv \Phi_\psi(\alpha,\bpT) \ .
\EE

This procedure is used in \cite{Hoyer} and the result is applied to 
calculation of the amplitudes \Ref{Mgam}. The result is discouraging,
since the $\psi'$ to $\Jpsi$ ratio of the photoproduction cross sections
are far too low in comparison with data. However, the oversimplified
dipole cross section $\sigma_{q\bar q}(\rT)\propto\rT^2$ has been used,
and what is even more essential, the important ingredient of Lorentz
transformations, the Melosh spin rotation, has been left out. The spin
transformation has also been left out in the recent publication \cite{Suzuki}
which repeats the calculations of \cite{Hoyer} with a more realistic
dipole cross section which levels off at large separations. This leads
to suppression of the node-effect (less cancelation) and enhancement
of $\Psi'$ photoproduction. Nevertheless, the calculated $\psi'$ to
$\Jpsi$ ratio is smaller than the data by a factor of two.

The 2-spinors $\chi_c$ and $\chi_{\bar c}$ describing $c$
and $\bar c$ respectively in the infinite momentum frame are known to be
related by Melosh rotation \cite{Melosh,Terent'ev} to the spinors 
$\bar\chi_c$ and $\bar\chi_{\bar c}$ in the rest frame:
\BA
  \nonumber
  \bf\overline{\chi}_c        &=& \widehat R(  \alpha, \bpT)\,\chi_c\ ,\\
  \bf\overline{\chi}_{\bar c} &=& \widehat R(1-\alpha,-\bpT)\,\chi_{\bar c}\ ,
  \label{Melosh}
\EA
where the matrix $\widehat R(\alpha,\bpT)$ has the form:
\BE
  \widehat R(\alpha,\bpT) = 
    \frac{  m_c+\alpha\,M - i\,[\vec\sigma \times \vec n]\,\bpT}
    {\sqrt{(m_c+\alpha\,M)^2+\pT^2}} \ .
\label{matrix}
\EE

Since the potentials we use in section~\ref{sec-wave} contain no spin-orbit
term, the $c\bar c$ pair is in $S$-wave. In this case spatial and spin 
dependences in the wave function factorize and we arrive at the following
light cone wave function of the $c\bar c$ in the infinite momentum frame
\BE
  \label{eq:lc-wf}
  \Phi^{(\mu,\bar\mu)}_\psi(\alpha,\bpT) =
     U^{(\mu,\bar\mu)}(\alpha,\bpT)\cdot\Phi_\psi(\alpha,\bpT)\ ,
\EE
where 
\BE
  U^{(\mu,\bar\mu)}(\alpha,\bpT) = 
    \chi_{c}^{\mu\dagger}\,\widehat R^{\dagger}(\alpha,\bpT)
    \,\vec\sigma\cdot\vec e_\psi\,\sigma_y
    \,\widehat R^*(1-\alpha,-\bpT)
    \,\sigma_y^{-1}\,\widetilde\chi_{\bar c}^{\bar\mu}
\EE
and $\widetilde\chi_{\bar c}$ is defined in \Ref{tildechi}.

Note that the wave function \Ref{eq:lc-wf} is different from one used in
\cite{Ryskin93,fs,Nemchik} where it was assumed that the vertex
$\psi\to c\bar c$ has the structure $\psi_{\mu}\,\bar u\,\gamma_{\mu}\,u$
like the for the photon $\gamma^*\to c\bar c$. The rest frame wave function
corresponding to such a vertex contains $S$ wave and $D$ wave. The weight
of the latter is dictated by the structure of the vertex and cannot be
justified by any reasonable nonrelativistic potential model for the
$c\bar c$ interaction.

Now we can determine the light-cone wave function in the mixed
longitudinal momentum - transverse coordinate representation:
\BE
  \label{lc-wf-r}
  \Phi^{(\mu,\bar\mu)}_\psi(\alpha,\brT) =
    \frac{1}{2\,\pi} 
    \int d^2\bpT\,e^{-i\bpT\brT}\,
    \Phi^{(\mu,\bar\mu)}_\psi(\alpha,\bpT)\ .
\EE

\subsection{Comparison with data}\label{data}

\begin{figure}[t]
  \centerline{
  \scalebox{0.7}{\includegraphics{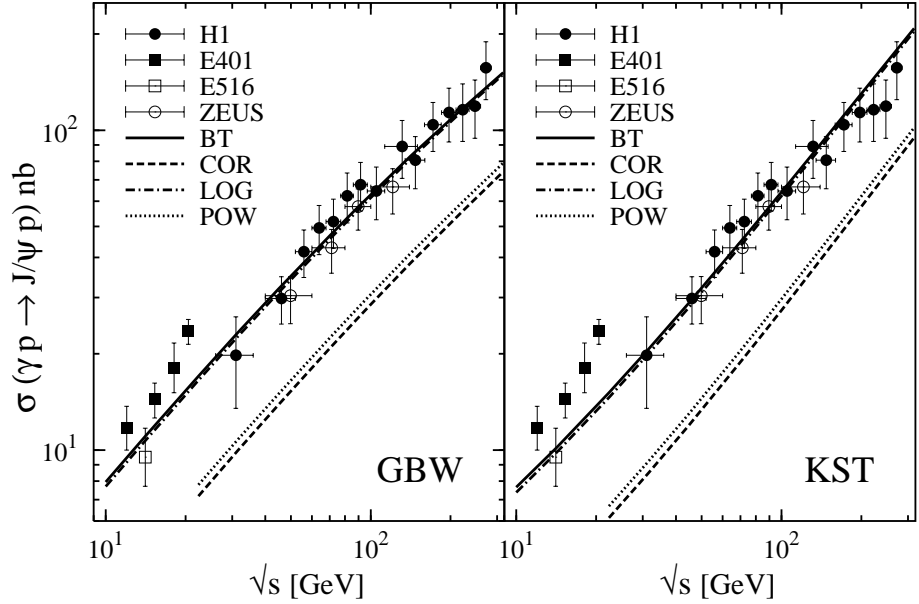}}}
\center{
{\caption{\em
  \label{Fig-s}
  Integrated cross section for elastic photoproduction $\gamma\,p%
  \rightarrow J/\psi\,p$ with real photons ($Q^2=0$) as a function
  of the energy calculated with GBW and KST dipole cross sections
  and for four potentials to generate $\Jpsi$ wave functions.
  Experimental data points from the H1~\cite{H1-s}, E401~\cite{E401-s},
  E516~\cite{E516-s} and ZEUS~\cite{ZEUS-s} experiments. }  
    }  }
\end{figure}

Having the light-cone wave function of charmonium, 
we are now in the position to calculate the cross section of charmonium
photoproduction using Eq.~(\ref{sigma-gp}).
The results for $J/\psi$ are compared with the data in
Fig.~\ref{Fig-s}. Calculations are performed with GBW and KST parameterizations
for the dipole cross section and for wave functions of the $\Jpsi$ calculated
from BT, LOG, COR and POW potentials. One observes
\begin{itemize}
 \item There are no major differences between different parameterizations
\cite{GBW,KST} of the dipole cross section.
 \item The use of different potentials to generate the wave functions
      of the $\Jpsi$ leads to two distinctly different behaviors. The
      potentials labeled BT and LOG (see sect. \ref{sec-wave}) describe
      the data very well, while the potentials COR and LOG underestimate
      them by a factor of two. The different behavior has been traced
      to the following origin: BT and LOG use $m_c \approx 1.5\GeV$,
      but COR and POW $m_c \approx 1.8\GeV$. While the bound state
      wave functions of $\Jpsi$ are little affected by this difference
      (see Fig.~\ref{Fig-y}), the photon wave function Eq.~\Ref{psi-g}
      depends sensitively on $m_c$ via the argument Eq.~\Ref{eps-Q} of
      the $\kzero$ function.
\end{itemize}

\subsection{Importance of spin effects for the $\psi'$ to $\Jpsi$
ratio}\label{ratio}

It turns out that the effects
of spin rotation have a gross impact on the cross section of elastic
photoproduction $\gamma\,p \to \Jpsi(\psi^\prime)p\,$. To demonstrate these
effects we present the results of our calculations at
$\sqrt{s}=90\,\GeV$ in Table~\ref{Tab-sR}.
\BT
\begin{center}
\begin{tabular}{|ll|c|c|c|c|}
\hline
\vphantom{\Bigg\vert}
  &
  & BT
  & LOG
  & COR
  & POW \\
\hline &&&&&\\[-3mm]
\framebox{$\sigma$} 
&
  GBW     & 52.01~(37.77) & 50.78~(36.63) & 23.13~(17.07) & 24.94~(18.64)\\[0.5mm]
& KST     & 49.96~(35.87) & 48.49~(34.57) & 21.05~(15.42) & 22.83~(16.92)\\[0.5mm]
& $\rT^2$ & 66.67~(47.00) & 64.07~(44.86) & 25.81~(18.71) & 28.23~(20.66)\\[1.5mm]
\hline &&&&&\\[-3mm]
\framebox{${\cal R}$} &
  GBW     & 0.147~(0.075) & 0.117~(0.060) & 0.168~(0.099) & 0.144~(0.085)\\[0.5mm]
& KST     & 0.147~(0.068) & 0.118~(0.054) & 0.178~(0.099) & 0.152~(0.084)\\[0.5mm]
& $\rT^2$ & 0.101~(0.034) & 0.081~(0.027) & 0.144~(0.070) & 0.121~(0.058)\\[1.5mm]
\hline
\end{tabular}
\end{center}
\center{\caption{\em
  \label{Tab-sR}
  The photoproduction $\gamma\,p \to \Jpsi\,p\,$ cross-section
  $\sigma(\Jpsi)$ in nb and the ratio ${\cal R}=\sigma(\psi')/\sigma(\Jpsi)$
  for the four different types of potentials (BT, LOG, COR, POW) and
  the three parameterizations (GBW, KST, $\rT^2$) for the dipole cross
  section $\sigma(\rT,s)$ at $\sqrt{s}=90\GeV$. The values in parentheses
  correspond to the case when the spin rotation is neglected. See
Ref.~\cite{HIKT} for a comparison with data.
}}
\ET
The upper half of the table shows the photoproduction cross sections for
$\Jpsi$ for different parameterizations of the dipole cross section (GBW,
KST, ``$\rT^2$'') and potentials (BT, COR, LOG, POW). The numbers in
parenthesis show what the cross section would be, if the spin rotation
effects were neglected. We see that these effects add 30-40\% to the
$\Jpsi$ photoproduction cross section. 

The spin rotation effects turn out to have a much more dramatic impact on
$\psi'$ increasing the photoproduction cross section by a factor 2-3. This
is visible in the lower half of the table which shows the ratio 
${\cal R}=\sigma%
(\psi')/\sigma(\Jpsi)$ of photoproduction cross sections, where the number
in parenthesis correspond to no spin rotation effects included. This spin
effects explain the large values of the ratio 
${\cal R}$ observed experimentally.
Our results for 
${\cal R}$ are about twice as large as evaluated in \cite{Suzuki}
and even more than in \cite{Hoyer}.

\subsection{Charmonium-nucleon total cross sections}\label{psi-n}

\begin{figure}[t]
  \centerline{
  \scalebox{0.7}{\includegraphics{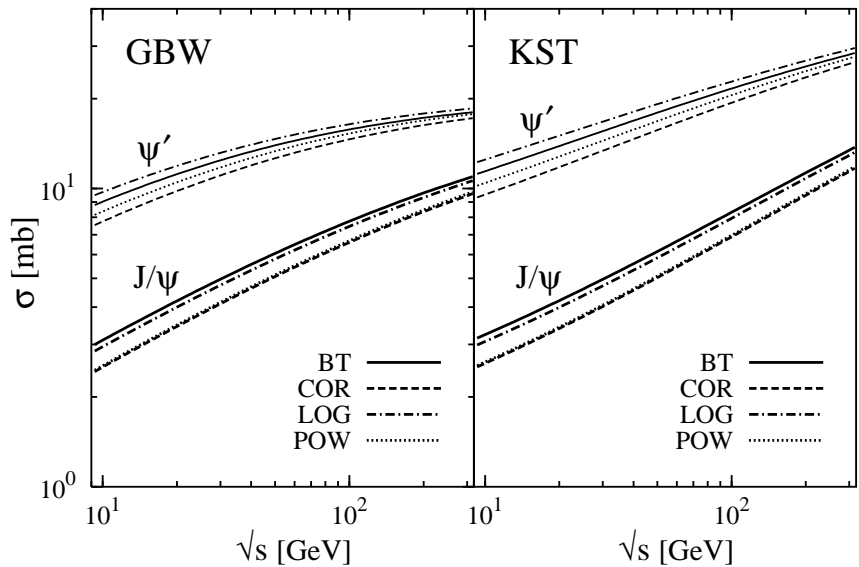}}}
\center{
{\caption{\em\label{Fig-S}
  Total $\Jpsi\,p$ (thick curves) and $\psi'\,p$ (thin curves)
  cross sections with the GBW and KST parameterizations for the
  dipole cross section. }  
    }  }
\end{figure}

After the light-cone formalism has been checked with the data for virtual
photoproduction we are in position to provide reliable predictions for
charmonium-nucleon total cross sections. The corresponding expressions are
given by Eq.~\Ref{Mpsi}) (compare with \cite{ZKL}). 
The calculated $\Jpsi$- and 
$\psi'$-nucleon total cross sections 
are plotted in Fig.~\ref{Fig-S} for for the GBW
and KST forms of the dipole cross sections and all four types of
the charmonium potentials. 

\section{Nuclear effects in exclusive leptoproduction of charmonia}

Charmonium production on nuclei can be exclusive, $\gamma^*A \to \Psi X$,
where $X=A$ (coherent) or $X=A^*$ (incoherent), and inclusive when $X$
includes pions. We skip the latter which is discussed in \cite{hkz} and
concentrate here on exclusive processes. In this case the following
phenomena are to be expected: color filtering, {\em i.e.} inelastic
interactions of the $c\bar c$ pair on its way through the nucleus is
expected to lead to a suppression of $\Psi$ production relative to $A
\sigma_{\gamma^*p\to\Psi p}$. Since the dipole cross section $\sqq$ also
depends on the gluon distribution in the target ($p$ of $A$), nuclear
shadowing of the gluon distribution is expected to reduce $\sqq$ in a
nuclear reaction relative to the one on the proton. Production of a
$c\bar c$ pair in a nucleus and its absorption are also determined by the
values of the coherence length $l_c$ and the formation length $l_f$
\cite{bm}.

Explicit calculations have been performed in Ref.~\cite{KZ} in the
approximation of a short coherence (or production) length, when one can
treat the creation of the colorless $c\bar c$ pair as instantaneous,
 \BE
  l_c = \frac{2\,\nu}{M_{c\bar c}^2} \approx 
        \frac{2\,\nu}{M_{\Jpsi}^2}\ \ll R_A,
  \label{10}
\EE
where $\nu$ is the energy of the virtual photon in the rest frame of the
nucleus. At the same time, the formation length may be long, comparable
with the nuclear radius $R_A$,
\BE
  l_f = \frac{2\,\nu}{M_{\psi'}^2 - M_{\Jpsi}^2} \sim R_A\ .
  \label{20}
\EE
In Ref.~\cite{KZ} the wave function formation is described by means of the
light-cone Green function approach summing up all possible paths of the
$c\bar c$ in the nucleus. The result has been unexpected. Contrary to
naive expectation, based on the larger size of the $\psi'$ compared to
$\Jpsi$, it has been found that $\psi'$ is not more strongly absorbed
than the $\Jpsi$, but may even be enhanced by the nuclear medium. This
is interpreted as an effect of filtering which is easy to understand
in the limit of long coherence length, $l_c\gg R_A$.
Indeed, the production rate of $\psi'$ on a proton target is small due
to strong cancelations in the projection of the produced $c\bar c$ wave
packet onto the radial wave function of the $\psi'$ which has a node.
After propagation through nuclear matter the transverse size of a
$c\bar c$ wave packet is squeezed by absorption and the projection of
the $\psi'$ wave function is enhanced \cite{KZ,Benhar} since the effect
of the node is reduced (see another manifestation of the node in 
\cite{brodsky1}).

However, the quantitative predictions of \cite{KZ} are not trustable
since the calculations have been oversimplified and quite some progress
has been made on the form of the dipole cross section $\sqq$ and the 
light cone wave functions for the charmonia. Therefore we take the 
problem up again and provide more realistic calculations for nuclear
effects in exclusive electroproduction of charmonia off nuclei relying on the
successful parameter free calculations which have been performed recently
in Ref.~\cite{HIKT} for elastic virtual photoproduction of
charmonia, $\gamma^*\,p \to \Psi\,p$ (see sect.~\ref{sec:cp}). 

Whenever one deals with high-energy reactions on nuclei, one cannot avoid
another problem of great importance: gluon shadowing. At small values of
$x$, gluon clouds overlap in longitudinal direction and may fuse. As a
result, the gluon density per one nucleon in a nucleus is expected to be
reduced compared to a free proton. Parton shadowing, which leads to an
additional nuclear suppression in various hard reactions (DIS, DY, heavy
flavor, high-$p_T$ hadrons, etc.) may be especially strong for exclusive
vector meson production like charmonium production which needs at least two
gluon exchange. Unfortunately, we have no experimental information for gluon
shadowing in nuclei so far, and we have to rely on the available theoretical
estimates, see {\em e.g.} Refs.~\cite{nestor,Mueller99,KST,KRTJ}.

\subsection{ Eikonal shadowing versus absorption for \boldmath$c\bar c$
pairs in nuclei \label{section-nuclei} }

Exclusive charmonium production off nuclei, $\gamma^* A \to \Psi X$ is
called coherent, when the nucleus remains intact, i.e. $X=A$, or
incoherent, when $X$ is an excited nuclear state which contains nucleons
and nuclear fragments but no other hadrons. The cross sections depend on
the polarization $\epsilon$ of the virtual photon (in all figures below
we will imply $\epsilon=1$),
\BE
  \sigma^{\gamma^*A}(s,Q^2) =
  \sigma^{\gamma^*_TA}(s,Q^2) + \epsilon\,
  \sigma^{\gamma^*_LA}(s,Q^2)~,
  \label{30}
\EE
where the indexes $T,L$ correspond to transversely or longitudinally 
polarized photons, respectively.

The cross section for exclusive production of charmonia off a nucleon 
target integrated over momentum transfer \cite{ZKL} is given by
\BE
  \sigma^{\gamma_{T,L}^*N}_{inc}(s,Q^2) =
  \left|\left\la\Psi\left|\sqq(r,s)
  \right|\gamma^{T,L}_{c\bar c}
  \right\ra\right|^2~,
  \label{35} 
\EE 
where $\Psi(\vec {r},\alpha)$ is the charmonium LC wave function which
depends on the transverse $c\bar c$ separation $\vec {r}$ and on the
relative sharing $\alpha$ of longitudinal momentum \cite{HIKT}. Both
variables are involved in the integration in the matrix element
Eq.~\Ref{35}. $\Psi(\vec {r},\alpha)$ is obtained by means of a Lorentz
boost applied the solutions of the Schr\"odinger equation. This procedure
involves the Melosh spin rotation \cite{Terent'ev,Melosh} which produces
sizable effects.
In addition,
$\gamma^{T,L}_{c\bar c}(\vec {r},\alpha,Q^2)$ is the
LC wave function of the $c\bar c$ Fock component of the photon. It
depends on the photon virtuality $Q^2$.  One can find the details in
Ref.~\cite{HIKT} including the effects of a nonperturbative $q\bar q$
interaction.

The cross sections for coherent and incoherent production on nuclei will
be derived under various conditions imposed by the coherence length
Eq.~(\ref{10}). At high energies the coherence length Eq.~\Ref{10} may
substantially exceed the nuclear radius. In this case the transverse size
of the $c\bar c$ wave packet is ``frozen'' by Lorentz time dilation, 
{\em i.e.}
it does not fluctuate during propagation through the nucleus, and the
expressions for the cross sections, incoherent ($inc$) or coherent
($coh$), are particularly simple \cite{KZ},
 \beq
  \sigma^{\gamma_{T,L}^*A}_{inc}(s,Q^2) =
  \int d^2b\,T_A(b)\,
  \left|\left\la\Psi\left|\sqq({r},s)\,
  \exp\left[ -{1\over2}\, \sqq({r},s)\,T_A(b)\right]
  \right|\gamma^{T,L}_{c\bar c}
  \right\ra\right|^2
  \label{40}\\
 \eeq
 \beq
  \sigma^{\gamma_{T,L}^*A}_{coh}(s,Q^2) =
  \int d^2b\,\left|\left\la\Psi\left|1\,-\,
  \exp\left[-{1\over2}\,\sqq({r},s)\,T_A(b)\right]
  \right|\gamma^{T,L}_{c\bar c}
  \right\ra\right|^2\ .
  \label{50}
 \eeq
Here $T_A(b)=\int_{-\infty}^{\infty}dz\,\rho_A(b,z)$ is the nuclear
thickness function given by the integral of the nuclear density along the
trajectory at a given impact parameter $b$. 

\begin{figure}[t]
  \centerline{
  \scalebox{0.65}{\includegraphics{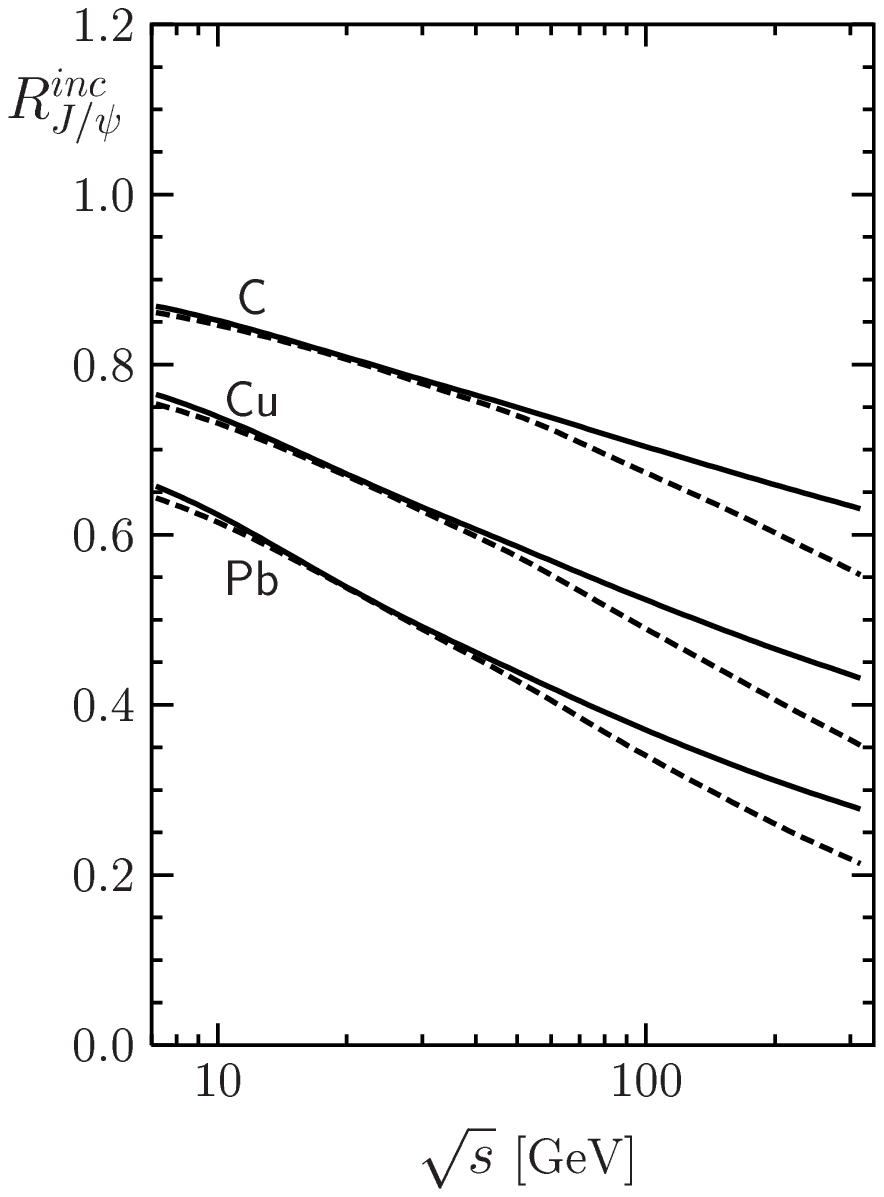}}~~
  \scalebox{0.65}{\includegraphics{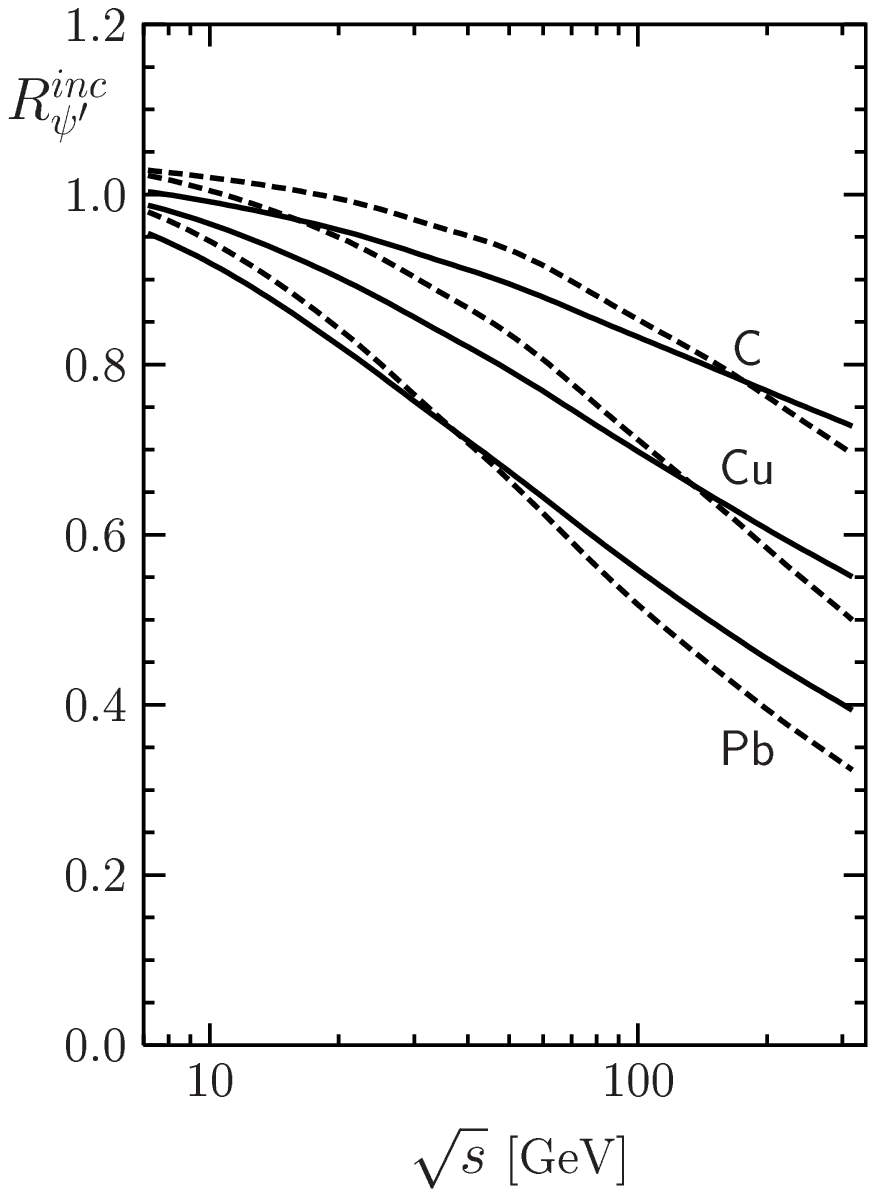}}}
\center{
{\caption{\em\label{s-inc}
  Ratios $R^{inc}_\Psi$ for $\Jpsi$ and $\psi'$ incoherent production
  on carbon, copper and lead as function of $\sqrt s$ and at $Q^2=0$.
  The solid curves refer to the GBW parameterization of $\sqq$ and
  dashed one refer to the KST parameterization. }  
    }  }
\end{figure}

The nuclear suppression
ratio for incoherent electroproduction of $\Jpsi$ and
$\psi'$ is shown in Fig.~\ref{s-inc} as a function of $\sqrt{s}$. We use
the GBW \cite{GBW} and KST \cite{KST} parameterizations for the dipole
cross section and show the results by solid and dashed curves,
respectively. Differences are at most $10-20\,\%$.
Analyzing the results shown in Fig.~\ref{s-inc}, we observe that nuclear
suppression of $\Jpsi$ production becomes stronger with energy. This is
an obvious consequence of the energy dependence of $\sqq({r},s)$, which
rises with energy (see sect.~\ref{sec-cross}). For $\psi'$ the suppression
is rather similar to the $\Jpsi$ case. In particular we do not see any
considerable nuclear enhancement of $\psi'$ which has been found earlier
\cite{KZ,KNNZ}, where the oversimplified form of the dipole cross
section, $\sqq({r})\propto {r}^2$ and the oscillator form of the wave
function had been used. Such a form of the cross section enhances the
compensation between large and small distances in the wave function of
$\psi'$ in the process $\gamma^*p\to\psi' p$. Therefore, the color
filtering effect which emphasizes the small distance part of the wave
function leads to a strong enhancement of the $\psi'$ production rate.
This is why using the more realistic ${r}$-dependence of $\sqq({r})$
leveling off at large ${r}$ leads to a weaker enhancement of the $\psi'$.
This effect becomes even more pronounced at higher energies since the 
dipole cross section saturates starting at a value ${r} \sim r_0(s)$
where $r_0(s)$ decreases with energy. This observation probably explains
why the $\psi'$ is less enhanced at higher energies as one can see from
Fig.~\ref{s-inc}.

Note that the ``frozen'' approximation is valid only for $l_c\gg R_A$ and
can be used only at $\sqrt{s} > 20-30\,\GeV$. Therefore, the low-energy
part of the curves depicted in Fig.~\ref{s-inc} should be corrected for
the effects related to the finiteness of $l_c$. 
This is done in Ref.~\cite{HIKT}.

One can change the effect of color filtering in nuclei in a controlled way
by increasing the photon virtuality $Q^2$ thereby squeezing the transverse
size of the $c\bar c$ fluctuation in the photon. For a narrower $c\bar c$ 
pair the cancelation which is caused by the node in the radial wave function
of $\psi'$ should be less effective. One expects that the $\psi'$ to $\Jpsi$
ratio on a proton target increases with $Q^2$, as is observed both in
experiment and calculation (Fig.~9 of \cite{HIKT}). A detailed investigation
of the $Q^2$ dependence of nuclear effects is published in Ref.~\cite{ikth}.

\begin{figure}[t]
  \centerline{
  \scalebox{0.65}{\includegraphics{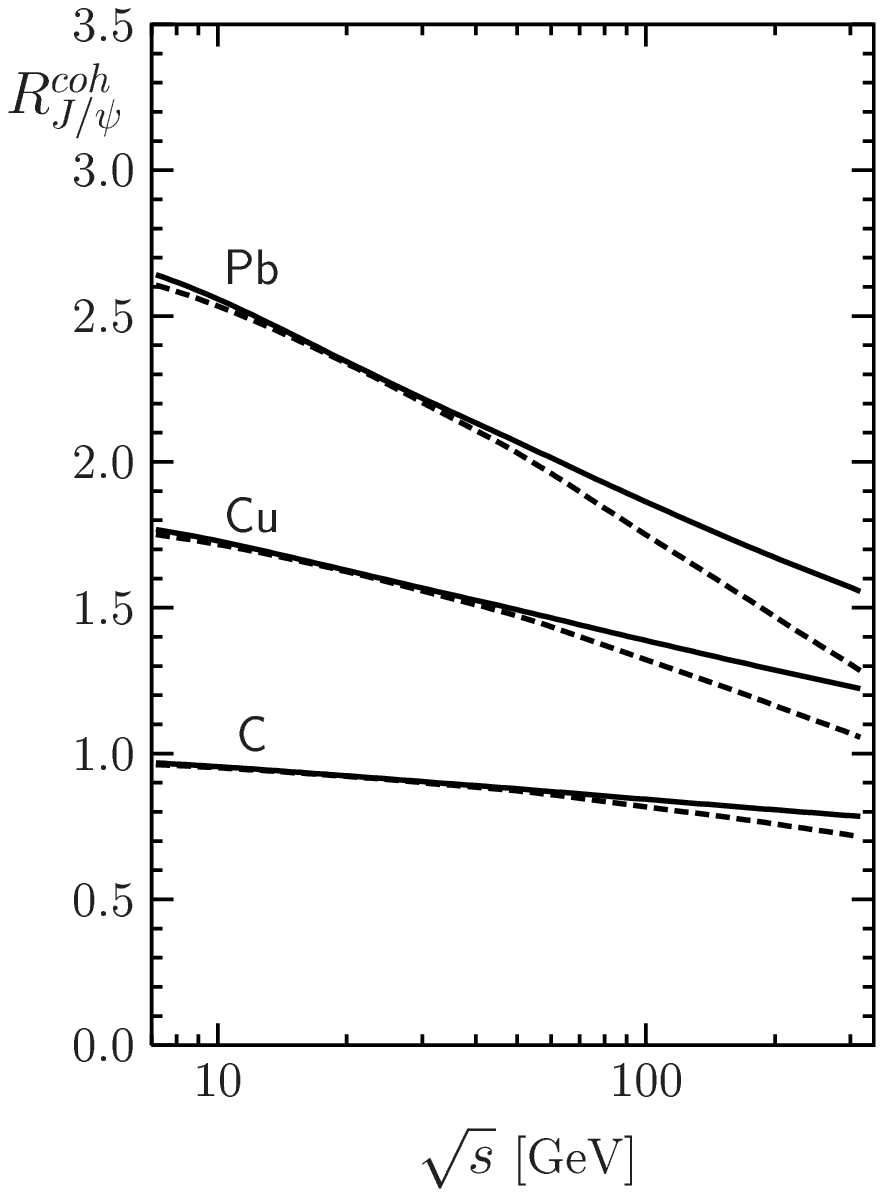}}~~
  \scalebox{0.65}{\includegraphics{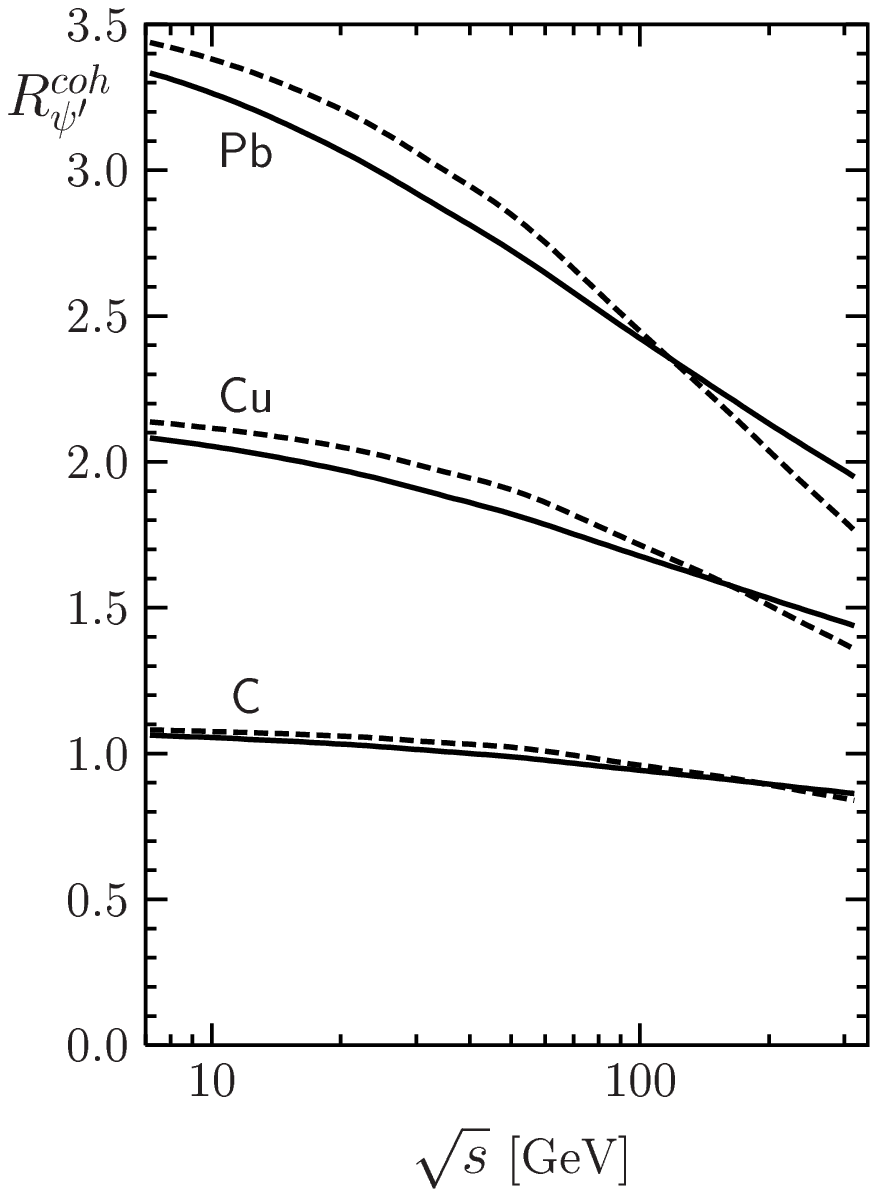}}}
\center{
{\caption{\em
  \label{s-coh}
  The ratios $R^{coh}_\Jpsi$ and $R^{coh}_{\psi'}$ for coherent
  production on nuclei as a function of $\sqrt s$. The meaning of
  the different lines is the same as in Fig.~\ref{s-inc}. }  
    }  }
\end{figure}

Cross sections for coherent production of charmonia on nuclei are calculated
analogously using Eq.~\Ref{50}. The results for the energy dependence are
depicted in Fig.~\ref{s-coh}.

It is not a surprise that the ratios exceed one. In the absence of
$c\bar c$ attenuation the forward coherent production would be
proportional to $A^2$, while integrated over momentum transfer,
the one depicted in Fig.~\ref{s-coh}, behaves as $A^{4/3}$. This is
a result of our definition  that $R^{coh}_{\Psi}$
exceeds one. 

\subsection{Gluon shadowing}\label{section-gluon}

The gluon density in nuclei at small Bjorken $x$ is expected to be
suppressed compared to a free nucleon due to interferences. This
phenomenon called gluon shadowing renormalizes the dipole cross section,
\BE
  \sqq({r},x) \Rightarrow \sqq({r},x)\,R_G(x,Q^2,b)\,.
  \label{glue}
\EE
where the factor $R_G(x,Q^2,b)$ is the ratio of the gluon density at $x$
and $Q^2$ in a nucleon of a nucleus to the gluon density in a free nucleon.
No data are available so far which could provide direct information about 
gluon shadowing. Currently it can be evaluated only theoretically.
In what follows we employ the technique developed in Ref.~\cite{KST}.

Note that the procedure Eq.~(\ref{glue}) differs from the prescription
in Ref.~\cite{fs}. The latter is based on QCD factorization applied to a
nuclear target and suggests to multiply by $R_G(x,Q^2,b)$ the whole
nuclear cross section. This approximation should not be used for charmonium
production which exposes according to above calculations a strong
deviation from factorization. Besides, gluon shadowing is overestimated
in Ref.~\cite{fs} as is discussed in Ref.~\cite{KST}.

The interpretation of the phenomenon of gluon shadowing depends very
much on the reference frame. It looks like glue-glue fusion in the
infinite momentum frame of the nucleus: although the nucleus is Lorentz
contracted, the bound nucleons are still well separated since they
contract too. However, the gluon clouds of the nucleons are contracted
less since they have a smaller momentum fraction $\sim x$. Therefore,
they do overlap and interact at small $x$, and gluons originating from
different nucleons can fuse leading to a reduction of the gluon density.

Although observables must be Lorentz invariant, the space-time
interpretation of shadowing looks very different in the rest frame of
the nucleus. Here it comes as a result of eikonalization of higher Fock
components of the incident particles. Indeed, the nuclear effect included
by eikonalization into Eqs.~(\ref{40})-(\ref{50}) corresponds to the
lowest $c\bar c$ Fock component of the photon. These expressions do not
include any correction for gluon shadowing, but rather correspond to
shadowing of sea quarks in nuclei, analogous to what is measured in
deep-inelastic scattering. Although the phenomenological dipole cross
section $\sqq(x,Q^2)$ includes all possible effects of gluon radiation,
the eikonal expressions Eqs.~(\ref{40})-(\ref{50}) assume that none of
the radiated gluons takes part in multiple interaction in the nucleus.
The leading order correction corresponding to gluon shadowing comes from
eikonalization of the next Fock component which contains the $c\bar c$
pair plus a gluon. One can trace on Feynman graphs that this is exactly
the same mechanism of gluon shadowing as glue-glue fusion in a different
reference frame.

Note that Eqs.~(\ref{40})-(\ref{50}) assume that for the coherence
length $l_c\gg R_A$. Even if this condition is satisfied for a $c\bar c$
fluctuation, it can be broken for the $c\bar cG$ component which is 
heavier. Indeed, it was found in \cite{KRT} that the coherence length
for gluon shadowing as about an order of magnitude shorter than the one
for shadowing of sea quarks. Therefore, one should not rely on the long 
coherence length approximation used in Eqs.~(\ref{40})-(\ref{50}), but take 
into account the finiteness of $l^G_c$. This can be done by using the 
light-cone Green function approach developed in \cite{krt1,KST}.

The factor $R_G(x,Q^2,b)$ has the form,
\BE
  R_G(x,Q^2,b) = 1-\frac{\Delta\sigma(\gamma^*A)}
  {T(b)\,\sigma(\gamma^*N)}~,
  \label{100}
\EE
where $\sigma(\gamma^*N)$ is the part of the total $\gamma^*N$ cross 
section related to a $c\bar c$ fluctuation in the photon,
\BE
  \sigma(\gamma^*N) = \int d^2{r}
  \int\limits_0^1 d\alpha\,
  \left|\Psi_{\gamma^*\to c\bar c}({r},\alpha,Q^2)\right|^2\,
  \sqq({r},x)\ .
  \label{105}
\EE
Here $\Psi_{\gamma^*\to c\bar c}({r},\alpha,Q^2)$ is the light-cone wave
function of the $c\bar c$ pair with transverse separation $\vec {r}$ and
relative sharing of the longitudinal momentum $\alpha$ and $1-\alpha$
(see details in sect.~\ref{sec:found}). 
The numerator $\Delta\sigma(\gamma^*A)$ in
(\ref{100}) reads \cite{KST},
\BA
  \Delta\sigma(\gamma^*A) &=&
    8\pi\,{\rm Re} \int\!dM^2\,
      \left.\frac{d^2\sigma(\gamma^*N\to c\bar cGN)}{dM^2\,dq_T^2}
      \right|_{q_T=0}
  \label{110}\\
  &\times&
  \int\limits_{-\infty}^{\infty}\!\!dz_1 
  \int\limits_{-\infty}^{\infty}\!\!dz_2\,
    \Theta(z_2-z_1)\,\rho_A(b,z_1)\,\rho_A(b,z_2)\,
    \exp\left[-i\,q_L\,(z_2-z_1)\right]~.
  \nonumber
\EA
Here the invariant mass squared of the $c\bar cG$ system is given by,
\BE 
  M^2=\sum\limits_i\frac{m_i^2+k_i^2}{\alpha_i}\ ,
  \label{115}
\EE
where the sum is taken over partons ($c\bar cG$) having mass $m_i$,
transverse momentum $\vec k_i$ and fraction $\alpha_i$ of the full
momentum. The $c\bar cG$ system is produced diffractively as an
intermediate state in a double interaction in the nucleus. $z_1$ and
$z_2$ are the longitudinal coordinates of the nucleons $N_1$ and $N_2$,
respectively, participating in the diffractive transition $\gamma^*\,N_1
\to c\bar cG\,N_1$ and back $c\bar cG\,N_2\to\gamma^*\,N_2$. The value
of $\Delta\sigma$ is controlled by the longitudinal momentum transfer
\BE
  q_L=\frac{Q^2+M^2}{2\,\nu}~,
  \label{120}
\EE
which is related to the gluonic coherence length $l^G_c=1/q_L$.
 
The Green
function $G_{c\bar cG}(\vec r_2,\vec\rho_2,z_2;\vec r_1,\vec\rho_1z_1)$
describes the propagation and interaction 
of the $c\bar cG$ system in the nuclear 
medium between the points $z_1$ and $z_2$.
Here, $\vec r_{1,2}$ and $\vec\rho_{1,2}$ are the transverse separations 
between the $c$ and $\bar c$ and between the $c\bar c$ pair and gluon
at the point $z_1$ and destination $z_2$ respectively. Then the Fourier
transform of the diffractive cross section in Eq.~\Ref{110},
\BE
  8\pi\int\!dM_X^2\,
  \left.\frac{d^2\sigma(\gamma^*N\to XN)}{dM_X^2\,dq_T^2}
  \right|_{q_T=0}\!\!\!\cos\left[q_L\,(z_2-z_1)\right]
\EE
can be represented in the form,
\BA
  &&{1\over2}\,{\rm Re}\int\!d^2\!r_2 d^2\!\rho_2 d^2\!r_1 d^2\!\rho_1
  \int d\alpha_q d\ln(\alpha_G)
  \label{130}\\
  &&\times\,
  F^{\dagger}_{\gamma^*\to c\bar cG}(\vec r_2,\vec\rho_2,\alpha_q,\alpha_G)~
  G_{c\bar cG}(\vec r_2,\vec\rho_2,z_2;\vec r_1,\vec\rho_1,z_1)~
  F_{\gamma^*\to c\bar cG}(\vec r_1,\vec\rho_1,\alpha_q,\alpha_G)~.
  \nonumber
\EA

Assuming that the momentum fraction taken by the gluon is small,
$\alpha_G \ll 1$, and neglecting the $c\bar c$ separation $r\ll\rho$
we arrive at a factorized form of the three-body Green function,
\BE
  G_{c\bar cG}(\vec r_2,\vec\rho_2,z_2;\vec r_1,\vec\rho_1,z_1)
  \Rightarrow
  G_{c\bar c}(\vec r_2,z_2;\vec r_1,z_1)\;
  G_{GG}(\vec\rho_2,z_2;\vec\rho_1,z_1)~,
  \label{140}
\EE
where $G_{GG}(\vec\rho_2,z_2;\vec\rho_1,z_1)$ describes propagation of the
$GG$ dipole (in fact the color-octet $c\bar c$ and gluon) in the nuclear
medium. This Green function satisfies the two dimensional Schr\"odinger
equation which includes the glue-glue nonperturbative interaction via the
light-cone potential $V(\vec\rho,z)$, as well as interaction with the
nuclear medium.
\BE
  i\,\frac{d}{dz_2} G_{GG}(\vec\rho_2,z_2;\vec\rho_1,z_1) =
  \left[-\frac{\Delta(\vec\rho_2)}{2\,\nu\,\alpha_G(1-\alpha_G)}
  +V(\vec\rho_2,z_2)\right]\,G_{GG}(\vec\rho_2,z_2;\vec\rho_1,z_1)~,
  \label{150}
\EE
where
\BE
  2\,{\rm Im}\,V(\vec\rho,z) =
  -\sigma_{GG}(\vec\rho)\,\rho_A(b,z)~,
  \label{160}
\EE
and the glue-glue dipole cross section is related to the $q\bar q$ one
by the relation,
\BE 
  \sigma_{GG}(r,x)={9\over4}\,\sqq(r,x)~. 
  \label{180}
\EE 

Following \cite{KST} we assume that the real part of the potential
has a form
\BE
  {\rm Re}\,V(\vec \rho,z) =
  \frac{b_0^4\,\rho^2}{2\,\nu\,\alpha_G(1-\alpha_G)}~.
  \label{170}
\EE
The parameter $b_0=0.65\GeV$ was fixed by the data on diffractive gluon 
radiation (the triple-Pomeron contribution in terms of Regge approach)
which is an essential part of Gribov's inelastic shadowing \cite{Gribov}.
The well known smallness of such a diffractive cross section explains why
$b_0$ is so large, leading to a rather weak gluon shadowing. In other words,
this strong interaction squeezes the glue-glue wave packet resulting in
small nuclear attenuation due to color transparency. 
 
\begin{figure}[t]
  \centerline{
  \scalebox{0.65}{\includegraphics{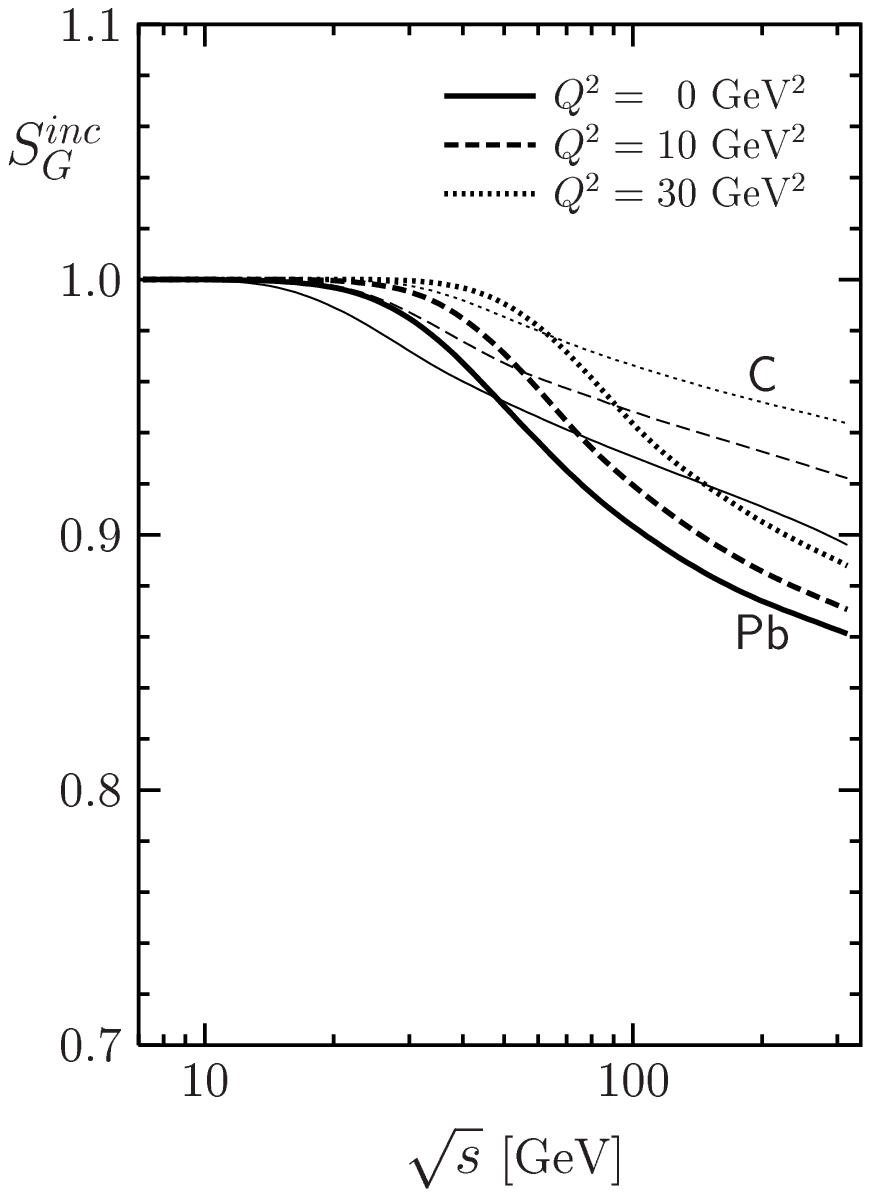}}~~
  \scalebox{0.65}{\includegraphics{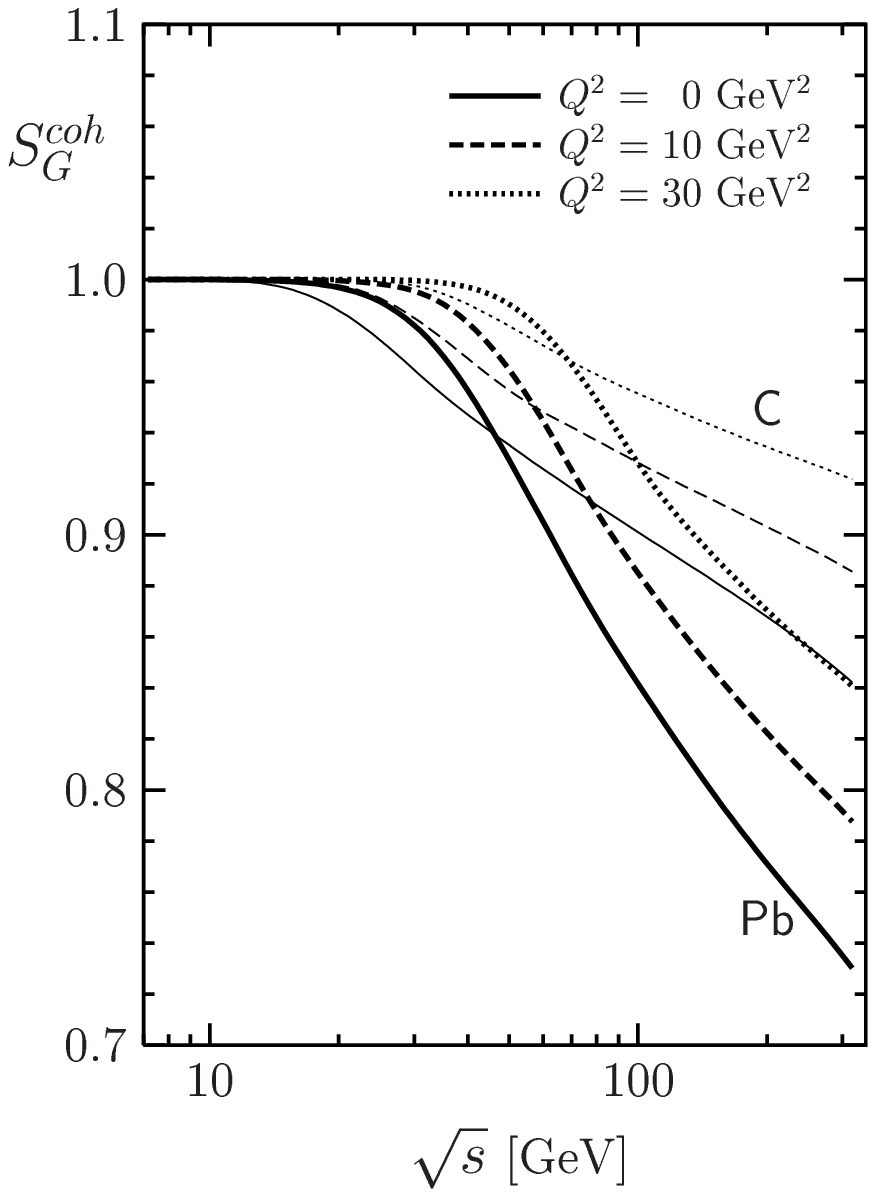}}}
\center{
{\caption{\em
  Ratios $S_G(s,Q^2)$ of cross sections calculated with and without
  gluon shadowing for incoherent and coherent charmonia production.
  We only plot ratios for $\Jpsi$ production, because ratios for
  $\psi'$ production are practically the same. All curves are
  calculated with the GBW parameterization of the dipole cross
  section $\sqq$.
  \label{SG-all} }  
    }  }
\end{figure}

Fig.~\ref{SG-all} shows the ratios of cross sections calculated with
and without gluon shadowing
for incoherent and coherent exclusive charmonium electroproduction.
We see that the onset of gluon shadowing happens at a {\em c.m.} 
energy of few
tens GeV. This onset is controlled by the longitudinal nuclear formfactor 
 \BE
F_A(q^G_c,b) = \frac{1}{T_A(b)}
\int\limits_{-\infty}^{\infty} dz\,
\rho_A(b,z)\,e^{iq_c z}\,
\label{f-factor}
 \EE
 where the longitudinal momentum transfer $q^G_c=1/l^G_c$.
For the onset of gluon shadowing $q^G_c\,R_A \gg 1$ one can keep only
the double scattering shadowing correction,
 \BE
S_G \approx 1-{1\over4}\,\sigma_{eff}
\int d^2b\, T^2_A(b)\,F_A^2(q^G_c,b)\ ,
\label{onset}
 \EE
 where $\sigma_{eff}$ is the effective cross section which depends on the dynamics
of interaction of the ${q\bar q}G$ fluctuation with a nucleon.

It was found in Ref.~\cite{KRT} 
that the coherence length for gluon shadowing is rather
short,
 \BE
  l^G_c \approx \frac{1}{10\,x\,m_N}~,
  \label{190} 
 \EE 
 where $x$ in our case should be an effective one,
$x=(Q^2+M_\Psi^2)/2m_N\nu$.  The onset of shadowing according to
Eqs.~(\ref{f-factor}) and
(\ref{onset}) should be expected at $q_c^2\sim 3/(R_A^{ch})^2$ corresponding to
 \BE
  s_G \sim 10 m_N R_A^{ch}(Q^2+M_\Psi^2)/\sqrt{3}~,
 \EE
where $(R_A^{ch})^2$ is the mean square of the nuclear charge radius.
This estimate is in a good agreement with Fig.~\ref{SG-all}. Remarkably,
the onset of shadowing is delayed with rising nuclear radius and $Q^2$.
This follows directly from Eq.~(\ref{onset}) and the fact that the formfactor
is a steeper falling function of $R_A$ for heavy than for light nuclei, 
provided
that $q_c^GR_A\gg1$.

\section{Hadroproduction of heavy quarks}
\label{sec:hadro}

\begin{figure}[t]
  \centerline{\scalebox{0.5}{\includegraphics{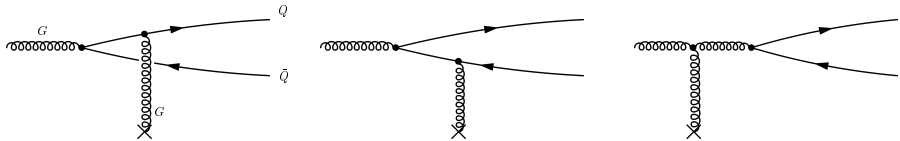}}}
    \center{
{\caption{\em
      \label{fig:3graphs}The three lowest order
graphs contributing to heavy quark
 production in the dipole approach.  
These graphs correspond to the gluon-gluon
fusion mechanism of heavy quark production in the parton model.}  
    }  }
\end{figure}

We now turn to open heavy flavor production in $pp$ collisions \cite{kt,pp}.
The color dipole formulation of this process was first introduced in 
Ref.~\cite{npz}.
In the
target rest frame, in which the dipole approach is formulated, 
heavy quark production looks like pair creation
in the target color field, Fig.~\ref{fig:3graphs}.
For a short time,
a gluon $G$ from the projectile hadron
can develop a fluctuation which contains a heavy quark
pair ($Q\bar Q$). Interaction with the 
color field of the target 
then may release these heavy quarks.  
Apparently, the mechanism depicted in
Fig.~\ref{fig:3graphs} corresponds to the gluon-gluon
fusion mechanism of heavy quark production in the leading order (LO)
parton model. This can be verified by explicit an calculation \cite{pp}.
The dipole formulation is therefore applicable only at low $x_2$, where
the gluon density of the target is much larger than all quark 
densities\footnote{We use standard kinematical
variables, $x_2=2P_{Q\bar Q}\cdot P_1/s$ and 
$x_1=2P_{Q\bar Q}\cdot P_2/s$,
where $P_1$ ($P_2$)
is the four-momentum of the projectile (target) hadron, and
$P_{Q\bar Q}$ is the four-momentum of the heavy quark pair.
In addition, $M_{Q\bar Q}$ is the invariant 
mass of the pair, and $s$ is the hadronic center of mass energy 
squared.}. 
The kinematical range where the dipole approach is valid can of 
course only be determined a posteriori. This is similar to determining the
minimal value of $Q^2$ for which perturbative QCD still works.
Note that while the mechanism for hadroproduction of heavy quark
boundstates is still subject to active theoretical and experimental
investigation, the mechanism for open heavy quark production 
is well established by now \cite{pm1,pm2,pm3}.

We shall now express the process depicted in Fig.~\ref{fig:3graphs} 
in terms of the cross section $\sig(r)$ for scattering a color neutral 
$q\bar q$-pair with transverse size $r$ off a nucleon.
The
$Q\bar Q$ pair can be produced
in three different color and spin states.  
These states are orthogonal and do not
interfere in the cross section. They include:
 \begin{enumerate}
 \item The color-singlet $C$-even $Q\bar Q$ state. The corresponding
amplitude is odd (O) relative to simultaneous permutation of spatial
and spin variables of the $Q\bar Q$ and has the form,
 \beq
A^{\bar\mu\mu}_{ij,a}(\vec\kappa,\vec k_T,\alpha) =
\sum\limits_{e=1}^8 {1\over6}\,
\delta_{ae}\,\delta_{ij}\ O^{\bar\mu\mu}_{e}
(\vec\kappa,\vec k_T,\alpha)\ .
\label{200}
 \eeq
 Here $\vec\kappa$ and $\vec k_T$ are the relative and total transverse
momenta of the $Q\bar Q$ pair respectively; $\mu,\bar\mu$ are spin indexes,
$a$ and $i,j$ are color indexes of the gluon and produced quarks,
respectively. We will classify such a state as $1^-$, which means a color
singlet with odd parity relative to index permutation. Note the $1^+$ cannot
be produced in the reaction shown in Fig.~\ref{fig:3graphs} . 

 \item Color-octet $Q\bar Q$ state with the production amplitude also
antisymmetric relative simultaneous permutation of spatial and spin variables
of the $Q\bar Q$ ($8^-$),
 \beq
B^{\bar\mu\mu}_{ij,a}(\vec\kappa,\vec k_T,\alpha) =
\sum\limits_{e,g=1}^8 {1\over2}\,
d_{aeg}\,\tau_{g}(ij)\ O^{\bar\mu\mu}_{e}
(\vec\kappa,\vec k_T,\alpha)\ .
\label{300}
 \eeq
 Here $\lambda_g = \tau_g/2$ are the Gell-Mann matrices.

 \item Color-octet $Q\bar Q$ with the amplitude symmetric relative
permutation of quark variables ($8^+$),
 \beq
C^{\bar\mu\mu}_{ij,a}(\vec\kappa,\vec k_T,\alpha) =
\sum\limits_{e,g=1}^8 {\imag\over2}\,
f_{aeg}\,\tau_{g}(ij)\ E^{\bar\mu\mu}_{e}
(\vec\kappa,\vec k_T,\alpha)\ .
\label{320}
 \eeq
\end{enumerate}

\noindent 
The two amplitudes in
Eq.~(\ref{200}) and (\ref{300}) contain the common 
factor
 \beq\label{eq:o}
O^{\bar\mu\mu}_{e}(\vec\kappa,\vec k_T,\alpha) = 
\int d^2r\,d^2s\,\euler^{\imag\vec\kappa\cdot\vec r -
\imag\vec k_T\cdot\vec s}\,
\Psi_{Q\bar Q}^{\bar\mu\mu}(\vec r)\,
\Bigl[\gamma^{(e)}(\vec s-\alpha\vec r) -
\gamma^{(e)}(\vec s+\bar\alpha\vec r)\Bigr]\ ,
\label{340}
 \eeq
which is odd (O) under permutation of the non-color variable of the 
quarks. Correspondingly, the even (E) factor in the amplitude 
Eq.~(\ref{320}) reads,
 \beqn\label{eq:e}\nonumber\lefteqn{
E^{\bar\mu\mu}_{e}(\vec\kappa,\vec k_T,\alpha) =}\\
&& 
\int d^2r\,d^2s\,\euler^{\imag\vec\kappa\cdot\vec r +
\imag\vec k_T\cdot\vec s}\,
\Psi_{Q\bar Q}^{\bar\mu\mu}(\vec r)\,
\Bigl[\gamma^{(e)}(\vec s-\alpha\vec r) +
\gamma^{(e)}(\vec s+\bar\alpha\vec r)
- 2\,\gamma^{(e)}(\vec s)\Bigr]\ ,
\label{360}
 \eeqn
 Here $\vec s$ and $\vec r$ is the position of the center of gravity and
the relative transverse separation of the $Q\bar Q$ pair, respectively.
It becomes evident from Eqs.~(\ref{eq:o}) and (\ref{eq:e}), that the 
{\em production} amplitude for the
$Q\bar Q$-pair depends on the difference between the {\em interaction}
amplitudes represented by the three
graphs in Fig.~\ref{fig:3graphs}. For example, the two
terms in the square bracket in Eq.~(\ref{eq:o}) correspond to the
two first graphs in  Fig.~\ref{fig:3graphs}. If the $Q$ and the $\bar Q$
would scatter at the same impact parameter, the production
amplitude would vanish
and nothing is produced. 
We stress that the interaction amplitudes represented by each of the
three graphs in  Fig.~\ref{fig:3graphs} is infrared divergent. 
This divergence, however, cancels in the production amplitude of the
heavy quark pair, and therefore one can express the cross section for 
heavy flavor production in terms of color neutral quantities, such as the
dipole cross section $\sig$.

The LC wave function $\Psi_{Q\bar Q}^{\bar\mu\mu}(\vec r)$ of the 
$Q\bar Q$ component of the incident gluon in Eqs.~(\ref{340})-(\ref{360}) 
reads,
 \beq
\Psi_{Q\bar Q}^{\bar\mu\mu}(\vec r) = 
\frac{\sqrt{2\,\alpha_s}}{4\pi}\,
\xi^\mu\,\hat\Gamma\,\tilde\xi^{\bar\mu}\,
\kzero(m_Q r)\ ,
\label{370}
 \eeq
 where the vertex operator has the form,
 \beq
\hat\Gamma = m_Q\,\vec\sigma\cdot\vec e +
\imag(1-2\alpha)\,(\vec\sigma\cdot\vec n)
(\vec e\cdot\vec\nabla) +
(\vec n\times\vec e)\cdot\vec\nabla\ ,
\label{370a}
 \eeq
 where $\vec\nabla = d/d\vec r$; $\alpha$ is the fraction of the
gluon light-cone momentum carried by the quark $Q$ and
 $\bar\alpha$ is the analogous quantity for the antiquark $\bar Q$; 
$\vec e$ is
the polarization vector of the gluon and $m_Q$ is the heavy
quark mass.

The profile function $\gamma^{(e)}(\vec s)$ in Eqs.~(\ref{340}) 
--(\ref{360}) is related by Fourier transformation to the amplitude
$F^{(e)}(\vec k_T,\{X\})$, of absorption of a real gluon by a nucleon,
$GN\to X$, which also can be treated as an "elastic"  (color-exchange)
gluon-nucleon scattering with momentum transfer $\vec k_T$,
 \beq
\gamma^{(e)}(\vec s) =
\frac{\sqrt{\alpha_s}}{2\pi\sqrt{6}}
\int \frac{d^2k_T}{k_T^2+\lambda^2}\,
\euler^{-\imag\vec k_T\cdot\vec s}\,
F^{(e)}_{GN\to X}(\vec k_T,\{X\})\ ,
\label{380}
 \eeq
 where the upper index $(e)$ shows the color polarization of the gluon, and
the variables $\{X\}$ characterize the final state $X$ including the color of
the scattered gluon. 

It is important for further consideration to relate the profile function
(\ref{380}) to the unintegrated gluon density ${\cal F}(k_T,x)$ and to the
dipole cross section $\sigma_{{q\bar q}}(r,x)$ ({\em cf.} 
sect.~\ref{sec:found}),
 \beqn
&& \int d^2b\,d\{X\}\,
\sum\limits_{e=1}^8 \left|\gamma^{(e)}(\vec s + \vec r) -  
\gamma^{(e)}(\vec s)\right|^2
\nonumber\\
&=& \frac{4\pi}{3}\,\alpha_s\,
\int \frac{d^2k_T}{k_T^2}\,
\left(1 - \euler^{\imag\vec k_T\cdot\vec r}\right)\,
{\cal F}(k_T,x_2) = \sigma_{{q\bar q}}(r,x_2)\ .
\label{390}
 \eeqn

Let us consider the production cross sections of a $Q\bar Q$ pair in each
of three states listed above, Eqs.~(\ref{200})--(\ref{320}). The cross
section of a color-singlet $Q\bar Q$ pair, averaged over polarization and
colors of the incident gluon reads,
 \beq
\sigma^{(1)} = \frac{1}{(2\pi)^4}
\sum\limits_{\mu,\bar\mu,i,j}\ 
\int\limits_0^1 d\alpha\int d^2\kappa\,
d^2k_T\, \overline{\Bigl|
A^{\bar\mu\mu}_{ij,a}(\vec\kappa,\vec k_T,\alpha)
\Bigr|^2}
\label{400}
 \eeq
 Using Eqs.~(\ref{340}), (\ref{370}) and (\ref{390})
this relation can be modified as,
 \beq
\sigma^{(1)} =
\sum\limits_{\mu,\bar\mu}\,\int\limits_0^1 
d\alpha\int d^2r\, \sigma_1(r,\alpha)\,
\Bigl|\Psi^{\mu\bar\mu}(\vec r,\alpha)
\Bigr|^2\ ,
\label{410}
 \eeq
 where
 \beq
\sigma_1(r,\alpha) = {1\over8}\,\sigma_{{q\bar q}}(r,x_2)\ ;
\label{420}
 \eeq
 \beq 
\sum\limits_{\mu,\bar\mu}\,
\Bigl|\Psi^{\mu\bar\mu}(\vec r,\alpha)
\Bigr|^2 \,=\,
\frac{\alpha_s}{(2\pi)^2}\,
\Bigl[m_Q^2\,\kzero^2(m_Q r) + 
(\alpha^2 + \bar\alpha^2)\,
m_Q^2\,\kone^2(m_Q r)\Bigr]\ .
\label{420a}
 \eeq

One finds in a similar way that
the cross sections of a color-octet $Q\bar Q$ pair production either 
in $8^-$ (Odd) or $8^+$ (Even) states has the form,
 \beq
\sigma^{(8)}_{O(E)} = \sum\limits_{\mu,\bar\mu}\,
\int\limits_0^1 d\alpha\int d^2r\,
\sigma^{(8)}_{O(E)}(r,\alpha)\,
\Bigl|\Psi^{\mu\bar\mu}(\vec r,\alpha)\Bigr|^2\ ,
\label{430}
 \eeq
 where
 \beqn
\sigma^{(8)}_{O}(r,\alpha,x_2) &=& 
{5\over16}\,\sigma_{{q\bar q}}(r,x_2)\ ;
\label{440}\\
\sigma^{(8)}_{E}(r,\alpha,x_2) &=&
{9\over16}\,\Bigl[ 2\sig(\alpha r,x_2) +
2\sig(\bar\alpha r,x_2) - \sig(r,x_2)\Bigr]\ .
\label{450}
 \eeqn

After summation over all three color states in which the 
$Q\bar Q$ pair in Fig.~\ref{fig:3graphs} can be produced, one obtains for the
partonic cross section \cite{kt},
\beq\label{eq:all}
\sigma(GN\to \{Q\bar Q\} X)
=\int_0^1 d\alpha \int d^2{r} 
\left|\Psi_{G\to Q\bar Q}(\alpha,{r})\right|^2
\sigma_{q\bar q G}(\alpha,{r}),
\eeq 
where $\sigma_{q\bar qG}$ is
the cross section for scattering a color neutral quark-antiquark-gluon
system on a nucleon \cite{kt},
\beq\label{eq:qqG}
\sigma_{q\bar qG}(\alpha,{r})
=\frac{9}{8}\left[\sigma_{q\bar q}(\alpha{r})
+\sigma_{q\bar q}({\bar\alpha}{r})\right]
-\frac{1}{8}\sigma_{q\bar q}({r}).
\eeq
In order to simplify the notation,
we do not explicitly write out the $x_2$ dependence of the dipole cross 
section.

The light-cone (LC) wavefunctions for the transition $G\to Q\bar Q$
can be calculated perturbatively and a very similar to the ones in 
leptoproduction, Eq.~(\ref{eq:lc-wf}),
\beqn\label{eq:lcwf}
\Psi_{G\to Q\bar Q}(\alpha,\vec{r}_1)
\Psi^*_{G\to Q\bar Q}(\alpha,\vec{r}_2)&=&
\frac{\alpha_s(\mu_R)}{(2\pi)^2}\Biggl\{m_Q^2\kzero(m_Q{r}_1)\kzero(m_Q{r}_2)
\Biggr. \\
\nonumber&+&\left.
\left[\alpha^2+{\bar\alpha}^2\right]m_Q^2
\frac{\vec{r}_1\cdot\vec{r}_2}{{r}_1{r}_2}
\kone(m_Q{r}_1)\kone(m_Q{r}_2)
\right\},
\eeqn
where $\alpha_s(\mu_R)$ is the strong coupling constant, which is probed
at a renormalization scale $\mu_R\sim m_Q$.

Eq.~(\ref{eq:all}) is a special case of the general rule that at high energy,
the cross section for the reaction $a+N\to\{b,c,\dots\}X$ can be expressed
as convolution of the LC wavefunction for the transition $a\to\{b,c,\dots\}$
and the cross section for scattering the color neutral
$\{{\rm anti-}a,b,c\dots\}$-system on the target nucleon $N$. 

Note that although 
the dipole cross section is flavor independent, the integral 
Eq.~(\ref{eq:all}) is not. Since the Bessel functions K$_{1,0}$ 
decay exponentially for large arguments, the largest values of ${r}$
which can contribute to the integral are of order $\sim 1/m_Q$. We
point out, that as a 
consequence of color transparency \cite{ZKL,ct}, the dipole 
cross section vanishes $\propto{r}^2$ for small ${r}$. Therefore, the 
$Q\bar Q$ production cross section behaves roughly like $\propto 1/m_Q^2$
(modulo logs and saturation effects).

We can estimate the relative yield of the $1^-$, $8^-$ and $8^+$
states we can rely upon the approximation $\sig(r)\propto r^2$ which is
rather accurate in the case of a $Q\bar Q$ pair, since its separation
$r\sim1/m_Q$ is small. We then derive,
 \beq
\sigma^{(1)}\ :\ \sigma^{(8)}_O\ :\ \sigma^{(8)}_E =
1\ :\ {5\over2}\ :\ {117\over70}\ .
\label{480}
 \eeq
 Thus, about $20\%$ of the produced $Q\bar Q$ pairs are in a color-singlet
state, the rest are color-octets.

In order to calculate the cross section for heavy quark pair production
in $pp$ collisions,
Eq.~(\ref{eq:all}) has to be weighted with the projectile gluon density,
\beq\label{eq:dy}
\frac{d\sigma(pp\to \{Q\bar Q\}X)}{dy}
=x_1G\left(x_1,\mu_F\right)\sigma(GN\to \{Q\bar Q\} X),
\eeq
where $y=\frac{1}{2}\ln(x_1/x_2)$ is the
rapidity of the pair and $\mu_F\sim m_Q$.
In analogy to the parton model, we call $\mu_F$ the factorization scale. 
Uncertainties arising from the choice of this scale 
will be investigated in section 
\ref{sec:phenom}.
Integrating over all kinematically allowed rapidities yields
\beq\label{eq:total}
\sigma_{\rm tot}(pp\to \{Q\bar Q\}X)=2\int_0^{-\ln(\frac{2m_Q}{\sqrt{s}})}dy\,
x_1G\left(x_1,\mu_F\right)\sigma(GN\to\{Q\bar Q\} X).
\eeq

A word of caution is in order, regarding the limits of the 
$\alpha$-integration in Eq.~(\ref{eq:all}). Since the invariant mass of
the $Q\bar Q$-pair is given by
\beq\label{eq:invmass}
M^2_{Q\bar Q}=\frac{\kappa_\perp^2+m_Q^2}{\alpha{\bar\alpha}},
\eeq
the endpoints of the $\alpha$-integration include configurations corresponding
to
arbitrarily large invariant masses, eventually exceeding the total available 
{\em cm.} energy. However, since ${r}$ and $\kappa_\perp$ 
(the single quark transverse momentum) are conjugate variables, the pair mass
is not defined in the mixed representation, nor are the integration limits
for $\alpha$. Fortunately, this problem is present only at the very edge
of the phase space and therefore numerically negligible. 

\subsection{Numerical results for hadroproduction of heavy quarks}
\label{sec:phenom}

Still the questions remain, how well does the dipole approach describe 
experimental data.
Since there are not many data for the total cross section, we shall
also compare predictions from the dipole approach to calculations
in the NLO parton model \cite{pm1,pm2,pm3}\footnote{ A FORTRAN program
for the NLO parton model calculation is available at
http://n.home.cern.ch/n/nason/www/hvqlib.html.}
For $\sig$, we use an improved version of the 
saturation model presented in Ref.~\cite{GBW}, which now also includes 
DGLAP evolution \cite{bgbk}. This improvement has no effect on open
charm, but is important for bottom production. 

In the dipole approach,
we use the one loop running coupling constant,
\beq
\alpha_s(\mu_R)=\frac{4\pi}
{\left(11-\frac{2}{3}N_f\right)\ln\!\left(\frac{\mu_R^2}{(200\,{\rm
MeV})^2}\right)}
\eeq 
at a renormalization scale $\mu_R\sim m_Q$, and
the number of light flavors is chosen to be $N_f=3$ for open charm
and $N_f=4$ for open bottom production. Furthermore, we use
the GRV98LO \cite{grv} gluon distribution
to model the gluon density in the projectile. 
We use a leading order parton distribution function (PDF), because of
its probabilistic interpretation.
Note that one could
attempt to calculate the projectile gluon distribution from 
the dipole cross section. However, the projectile distribution functions
are needed mostly at large momentum fraction $x_1$, where the dipole cross
section is not constrained by data.  

\begin{figure}[t]
  \centerline{\scalebox{0.35}{\includegraphics{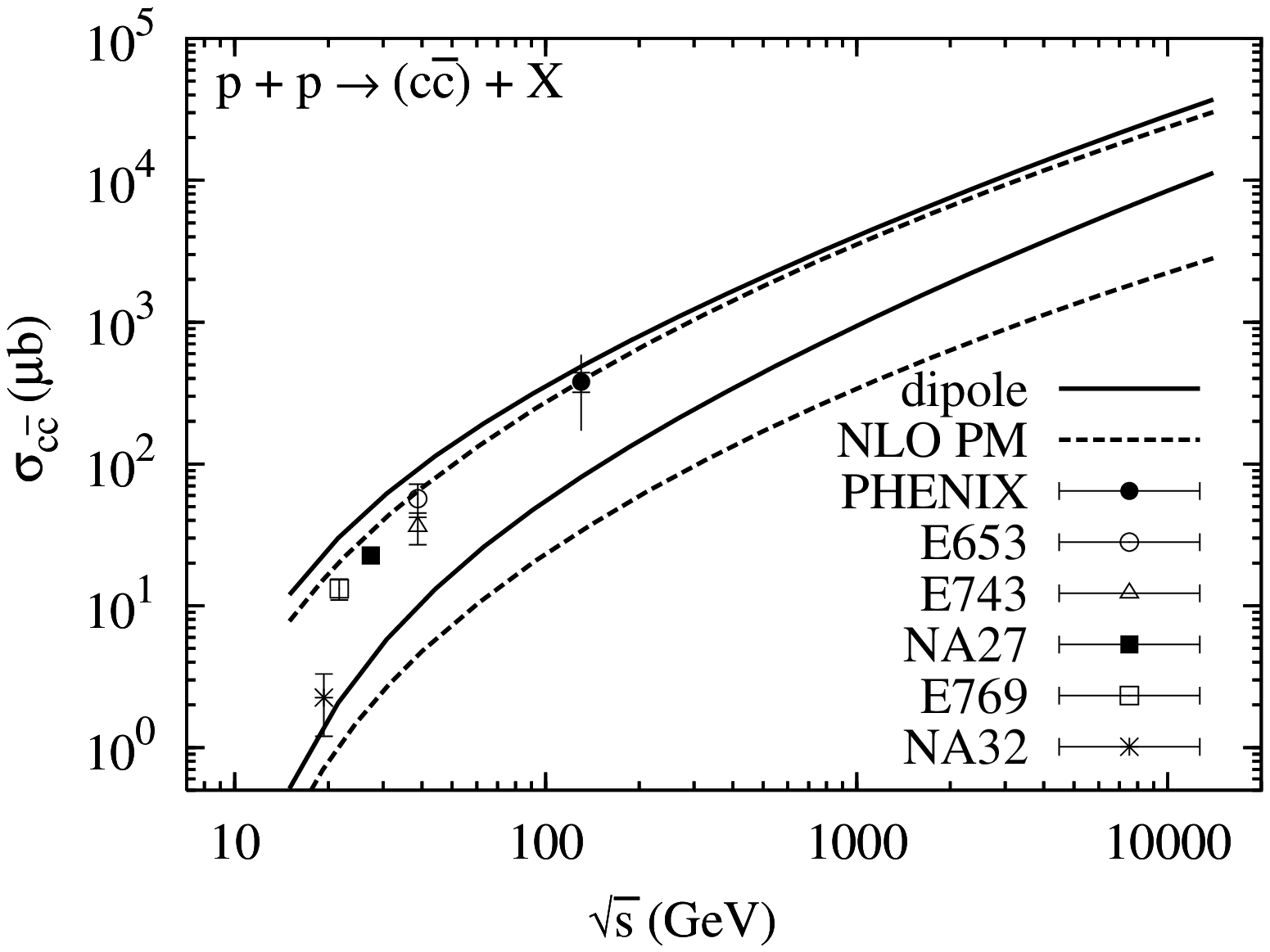}}
	      \scalebox{0.35}{\includegraphics{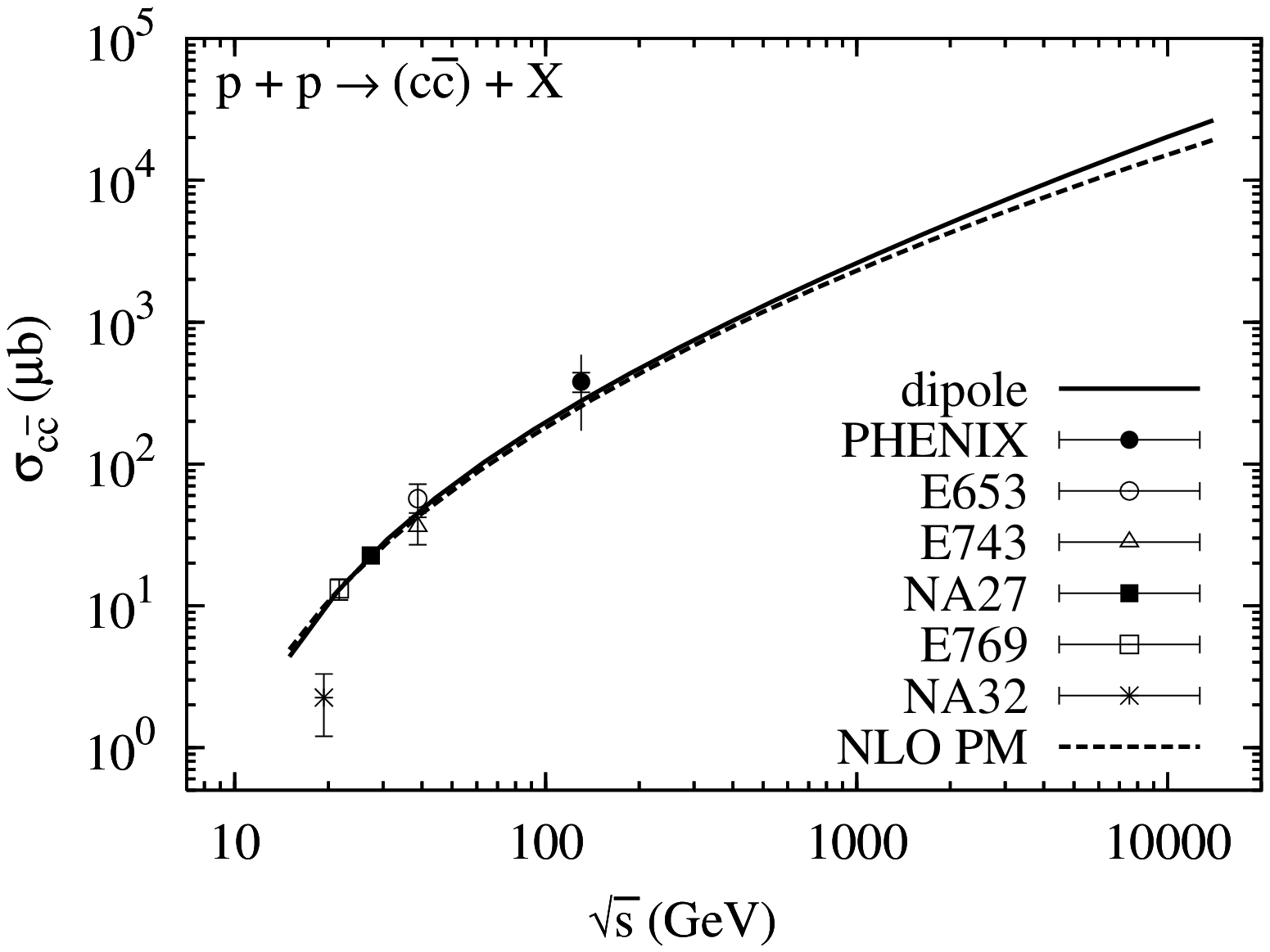}}}
\center{
{\caption{\em
      \label{fig:tcharm}
Results for the total open charm pair cross section as function 
of cm.\ energy. Varying free parameters in dipole approach (solid lines)
and in the 
parton model (dashed lines)
gives rise to the uncertainties shown on the left. In the
figure on the right, 
parameters in both models
have been adjusted so that experimental data \cite{ft,PHENIX} are 
described.}  
    }  }
\end{figure}

Our results for the total charm pair cross section 
in proton-proton ($pp$) collisions is shown in 
Fig.~\ref{fig:tcharm} as function of center of mass energy. 
The left panel shows the uncertainties of both approaches by varying
quark mass $m_c$ and renormalization scale $\mu_R$ 
in the intervals $1.2\GeV\le m_c \le 1.8\GeV$ and
$m_c\le \mu_R\le 2m_c$, respectively. 
The factorization scale is kept fixed at
$\mu_F=2m_c$, because in our opinion, the charm quark mass is too low for 
DGLAP evolution. 
A large fraction of the
resulting uncertainty originates 
from different possible choices of the charm
quark mass, since the total cross section behaves approximately 
like $\sigma_{\rm tot}\propto m_Q^{-2}$. 

Note that the mean value of $x_2$ increases with decreasing energy. At 
$\sqrt{s}=130\GeV$ one has $x_2\sim0.01$. For lower energies, our calculation 
is an extrapolation of the saturation model. For the highest fixed target 
energies of $\sqrt{s}\approx 40\GeV$, values of $x_2\sim0.1$ become important.
Unlike in the Drell-Yan case, which was studied in \cite{rpn}, the
dipole approach to heavy quark production does not show any unphysical 
behavior when extrapolated to larger~$x_2$. One reason for this is that
the new saturation model \cite{bgbk} assumes a realistic behavior of the
gluon density at large $x_2$. In addition, even at energies as low as
$\sqrt{s}=15\GeV$, the gluon-gluon fusion process is the 
dominant contribution to the cross section.
 
Because of the wide uncertainty bands, one can adjust $m_c$ and
$\mu_R$ in both approaches
so that experimental data are reproduced.
Then, dipole
approach and NLO parton model yield almost identical results.
However, the predictive power of the theory is rather small.
In 
Fig.~\ref{fig:tcharm} (right), we used $m_c=1.2\GeV$ and $\mu_R=1.5m_c$
for the NLO parton model calculation and $m_c=1.4\GeV$, $\mu_R=m_c$
in the dipole approach. The data points tend to
lie at the upper edge of
the uncertainty bands, so that rather small values of $m_c$ are needed
to describe them.

There are remaining uncertainties which are not shown
in Fig.~\ref{fig:tcharm} (right), because different combinations 
of $m_c$ and $\mu_R$ can also yield a good description of the data. In
addition, different PDFs will lead to different values of the
cross section at high energies, since the heavy quark cross section
is very sensitive to the low-$x$ gluon distribution. In \cite{rvt},
it was found that an uncertainty of a factor of $\sim 2.3$ remains at
$\sqrt{s}=14$ TeV (in the NLO parton model), 
even after all free parameters had been fixed to
describe total cross section data at lower energies.
It is interesting to see that 20 -- 30\% of the total $pp$
cross section at LHC ($\sqrt{s}=14$ TeV) goes into open 
charm \footnote{The Donnachie-Landshoff parameterization of the
total $pp$ cross section \cite{DL} predicts 
$\sigma_{\rm tot}^{pp}(\sqrt{s}=14\TeV)=100$~mb.}.

\begin{figure}[t]
  \centerline{\scalebox{0.5}{\includegraphics{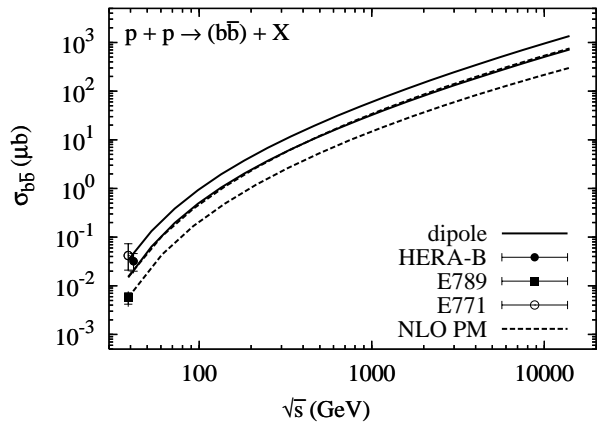}}}
{\caption{\em
      \label{fig:btotal} Uncertainties of
  open $b\bar b$ pair production calculated in the dipole approach (solid)
and in the NLO parton model (dashed). 
The dipole approach seems to provide a better 
description of the data, even though HERA-B energy is too low for the dipole
approach.  
    }} 
\end{figure}

Next, we calculate the total $b\bar b$-pair cross section as function of
center of mass energy, see Fig.~\ref{fig:btotal}. 
In order to quantify the theoretical uncertainties, we vary the free
parameters over the ranges $4.5\GeV\le m_b\le 5\GeV$ and
$m_b\le\mu_R,\mu_F\le 2m_b$
Because of the large $b$-quark mass, 
uncertainties are much smaller than for open charm production. One can see 
that the dipole approach tends to predict higher values than the NLO parton
model, even though the energy dependence expected in both approaches is very
similar. In fact, the results calculated in the dipole approach with
$m_b=5\GeV$ agree almost exactly with the NLO parton model 
calculation with $m_b=4.5\GeV$. For all other 
values of $m_b$, the uncertainty bands of
the two approaches do not overlap, in contrast to
the case for open charm production.

Three measurements of
open $b\bar b$ production
are published in the literature \cite{e789,e771,herab}. 
The two values of the open bottom cross section  
measured at Fermilab \cite{e789,e771} at {\em cm.} energy
$\sqrt{s}=38.8\GeV$
differ by almost three standard deviations. 
The HERA-B measurement at slightly larger {\em cm.} energy
$\sqrt{s}=41.6\GeV$ \cite{herab} is consistent with 
the E771 \cite{e771} value.
These two points seem to be better described by the dipole approach,
though the NLO parton model (with $m_b=4.5\GeV$)
still touches the HERA-B error bar.
Note that also a different set of PDFs would not significantly
pull up the parton model curve \cite{rvt}, as a lower value of the
$b$-quark mass would do. 
With a resummation of terms from higher order corrections \cite{thresh}, 
however, the parton model can reproduce each of the three 
measurements within theoretical uncertainties, see Ref.~\cite{herab}. 
On the other hand, typical values of $x_2$ which are important for $b\bar b$
production at HERA-B energy are of order $x_2\sim 0.2$, while
the parameterization
\cite{bgbk} of the dipole cross section is constrained only by DIS data
with $x_{Bj}\le 0.01$.

While it is an advantage of the dipole formulation
to provide very simple formulas that allow one to absorb much of the higher
order corrections into a phenomenological parameterization of
$\sigma_{q\bar q}(x_2,{r})$, one cannot clarify the origin of the 
discrepancy in 
normalizations without a systematic calculation of higher orders
in this approach.

\section{Nuclear effects in hadroproduction of open charm}

It is still unclear whether available data from fixed target experiments
demonstrate any nuclear effects for open charm production
\cite{mike,e769,wa82}.  Naively one might expect no effects at all, since a
heavy quark should escape the nucleus without attenuation or reduction of its
momentum. In fact, this is not correct even at low energies as is explained
below.  Moreover, at high energies one cannot specify any more initial or
final state interactions. The process of heavy flavor production takes a time
interval longer than the nuclear size, and the heavy quarks are produced
coherently by many nucleons which compete with each other.  As a result the
cross section is reduced, and this phenomenon is called shadowing. 

In terms of parton model the same effect is interpreted in the infinite
momentum frame of the nucleus as reduction of the nuclear parton density
due to overlap and fusion of partons at small Bjorken $x$. The
kinematic condition for overlap is the same as for coherence in the
nuclear rest frame. Thus, heavy quark via gluon fusion can be shadowed in the 
leading twist, if the gluon density in nuclei is reduced due 
to gluon shadowing.

There are well known examples of shadowing observed in hard reactions, like
deep-inelastic scattering (DIS) \cite{nmc} and the
Drell-Yan process (DY) 
\cite{e772} demonstrating a sizable reduction of the density of light sea
quarks in nuclei. Shadowing is expected also for gluons, although there is
still no experimental evidence for that.

Shadowing for heavy quarks is a higher twist effect, and although its
magnitude is unknown within the standard parton model approach, usually it is
neglected for charm and beauty production. However, this correction is
proportional to the gluon density in the proton and steeply rises with
energy. Unavoidably, such a correction should become large at high energies. 
In some instances, like for charmonium production, this higher twist effect
gains a large numerical factor and leads to a rather strong suppression even
at energies of fixed target experiments (see below).

On the other hand, gluon shadowing which is a leading twist effect, is
expected to be the main source of nuclear suppression for
heavy flavor
production at high energies.  This is why this process is usually
considered as a sensitive probe for the gluon density in hadrons and
nuclei.  If one neglects terms suppressed
by a power of $1/m_Q^2$, 
the cross section of heavy ${Q\bar Q}$ production in $pA$
collision is suppressed by the gluon shadowing factor $R^G_A$ compared to
the sum of $A$ nucleon cross sections,
 \beq
\sigma^{{Q\bar Q}}_{pA}(x_1,x_2) = 
R^G_A(x_1,x_2)\, A\,\sigma^{{Q\bar Q}}_{pN}(x_1,x_2)\ .
\label{2}
 \eeq
 Here 
 \beq 
R^G_A(x_1,x_2) = {1\over A}\int d^2b\,R^G_A(x_1,x_2,b)\,T_A(b)\ , 
\label{3}
 \eeq   
 where $R^G_A(x_1,x_2,b)$ is the (dimensional) gluon shadowing factor at
impact parameter $b$;  $T_A(b) = \int_{-\infty}^{\infty}\,dz\,\rho_A(b,z)$ is
the nuclear thickness function, and  $x_1,\ x_2$ are the
Bjorken variables of the gluons participating in ${Q\bar Q}$ production from
the colliding proton and nucleus. 

The 
parton model cannot predict shadowing, but only its evolution at high $Q^2$,
while the main contribution originates from the soft part of the interaction. 
The usual approach is to fit data at different values of $x$ and $Q^2$
employing the DGLAP evolution and fitting the distributions of different
parton species parametrized at some intermediate scale \cite{eks,kumano}.
However, the present accuracy of data for DIS on nuclei do not allow to fix
the magnitude of gluon shadowing, which is found to be compatible with zero
\footnote{Gluon shadowing was guessed in \cite{eks} to be the same as for
$F_2(x,Q^2)$ at the semi-hard scale.}. Nevertheless, the data exclude some
models with too strong gluon shadowing \cite{eks-new}. 

Another problem faced by the parton model is the
impossibility to predict gluon
shadowing effect in nucleus-nucleus collisions even if the shadowing factor
Eq.~(\ref{2}) in each of the two nuclei was known. Indeed, the cross section
of ${Q\bar Q}$ production in collision of nuclei $A$ and $B$ at impact
parameter $\vec b$ reads,
 \beq
\frac{d\sigma^{{Q\bar Q}}_{AB}(x_1,x_2)}{d^2b} = 
R^G_{AB}(x_1,x_2,b)\,AB\,
\sigma^{{Q\bar Q}}_{NN}(x_1,x_2)\ ,
\label{4}
 \eeq
 where 
 \beq
R_{AB}^G(x_1,x_2,b) = \frac{1}{AB}
\int d^2s\,R^G_A(x_1,\vec s)\,T_A(s)\ 
R^G_B(x_2,\vec b-\vec s)\,T_B(\vec b-\vec s)\ .
\label{6}
 \eeq
 In order to calculate the nuclear suppression factor Eq.~(\ref{6})  one
needs to know the impact parameter dependence of gluon shadowing,
$R^G_A(x_1,\vec b)$, while only integrated nuclear shadowing Eq.~(\ref{3}) 
can be extracted from lepton- or hadron-nucleus data\footnote{One can get
information on the impact parameter of particle-nucleus collision measuring
multiplicity of produced particles or low energy protons (so called grey
tracks). However, this is still a challenge for experiment.}. Note that
the parton model prediction of shadowing effects for minimum bias events
integrated over $b$ suffers the same problem.  Apparently, QCD factorization
cannot be applied to heavy ion collisions even at large scales. The same is
true for quark shadowing expected for Drell-Yan process in heavy ion
collisions \cite{hir,KRTJ,gay}.
            
Nuclear shadowing can be predicted within the light-cone (LC) dipole approach
which describes it via simple eikonalization of the dipole cross section. It
was pointed out in 
Ref.~\cite{ZKL} that quark configurations (dipoles) with fixed
transverse separations are the eigenstates of interaction in QCD, therefore
eikonalization is an exact procedure. In this way one effectively sums up the
Gribov's inelastic corrections to all orders \cite{ZKL}. 

The advantage of this formalism is that it does not need any $K$-factor. 
Indeed, it was demonstrated recently in Ref.~\cite{rpn} that the
simple dipole formalism for Drell-Yan process \cite{hir,bhq,kst1} precisely
reproduces the results of very complicated next-to-leading calculations at
small $x$.  The LC dipole approach also allows to keep under control
deviations from QCD factorization.  In particular, we found a substantial
process-dependence of gluon shadowing due to the 
existence of a semi-hard scale
imposed by the strong nonperturbative interaction of light-cone gluons
\cite{KST}. For instance gluon shadowing for charmonium production off
nuclei was found in Ref.~\cite{kth} to be much stronger than in deep-inelastic
scattering \cite{KST}. 

The LC dipole approach also provides effective tools for calculation of
transverse momentum distribution of heavy quarks, like it was done for
radiated gluons in Ref.~\cite{kst1,jkt}, or Drell-Yan pairs in 
Ref.~\cite{KRTJ}.
Nuclear broadening of transverse momenta of the heavy quarks also is an
effective way to access the nuclear modification of the transverse
momentum distribution of gluons, {\em i.e.} the
so called phenomenon of color glass
condensate or gluon saturation \cite{mv,al}. We consider only integrated
quantities here.

In what follows we find sizable deviations from QCD factorization for heavy
quark production off nuclei. First of all, for open charm production
shadowing related to propagation of a ${c\bar c}$ pair through a nucleus is not
negligible, especially at the high energies of RHIC and LHC, in spite of
smallness of ${c\bar c}$ dipoles. Further, higher Fock components containing
gluons lead to gluon shadowing which also deviates from factorization and
depends on quantum numbers of the produced heavy pair ${c\bar c}$.

\subsection{Higher twist shadowing for 
\boldmath$c\bar c$ production}\label{eikonal}

An important advantage of the LC dipole approach is the simplicity of
calculations of nuclear effects. Since partonic dipoles are the eigenstates
of interaction one can simply eikonalize the cross section on a nucleon
target \cite{ZKL} provided that the dipole size is ``frozen'' by Lorentz time
dilation. Therefore, the cross section of a ${c\bar c}$ pair production off a
nucleus has the form \cite{npz,kt},
 \beqn
\sigma(GA\to {c\bar c}X) &=& 
2\,\sum\limits_{\mu,\bar\mu}\,\int d^2b\int d^2r
\int\limits_0^1 d\alpha\,
\Bigl|\Psi^{\mu\bar\mu}(\vec r,\alpha)\Bigr|^2\nonumber\\
&\times&
\left\{1\,-\,\exp\left[-{1\over2}\,
\sigma_{q\bar qG}(r,\alpha,x_2)\,T_A(b)\right]\right\}\ ,
\label{500}
 \eeqn
 where $\sigma_{q\bar qG}(r,\alpha,x_2)$ is the cross section 
of interaction of a
${c\bar c}G$ three particle state with a nucleon, see sect.~\ref{sec:hadro}. 

 Apparently, this expression leads to shadowing correction which is a
higher twist effect and vanishes as $1/m_c^2$.  Indeed, it was found in
\cite{npz} that in the kinematic range of fixed target experiments at the
Tevatron, Fermilab, $x_2\sim 10^{-2},\ x_F\sim0.5$, the shadowing effects
are rather weak even for heavy nuclei,
 \beq
1-R_A \lsim 0.05\ ,
\label{510}
 \eeq
 where $R_A$ is defined in (\ref{2}).

On the other hand, a substantial shadowing effect, several times stronger than
in Eq.~(\ref{510}) 
was found in Ref.~\cite{kth} for charmonium production (see below),
although it is also a higher twist effect.  In the case of open charm
production there are additional cancelations which grossly diminish shadowing.
The smallness of the effect maybe considered as a justification for the parton
model prescription to neglect this correction as a higher twist effect.
However, the dipole cross section $\sig(r,x_2)$ steeply rises with $1/x_2$
especially at small $r$ and and the shadowing corrections increase, reaching
values of about $10\%$ at $x_2=10^{-3}$, and about $30\%$ at $x_2=10^{-5}$.

\subsection{Process dependent gluon shadowing}
\label{shadowing}

The phenomenological dipole cross section which enters the exponent in
Eq.~(\ref{500}) is fitted to DIS data. Therefore it includes effects of gluon
radiation which are in fact the source of rising energy ($1/x$) dependence of
the $\sig(r,x)$. However, a simple eikonalization in Eq.~(\ref{500}) 
corresponds to the Bethe-Heitler approximation assuming that the whole
spectrum of gluons is radiated in each interaction independently of other
rescatterings. This is why the higher order terms in expansion of (\ref{500})
contain powers of the dipole cross section.  However, gluons radiated due to
interaction with different bound nucleons can interfere leading to damping of
gluon radiation similar to the Landau-Pomeranchuk \cite{lp} effect in QED. 
Therefore, the eikonal expression Eq.~(\ref{500}) needs corrections which are
known as gluon shadowing.

Nuclear shadowing of gluons is a leading twist effect since the cloud of
massless gluons has a larger size than the source which is a small size
$bar cc$ pair. Gluon shadowing is treated by the parton model in the
infinite momentum frame of the nucleus as a result of glue-glue fusion.
On the other hand, in the nuclear rest frame the same phenomenon is
expressed in terms of the Glauber like shadowing for the process of gluon
radiation \cite{mueller}.  In impact parameter representation one can
easily sum up all the multiple scattering corrections which have the
simple eikonal form \cite{ZKL}. Besides, one can employ the well
developed color dipole phenomenology with parameters fixed by data from
DIS. Gluon shadowing was calculated employing the light-cone dipole
approach for DIS \cite{KST} and production of charmonia \cite{kth}, and
a substantial deviation from QCD factorization was found. Here we
calculate gluon shadowing for ${c\bar c}$ pair production.

First of all, one should develop a dipole approach for gluon radiation
accompanying production of a ${c\bar c}$ pair in gluon-nucleon collision. 
Then nuclear effects can be easily calculated via simple eikonalization.
This is done in Ref.~\cite{kt}.

  According to the general prescription \cite{hir} the dipole cross section
which enters the factorized formula for the process of parton $a$-nucleon
collision leading to multiparton production, $a\,N \to b+c+\dots+d\ X$, is the
cross section for the colorless multiparton ensemble $|\bar abc\dots d\ra$. The
same multiparton dipole cross section is responsible for nuclear shadowing.
Indeed, in the case of the process $G\,N\to{c\bar c}\,X$ it was the cross section
$\sigma_{q\bar qG}$, Eq.~(\ref{eq:qqG}), 
which correspond to a state $|{c\bar c}G\ra$
interacting with a nucleon.

Correspondingly, in the case of additional gluon production, $G\to\bar
cc\,G$, it is a 4-parton, $|{c\bar c}GG\ra$, cross section $\sigma_4(\vec
r,\vec\rho,\alpha_1,\alpha_2,\alpha_3)$. Here $\vec r$ and $\vec\rho$ are the
transverse ${c\bar c}$ separation and the distance between the ${c\bar c}$ center
of gravity and the final gluon, respectively. Correspondingly,
$\alpha_1=\alpha_c$, $\alpha_2=\alpha_{\bar c}$, and $\alpha_3=\alpha_G$.
Treating the charm quark mass as a large scale, one can neglect
$r\ll\rho$, then the complicated expression for $\sigma_4$ becomes rather
simple. One can find details in Ref.~\cite{kt}.

 \begin{figure}[t]
\includegraphics{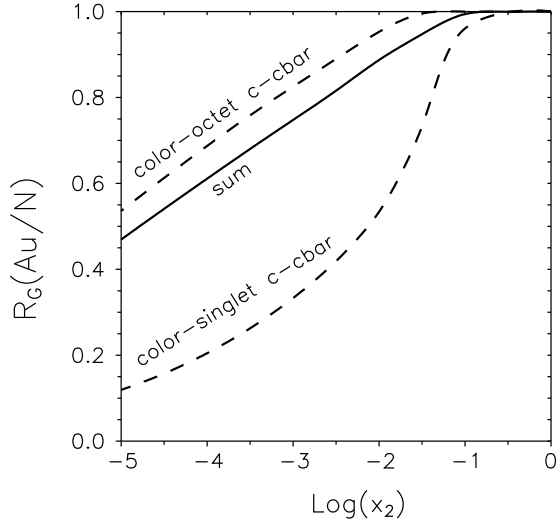}
\begin{center}
\vspace{7.5cm}
 {\caption[Delta]
 {\sl Ratio of gluon densities $R_G(Au/p)=G_{Au}(x_2)/195\,G_{p}(x_2)$
for color octet-octet and singlet-octet states $({c\bar c})-G$ (dashed curves).
The averaged gluon shadowing is depicted by the solid curve.}
 \label{gl-shad}}
\end{center}
 \end{figure}

One can treat partons as free only if their transverse momenta are
sufficiently large, otherwise the nonperturbative interaction between partons
may generate power corrections \cite{KST}. Apparently, the softer the
process is, the more important are these corrections. In particular,
diffraction and nuclear shadowing are very sensitive to these effects.
Indeed, the cross section of diffractive dissociation to large masses (so
called triple-Pomeron contribution)  is proportional to the fourth power of
the size of the partonic fluctuation.  Therefore, the attractive
nonperturbative interaction between the partons squeezes the fluctuation and
can substantially reduce the diffractive cross section. Smallness of the
transverse separation in the quark-gluon fluctuation is the only known
explanation for the observed suppression of the diffractive cross section,
which is also known as the problem of smallness of the triple-Pomeron
coupling.  While no data sensitive to gluon shadowing are available yet, a
vast amount of high accuracy diffraction data can be used to fix the
parameters of the nonperturbative interaction.

It turns out \cite{kt} that the color interaction between the gluon and 
the $c\bar c$ pair depends 
on color states of the latter. 
If the $c\bar c$ pair is in one of the two color octet states, 
the nonperturbative interaction between the pair and the gluon is strong, 
and the $|q\bar q G\ra$ system cannot become larger than a typical 
constituent quark radius $\sim 0.3\fm$. For these small configurations,
shadowing is rather small, see  Fig.~\ref{gl-shad}.
If, on the other hand, the $c\bar c$ pair is in a color singlet state,
the color charges of the $c$ and the $\bar c$ screen each other, so
that the pair cannot interact strongly with the radiated gluon,
{\em i.e.} the value of $b_0$ 
(see  sect.~\ref{section-gluon}) is much smaller than
$0.65\GeV$. 
The transverse size of these configurations is limited only by confinement,
hence they can become as large as a typical hadron. Therefore,
gluon shadowing is much stronger in the color singlet channel \cite{kth}.  

\subsection{Numerical results}

\begin{figure}[tbh]
\includegraphics{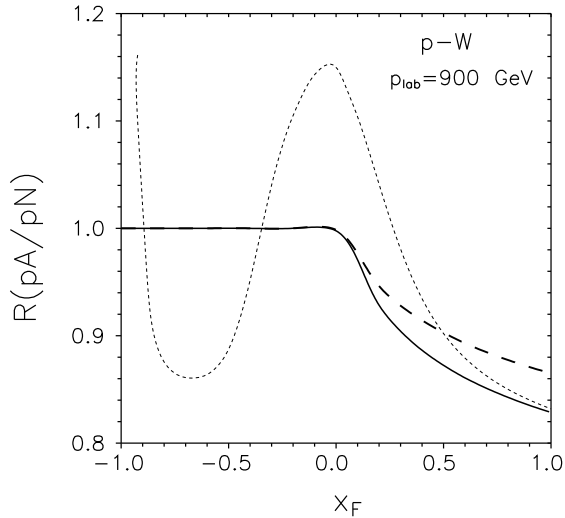}
\begin{center}
\vspace{7.5cm}
 {\caption[Delta]
 {\sl Nuclear effects for open charm production in $p--W$ collisions at
$900\GeV$ beam energy. The contribution of gluon shadowing is
shown by dashed curve. The solid curve represents the full shadowing
effect including the higher twist contribution given by Eq.~(\ref{500}).
Possible medium effects including antishadowing and EMC-suppression for
gluons \cite{eks} are also added and the result is represented by the
dotted curve.}
 \label{pa-xf}}
\end{center}
 \end{figure}

To observe the shadowing effects in open charm production, one must access the
kinematic region of sufficiently small $x_2 \lsim 0.1$. With fixed targets it
can be achieved at highest energies at Fermilab and in the experiment HERA-B
at DESY. We apply the results of the
previous section for gluon shadowing to
${c\bar c}$ pair production in proton-nucleus collisions. We assume that the
${c\bar c}$ is produced with Feynman $x_F$ corresponding to
$x_2=(-x_F+\sqrt{x_F^2+4M_{{c\bar c}}^2/s})/2$, where we fix
$M_{c\bar c}=4\GeV$. The contribution of gluon shadowing to nuclear effects in
proton-tungsten collision at $p_{lab}=900\GeV$ is depicted by the dashed
curve in Fig.~\ref{pa-xf}. 

The higher twist shadowing correction, which corresponds 
to the eikonalized dipole
cross section $\sigma_{q\bar qG}$ in Eq.~(\ref{500}), 
is also a sizable effect and should
be added. It is diminished, however, due to the strong gluon shadowing which
also reduces the amount of gluons available for multiple interactions
compared to the eikonal approximation Eq.~(\ref{500}). We take this reduction
into account multiplying $\sigma_{q\bar qG}$ in Eq.~(\ref{500})  
by $R_G(x_2,M_{\bar cc})$. 
This procedure is justified at small transverse separations, since
$\sigma(r,x) = (\pi^2/3)\,\alpha_s/\,r^2\,G(x,Q^2\sim 1/r^2)$ \cite{fs}.
For large separations see discussion in Ref.~\cite{kth}.
The summed shadowing suppression of ${c\bar c}$ production is depicted
in Fig.~\ref{pa-xf} by the solid curve. 

Besides shadowing, other nuclear effect are possible. The EMC effect,
suppression of the nuclear structure function $F_2^A(x,Q^2)$ at large $x$, as
well as the enhancement at $x\sim 0.1$ should also lead to similar
modifications in the gluon distribution function $G^A(x,Q^2)$. These effects
are different from shadowing which is a result of coherence. A plausible
explanation relates them with medium effects, like swelling of bound nucleons
\cite{michele}. To demonstrate a possible size of the medium effects on gluon
distribution we parametrize and apply the effect of gluon enhancement and
suppression at large $x$ suggested in 
Ref.~\cite{eks}. Although it is based on ad
hoc gluon shadowing and underestimated shadowing for valence quarks (see 
discussion in
Ref.~\cite{KRTJ}), 
it demonstrates the scale of possible effects missed in our
analysis. 

There are still other effects missed in our calculations.  At this energy, the
effect of energy loss due to initial state interactions \cite{eloss} 
causes additional nuclear suppression at large $x_F$ (compare with
Ref.~\cite{kth}). 
Another correction is related to the observation that detection of
a charm hadron at large $|x_F|$ does not insure that it originates from a charm
quark produced perturbatively with the same $x_F$. Lacking gluons with
$x_{1,2}\to 1$ one can produce a fast charm hadron via a fast projectile
(usually valence) quark which picks up a charm quark created at smaller
$|x_F|$. This is actually the mechanism responsible for the observed $D/\bar D$
asymmetry. It provides a rapidity shift between the parent charm quark and the
detected hadron.  Therefore, it may reduce shadowing effects at largest
$|x_F|$. We leave this problem open for further study.

\begin{figure}[tbh]
\includegraphics{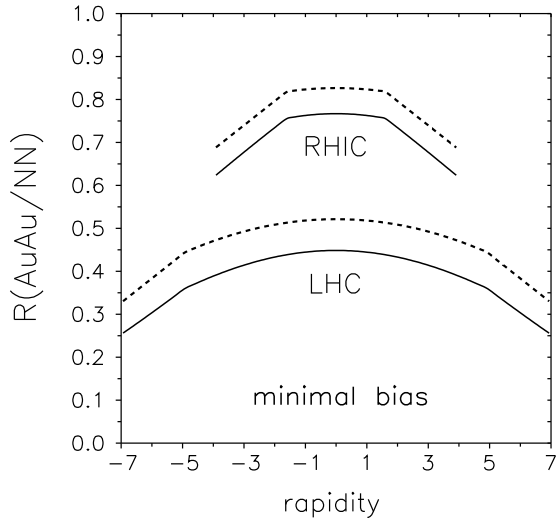}
\begin{center}
\vspace{7.5cm}
 {\caption[Delta]
 {\sl Nuclear shadowing for open charm production in minimal bias gold-gold
collision. Dotted curves show the net effect of gluon shadowing, while
solid curves include both effects of gluon shadowing and the higher twist
correction related to the nonzero separation of the ${c\bar c}$. The top (RHIC)
and bottom (LHC) curves correspond to $\sqrt{s}=200\GeV$ and $5500\GeV$
respectively.}
 \label{aa-mb}}
\end{center}
 \end{figure}

To predict shadowing effects in heavy ion collisions we employ QCD
factorization, which we apply only for a given impact parameter. For minimal
bias events
 \beq
R_{AB}(y)=R_A(x_1)\,R_B(x_2)\ ,
\label{170b}
 \eeq
 where $y=\ln(x_1/x_2)/2$ is the rapidity of the ${c\bar c}$ pair.
Our predictions for RHIC ($\sqrt{s}=200\GeV$) and LHC ($\sqrt{s}=5500\GeV$)
are depicted in Fig.~\ref{aa-mb} separately for net gluon shadowing (dashed
curves) and full effect including higher twist quark shadowing (solid curves).
Although shadowing of charmed quarks is a higher twist effect, its
contribution is about $10\%$ at RHIC and rises with energy.

One might be surprised by the substantial magnitude of shadowing expected at
the energy of RHIC. Indeed, the value of $x_{1,2}\approx 0.02$ at
mid-rapidity is rather large, and no gluon shadowing would be expected for
DIS \cite{KST}.  However, the process of charm production demonstrates a
precocious onset of gluon shadowing as was discussed above.  Besides, the
nuclear suppression is squared in $AA$ collisions.

Another interesting observation made in 
Ref.~\cite{kt} is
that shadowing is the same for 
central and minimal bias events. This has indeed been 
observed (within large error bars) by the PHENIX experiment 
\cite{PHENIX}.

\section{The light-cone dipole formalism for charmonium production 
off a nucleon}\label{LC}
 
 The important advantage of the light-cone (LC) dipole approach is its
simplicity in the calculations of nuclear effects. It has been suggested
two decades ago \cite{ZKL} that quark configurations (dipoles) with fixed
transverse separations are the eigenstates of interaction in QCD.
Therefore the amplitude of interaction with a nucleon is subject to
eikonalization in the case of a nuclear target. In this way one
effectively sums the Gribov's inelastic corrections in all orders.
 
\begin{figure}[t]
  \centerline{\scalebox{0.5}{\includegraphics{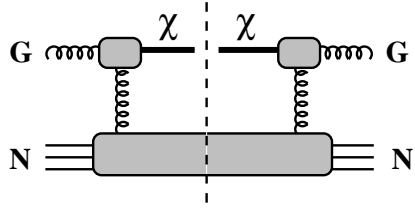}}}
{\caption{\em\label{fig1}
      Perturbative QCD mechanism of
production of the $\chi$ states
 in a gluon-nucleon collision. 
    }} 
\end{figure}

 Assuming that the produced ${c\bar c}$ pair is sufficiently small so that
multigluon vertices can be neglected, we can write the cross section for
$G\,N\to \chi\,X)$ as (see Fig.~\ref{fig1}),
 \beq
 \sigma(GN\to\chi X)=\frac{\pi}{2(N_c^2-1)}\,
 \sum\limits_{a,b}
\int \frac{d^2k_T}{k_T^2}\,
 \alpha_s(k_T^2)\,{\cal F}(x,k_T^2)\,
\Bigl|M_{ab}(\vec k_T)\Bigr|^2\ ,
\label{1.0}
 \eeq
 where ${\cal F}(x,k_T^2)=\partial G(x,k_T^2)/\partial (k_T^2)$
 is
the unintegrated gluon density, $G(x,k_T^2)=x\,g(x,k_T^2)$
($x=M_{\chi}^2/\hat s$); $M_{ab}(\vec k_T)$ is the fusion amplitude
$G\,G\to\chi$ with $a,\ b$ being the gluonic indices.

 In the rest frame of the nucleon the amplitude can be represented in
terms of the ${c\bar c}$ LC wave functions of the projectile gluon and
 ejectile
charmonium,
 \beq
 M_{ab}(\vec k_T) = \frac{\delta_{ab}}{\sqrt{6}}\,
\int\limits_0^1 d\alpha\,\int d^2 {r}\,
 \sum\limits_{\bar\mu\mu}\,
\Bigl(\Phi_{\chi}^{\bar\mu\mu}(\vec {r},\alpha)\Bigr)^*\,
 \left[e^{i\vec
k_T\cdot\vec r_1} -
 e^{i\vec k_T\cdot\vec r_2}\right]\,
\Phi_{G}^{\bar\mu\mu}(\vec {r},\alpha)\ ,
\label{1.00}
\eeq
 where
 \beq
 \vec r_1=(1-\alpha)\,\vec {r},\ \ \ \ \ \
 \vec r_2=
 - \vec\alpha\,\vec {r}\ .
\label{1.00a}
\eeq
 
 For the sake of simplicity, we separate the  color
parts
 $\la{c\bar c},{\{8\}_a}|$ and $\la{c\bar c},\{1\}|$ from the LC
wave
 function of the gluon and charmonium respectively, and calculate the
matrix element,
 \beq
 \Bigl\la{c\bar c},{\{8\}_a}\Bigr|
{1\over2}\,\lambda_b\Bigl|{c\bar c},\{1\}\Bigr\ra =
\frac{\delta_{ab}}{\sqrt{6}}\ ,
\label{1.01}
\eeq
 which is shown explicitly in Eq.~(\ref{1.00}). Thus, the functions
 $\Phi_{G(\chi)}^{\bar\mu\mu}(\vec {r},\alpha)$ in Eq.~(\ref{1.00}) represent
 only the spin- and coordinate dependent parts of the corresponding full
 wave
 functions.
 
 The gluon wave function differs only by a factor from the
 photon one,
 \beq
 \Phi_{G}^{\bar\mu\mu}(\vec {r},\alpha) =
 \frac{\sqrt{2\,\alpha_s}}{4\pi}\,
 \Bigl(\xi^\mu_c\Bigr)^\dagger\,\hat O\,
 \tilde\xi^{\bar\mu}_{\bar c}\,
 K_0(\epsilon\,{r})\ ,
\label{1.02}
\eeq
 where $\xi^{\mu}_{c}$ is the $c$-quark spinor, and
 \beqn
 \tilde\xi^{\bar\mu}_{\bar c} &=&
 i\,\sigma_y\,\xi^{\bar\mu}_{\bar
 c}\Bigr.^*\ ,
 \label{1.02a}\\
 \hat O &=& m_c\,\vec\sigma\cdot\vec e +
 i(1-2\alpha)\,(\vec\sigma\cdot\vec n)\,
 (\vec e\cdot\vec\nabla) +
 (\vec
 e\times\vec n)\vec\nabla\ ,
\label{1.02b}\\
 \epsilon^2 &=& Q^2\alpha(1-\alpha)+m_c^2\ ,
 \label{1.02c}\\
 \vec\nabla &=& \frac{d}{d\,\vec {r}}\ .\nonumber
\label{1.03}
 \eeqn
 The gluon has virtuality $Q^2$ and polarization vector $\vec e$
 and is moving along the unit vector $\vec n$ (in what follows we consider
 only transversely polarized gluons, $\vec e\cdot\vec n=0$).
 
The expression for the LC wave function of a charmonium 
and the wavefunction of the charmonium in its rest frame are related
in a somewhat complicated way by Lorentz transformation, as discussed in 
sect.~\ref{sec:cp}. 
This complexity is a consequence of the
nonlocal relation between the LC variables ($\vec {r}$, $\alpha$) and the
components of the 3-dimensional relative ${c\bar c}$ radius-vector
in the rest frame of the charmonium. Also the Melosh spin rotation leads
to a nontrivial relations between the two wave functions (see {\it e.g.}
in \cite{HIKT}). This is a relativistic effect, it vanishes in the limit
of small velocity $v\to 0$ of the quarks in the charmonium rest frame.

A word of caution is in
 order. In some cases the Melosh
 spin rotation is important even in the limit
 of vanishing quark velocity
 $v\to 0$. An example is the Landau-Yang theorem
 \cite{landau} which
 forbids production of the $\chi_1$ state by two massless
 gluons. However,
 the LC approach leads to creation of the $\chi_1$ even in
 the limit $v\to
 0$ if the effect of spin rotation is neglected. It is
 demonstrated in
Ref.~\cite{kth} that the Landau-Yang theorem is restored only if
 the Melosh
 spin rotation is included. Such a cancelation of large values is
 a kind
 of fine tuning and is a good support for the procedure of Lorentz
 boosting which we apply to the charmonium wave functions.
 
 Since the
 gluon LC wave function smoothly depends on $\alpha$
 while the charmonium
 wave function peaks at $\alpha=1/2$ with a tiny width
 estimated in Ref.~\cite{kth}, $\la(\alpha-1/2)^2\ra=0.01$,
 we can replace the charmonium wave function
 in the matrix element in
 Eq.~(\ref{1.00}) with
 \beq
 \Phi^{\bar\mu\mu}_{\chi}({r},\alpha) \approx
 \delta\left(\alpha-{1\over2}\right)\,
 \int
 d\alpha\,\Phi^{\bar\mu\mu}_{\chi}({r},\alpha)\ .
\label{1.06}
\eeq

 It is convenient to expand the LC charmonium wave function in
 powers
 of $v$. The result depends on the total momentum $J$ and its
 projection $J_z$ on the direction $\vec n$.
 The charmonium wave function
 integrated over $\alpha$ has
 the form,
 \beqn
 \int d\alpha\,
 \Phi^{\bar\mu\mu}_{\chi}({r},\alpha) &=&
 \Bigl(\xi^\mu\Bigr)^{\dagger}\,
 \left[\vec\sigma\cdot\vec e_{\pm} +
 {1\over m}\,(\vec e_{\pm}\times\vec
 n)\cdot\vec\nabla\right. \nonumber\\
&-& \left.
 \frac{1}{2\,m_c^2}\,(\vec e_{\pm}\cdot\vec\nabla)\,
 (\vec\sigma\cdot\vec\nabla)\right]\,
 \tilde\xi^{\bar\mu}\,W +
 O(v^4)\ ,
\label{1.07}
\eeqn
 where
 \beq
 W=\frac{\vec e_{\pm}\cdot\vec {r}}{{r}}\,
 \left[R({r}) + \frac{3}{4\,m_c^2}\,R^{\prime\prime}({r}) +
 O(v^4)\right]\  ,
\label{1.08}
\eeq
 and $R(r)$ is the radial part of the P-wave charmonium
 in its rest
frame (see derivation in Appendix A of \cite{kth}).
 The new notations for the polarization
vectors are,
 \beqn
 \vec e_+ &=& - \frac{\vec e_x+i\vec e_y}
{\sqrt{2}}\ ,\nonumber\\
 \vec e_- &=& \frac{\vec e_x-i\vec e_y}
 {\sqrt{2}}\
.
\label{1.09}
\eeqn
 
 In what follows we use the LC wave functions of gluons and
charmonium in order to calculate matrix elements of operators which
 depend
only on the LC variables ${r}$ and $\alpha$. Therefore,
 for the sake of
simplicity we can drop off the indexes
 $\mu,\bar\mu$ and summation over them,
{\it i.e.} replace
 \beq
\sum\limits_{\mu\bar\mu}\Bigl(\Phi^{\mu\bar\mu}_{\chi}
 (\vec
{r},\alpha)\Bigr)^*\,
 \Phi^{\mu\bar\mu}_{G}(\vec {r},\alpha) \Rightarrow
\Phi^*_{\chi}(\vec {r},\alpha)\,
 \Phi_{G}(\vec {r},\alpha)
\label{1.10}
\eeq
 With this convention we can rewrite the cross section
 Eq.~(\ref{1.00}) as,
 \beqn
 && \sigma(GN\to\chi X) =
 \int\limits_0^1
 d\alpha
 \int\limits_0^1 d\alpha'
 \int d^2{r}\,d^2{r}^{\prime}\nonumber\\
 &\times& \Bigl\{\Phi_{\chi}^*(\vec {r},\alpha)\,
 \Phi_{\chi}(\vec
 {r}^{\,\prime},\alpha^{\prime})\,
 \Sigma^{tr}(\vec {r},\vec
 {r}^{\prime},\alpha,\alpha^{\prime})\,
 \Phi_{G}(\vec {r},\alpha)\,
 \Phi_{G}^*(\vec {r}^{\,\prime},\alpha^{\prime})\Bigr\}\ ,
\label{1.1}
\eeqn
 where the transition cross section $\Sigma^{tr}$ is a
 combination of
 dipole cross sections,
 \beq
 \Sigma^{tr}(\vec {r},\vec
 {r}^{\,\prime},\alpha,\alpha^{\prime})
 = {1\over16}\,\Bigl[\sigma_{\bar
 qq}(\vec r_1-\vec
 r_2^{\,\prime})+
 \sigma_{{q\bar q}}(\vec r_2-\vec
 r_1^{\,\prime}) -
 \sigma_{{q\bar q}}(\vec r_1-\vec
 r_1^{\,\prime}) -
 \sigma_{{q\bar q}}(\vec r_2-\vec r_2^{\,
 \prime})\Bigr]\ ,
\label{1.2}
 \eeq
 and $\vec r_1,\ \vec r_2'\ \vec r_1^\prime$ and $\vec r_2^\prime$ are
 defined like in Eq.~(\ref{1.00a}). 

\section{Charmonium hadroproduction off nuclei}

 Nuclear effects in charmonium production have drawn much attention during
the last two decades since the NA3 experiment at CERN \cite{na3} has
found a steep increase of nuclear suppression with rising Feynman $x_F$.
This effect has been confirmed later in the same energy range
\cite{katsanevas}, and at higher energy recently by the most precise
experiment E866 at Fermilab \cite{e866}. No unambiguous explanation for
these observations has been provided yet. With the advent of RHIC new
data are expected soon in the unexplored energy range. Lacking a
satisfactory understanding of nuclear effects for charmonium production
in proton-nucleus collisions it is very difficult to provide a convincing
interpretation of data from heavy ion collisions experiments
\cite{na38,na50} which are aimed to detect the creation of a quark-gluon
plasma using charmonium as a sensitive probe. Many of existing analyses
rely on an oversimplified dynamics of charmonium production which fail
to explain even data for $pA$ collisions, in particular the observed
$x_F$ dependence of $J/\Psi$ suppression. Moreover, sometimes even
predictions for RHIC employ those simple models. It is our purpose to
demonstrate that the dynamics of charmonium suppression strikingly
changes between the SPS and RHIC energies. We perform full QCD
calculations of nuclear effects within the framework of the light-cone Green
function approach aiming to explain observed nuclear effects without
adjusting any parameters, and to provide realistic predictions for RHIC.

To avoid a confusion, we should make it clear that we will skip
discussion of any mechanisms of charmonium suppression caused by the
interaction with the produced comoving matter, although it should be an
important effect in central heavy ion collisions. Instead, we consider
suppression which originates from the production process and propagation
of the ${c\bar c}$ pair through the nucleus.  It serves as a baseline for
search for new physics in heavy ion collisions.

We focus here on coherence phenomena which are still a rather small
correction for charmonium production at the SPS, but whose onset has
already been observed at Fermilab and which are expected to become a
dominant effect at the energies of RHIC and LHC. One realizes the
importance of the coherence effects treating charmonium production in an
intuitive way as a hard ${c\bar c}$ fluctuation that loses coherence with
the projectile ensemble of partons via interaction with the target, and
is thus liberated. In spite of the hardness of the fluctuation, its
lifetime in the target rest frame increases with energy and eventually
exceeds the nucleus size. Apparently, in this case the ${c\bar c}$ pair is
freed by interaction with the whole nucleus, rather than with an
individual bound nucleon as it happens at low energies. Correspondingly,
nuclear effects become stronger at high energies since the fluctuation
propagates through the whole nucleus, and different nucleons compete with
each other in freeing the ${c\bar c}$. In terms of the conventional Glauber
approach it leads to shadowing. In terms of the parton model it is
analogous to shadowing of $c$-quarks in the nuclear structure function.  
It turns out (see Sect.~\ref{gluons}) that the fluctuations containing
gluons in addition to the ${c\bar c}$ pair are subject to especially strong
shadowing. Since at high energies the weight of such fluctuations rises,
as well as the fluctuation lifetime, it becomes the main source of
nuclear suppression of open and hidden charm at high energies, in
particular at RHIC. In terms of the parton model, shadowing for such
fluctuations containing gluons correspond to gluon shadowing.

The parton model interpretation of charmonium production contains no
explicit coherence effects, but they are hidden in the gluon distribution
function of the nucleus which is supposed to be subject to QCD
factorization. There are, however, a few pitfalls on this way. First of
all, factorization is exact only in the limit of a very hard scale. That
means that one should neglect the effects of the order of the inverse
$c$-quark mass, in particular the transverse ${c\bar c}$ separation $\la
{r}^2\ra \sim 1/m_c^2$. However, shadowing and absorption of ${c\bar c}$
fluctuations is a source of a strong suppression which is nearly factor
of $0.5$ for heavy nuclei (see Fig.~\ref{e-dep}). QCD factorization
misses this effect. Second of all, according to factorization gluon
shadowing is supposed to be universal, {\em i.e.} one can borrow it from
another process (although we still have no experimental information about
gluons shadowing, it only can be calculated) and use to predict nuclear
suppression of open or hidden charm. Again, factorization turns out to be
dramatically violated at the scale of charm and gluon shadowing for
charmonium production is much stronger than it is for open charm or
deep-inelastic scattering (DIS) (compare gluon shadowing exposed in
Fig.~\ref{glue-shad} with one calculated in Ref.~\cite{KST} for DIS). All
these important, sometimes dominant effects are missed by QCD
factorization. This fact once again emphasizes the advantage of the
light-cone dipole approach which does reproduce QCD factorization in
the limit where it is expected to  work, and which is also able to
calculate the deviations from factorization in a parameter free way.

Unfortunately, none of the existing models for $J/\Psi$ or $\Psi^\prime$
production in $NN$ collisions is fully successful in describing all the
features observed experimentally. In particular, the $J/\Psi$,
$\Psi^\prime$ and $\chi_1$ production cross sections in $NN$ collisions
come out too small by at least an order of magnitude \cite{bhtv}. Only
data for production of $\chi_2$ whose mechanism is rather simple seems to
be in good accord with the theoretical expectation based on the color
singlet mechanism (CSM) \cite{csm,tv} treating $\chi_2$ production via
glue-glue fusion. The contribution of the color-octet mechanism is an
order of magnitude less that of CSM \cite{tv}, and is even more
suppressed according to Ref.~\cite{schaefer}. The simplicity of the production
mechanism of $\chi_2$ suggests to use this process as a basis for the
study of nuclear effects. Besides, about $40\%$ of the $J/\Psi$s have
their origin in $\chi$ decays. We drop the subscript of $\chi_2$ in what
follows unless otherwise specified.

\subsection{Interplay of formation and coherence time scales and related 
phenomena}

A lot of work has been done and considerable progress has been
achieved in the understanding of many phenomena related to the dynamics
of the charmonium production and nuclear suppression. 

$\bullet$ Relative nuclear suppression of $J/\Psi$ and $\Psi^\prime$ has
attracted much attention.  The $\Psi^\prime$ has twice as large a
radius as the
$J/\Psi$, therefore should attenuate in nuclear matter much stronger. However,
formation of the wave function of the charmonia takes time, one cannot
instantaneously distinguish between these two levels.  This time interval or so
called formation time (length) is enlarged at high energy $E_{\Psi}$ by Lorentz
time dilation,
 \beq
t_f=\frac{2\,E_{\Psi}}{M_{\Psi^\prime}^2-M_{J/\Psi}^2}\ ,
\label{1}
 \eeq
 and may become comparable to or even longer than the nuclear radius. 
In this case neither $J/\Psi$, nor $\Psi^\prime$ propagates through
the nuclear medium, but a pre-formed ${c\bar c}$ wave packet \cite{bm}. 
Intuitively, one might even expect a universal nuclear suppression,
indeed supported by data \cite{e772,na38,e866}. However, 
a deeper insight shows that such a point of view
is oversimplified, namely, the mean transverse size of the ${c\bar c}$ 
wave packet propagating through the nucleus varies depending on the wave 
function of the final meson on which the ${c\bar c}$ is projected.
In particular, the nodal structure of the $2S$ state substantially enhances
 the yield of $\Psi^\prime$ \cite{KZ,Benhar} (see in \cite{kh-prl,kh-z} a
complementary interpretation in the hadronic basis). 
 
$\bullet$
The next phenomenon is related to the so called coherence time.
Production of a heavy ${c\bar c}$ is associated with a 
longitudinal momentum transfer $q_c$ which decreases with energy.
Therefore the production amplitudes on different nucleons add up coherently
and interfere if the production points are within the interval
$l_c=1/q_c$ called coherence length or time,
 \beq
t_c=\frac{2\,E_{\Psi}}{M_{J/\Psi}^2}\ .
\label{eq:2}
 \eeq
This time interval is much shorter than the formation time Eq.~(\ref{1}).
One can also interpret it in terms of the uncertainty principle as the mean
lifetime of a ${c\bar c}$ fluctuation.
If the coherence time is long compared to the nuclear radius,  
$t_c\gsim R_A$, different nucleons compete with each
 other in producing the charmonium. Therefore, the amplitudes interfere 
destructively leading to an additional suppression called shadowing.
Predicted in Ref.~\cite{KZ}, this effect was confirmed by the NMC
measurements of exclusive $J/\Psi$ photoproduction off nuclei \cite{nmc} 
(see also Ref.~\cite{Benhar}). 
The recent precise data from the HERMES experiment 
\cite{hermes} for electroproduction of
$\rho$ mesons also confirms the strong effect of coherence time \cite{hkn}.

Note that the coherence time Eq.~(\ref{eq:2}) is relevant only for the
lightest fluctuations $|{c\bar c}\ra$. Heavier ones which contain
additional gluons have shorter lifetime. However, at high energies they
are also at work and become an important source of an extra suppression
(see Ref.~\cite{KST} and Sect.~\ref{gluons}). They correspond to shadowing
of gluons in terms of parton model.  In terms of the dual parton model
the higher Fock states contain additional ${q\bar q}$ pairs instead of
gluons. Their contribution is enhanced on a nuclear target and leads 
to softening
of the $x_F$ distribution of the produced charmonium. This mechanism has
been used in Ref.~\cite{capella} to explain the $x_F$ dependence of charmonium
suppression. However the approach was phenomenological and data were
fitted.

The first attempt to implement the coherence time effects into the
dynamics of charmonium production off nuclei has been made in
Ref.~\cite{hir2}. However, the approach still was phenomenological and data
also were fitted. Besides, gluon shadowing (see Sect.~\ref{gluons}) had
been missed.

$\bullet$ The total $J/\Psi$-nucleon cross section steeply rises with
energy, approximately as $s^{0.2}$. This behavior is suggested by the
observation of a steep energy dependence of the cross section of $J/\Psi$
photoproduction at HERA. This fact goes well along with observation of
the strong correlation between $x_{Bj}$ dependence of the proton
structure function at small $x_{Bj}$ and the photon virtuality $Q^2$:  
the larger is $Q^2$ (the smaller is its ${q\bar q}$ fluctuation), the
steeper the rises $F_2(x_{Bj},Q^2)$ with $1/x_{Bj}$. Apparently, the
cross section of a small size charmonium must rise with energy faster
than what is known for light hadrons. The $J/\Psi$-nucleon cross section
has been calculated recently in Ref.~\cite{HIKT} employing the light-cone
dipole phenomenology, realistic charmonium wave functions and
phenomenological dipole cross section fitted to data for $F_2(x,Q^2)$
from HERA.  The results are in a good accord with data for the
electroproduction cross sections of $J/\Psi$ and $\Psi^\prime$ and also
confirm the steep energy dependence of the charmonium-nucleon cross
sections (see sect.~\ref{sec:cp}). 
Knowledge of these cross sections is very important for
the understanding of nuclear effects in the production of charmonia. A new
important observation made in Ref.~\cite{HIKT} is a strong effect of spin
rotation associated with boosting the ${c\bar c}$ system from its rest
frame to the light cone.  It substantially increases the $J/\Psi$ and
especially $\Psi^\prime$ photoproduction cross sections. The effect of
spin rotation is also implemented in our calculations below and it is
crucial for restoration of the Landau-Yang theorem \cite{kth,landau}.

$\bullet$
 Initial state energy loss by partons traveling through the nucleus
affects the $x_F$ distribution of produced charmonia \cite{kn1}
especially at medium high energies. A shift in the effective value of
$x_1$, which is the fraction of the incident momentum carried by the
produced charmonium, and the steep $x_1$-dependence of the cross section
of charmonium production off a nucleon lead to a dramatic nuclear
suppression at large $x_1$ (or $x_F$) in a good agreement with data
\cite{na3,katsanevas}. The recent analyses \cite{eloss} of data
from the E772 experiment for Drell-Yan process on nuclei reveals for the
first time a nonzero and rather large energy loss.

\subsection{Higher twist nuclear effects}\label{quarks}

Nuclear effects in the production of a $\chi$ are controlled by the
coherence and formation lengths which are defined in Eqs.~(\ref{1}),
(\ref{eq:2}). One can identify two limiting cases. The first one corresponds
to the situation where both $l_c$ and $l_f$ are shorter that the mean
spacing between bound nucleons. In this case one can treat the process
classically, the charmonium is produced on one nucleon inside the nucleus
and attenuates exponentially with an absorptive cross section which is
the inelastic $\chi-N$ one. This simplest case is described in
Refs.~\cite{kn1,reviews}.
 
In the limit of a very long coherence length $l_c\gg R_A$ one can think
about a ${c\bar c}$ fluctuation which emerges inside the incident hadron
long before the interaction with the nucleus. Different bound nucleons
compete and shadow each other in the process of liberation of this
fluctuation. This causes an additional attenuation in addition to
inelastic collisions of the produced color-singlet ${c\bar c}$ pair on its
way out of the nucleus. Since $l_c \ll l_f$ an intermediate case is also
possible where $l_c$ is shorter than the mean internucleon separation,
while $l_f$ is of the order or longer than the nuclear radius.

\begin{figure}[t]
  \centerline{\scalebox{0.5}{\includegraphics{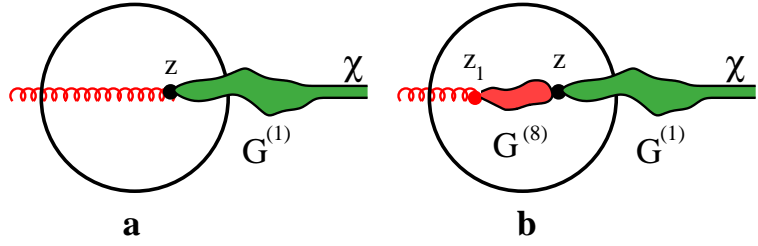}}}
{\caption{\em
      \label{general-fig} The incident gluon can either produce
the colorless
 ${c\bar c}$ pair with quantum numbers of $\chi$ at the point
$z$ ({\bf a}), or it produces diffractively
a color-octet ${c\bar c}$ with the quantum
 numbers of the gluon at the point $z_1$ which is then 
converted into a color singlet state at $z$ 
({\bf b}). Propagation of a
 color-singlet or octet ${c\bar c}$ is described by the
Green
 functions $G_{{c\bar c}}^{(1)}$ and $G_{{c\bar c}}^{(8)}$, respectively.
    }} 
\end{figure}

 The transition between the limits of very short and very long coherence
lengths is performed using the prescription suggested in Ref.~\cite{hkz} for
inelastic photoproduction of $J/\Psi$ off nuclei. The amplitude of
$\chi$ production off a nucleus can be represented in the form,
 \beq
 A^{(\lambda)}(b,z) =
 \int\limits_0^1 d\alpha \int d^2{r}\int
 d^2{r}^{\,\prime}\,
\Phi_{\chi}^*(\vec {r},\alpha)\,
 \hat D^{(\lambda)}(\vec {r},\vec
{r}^{\,\prime},
 \alpha;b,z)\, \Phi_G(\vec {r}^{\,\prime},\alpha)\ ,
\label{3.1}
 \eeq
 where $\hat D^{(\lambda)}(\vec {r},\vec {r}^{\,\prime}, \alpha;b,z)$ is 
the amplitude of production of a colorless ${c\bar c}$ pair which
reaches a separation $\vec {r}$ outside the
 nucleus. It is produced at the
 point $(\vec b,z)$ by a
 color-octet ${c\bar c}$ with separation
 ${r}^{\,\prime}$.
 The amplitude consists of two terms,
 \beq
 \hat
 D^{(\lambda)}(\vec {r},\vec
 {r}^{\,\prime},\alpha;b,z) =
 \hat
 D_1^{(\lambda)}(\vec {r},\vec
 {r}^{\,\prime},\alpha;b,z) +
 \hat
 D_2^{(\lambda)}(\vec {r},\vec
 {r}^{\,\prime},\alpha;b,z)\ .
\label{3.2}
 \eeq
 Here the first term reads,
 \beq
 \hat D_1^{(\lambda)}(\vec
 {r},\vec
 {r}^{\,\prime},\alpha;b,z) =
 G_{{c\bar c}}^{(1)}(\vec {r},z_+;\vec
 {r}^{\,\prime},z)\,
 \vec e^{\,(\lambda)}\cdot \vec d^{\,\prime}\,
 e^{iq_Lz}\ ,
\label{3.3}
 \eeq
 where $G_{{c\bar c}}^{(1)}(\vec {r},z_+;\vec {r}^{\,\prime},z)$
 is the
color-singlet Green function describing evolution
 of a ${c\bar c}$ wave packet
with initial separation
 $\vec {r}^{\,\prime}$ at the point $z$ up to the
final
 separation $\vec {r}$ at $z_+\to\infty$. This term
 is illustrated in
Fig.~\ref{general-fig}a.

 There is also a possibility for the projectile gluon to
experience diffractive interaction with production of
 color-octet ${c\bar c}$
with the same quantum numbers of the
 gluon at the point $z_1$. This pair
propagates from
 the point $z_1$ to $z$ as is described by the corresponding
color-octet Green function $G_{{c\bar c}}^{(8)}$ and produces the final
colorless pair which propagation is described by the
 color-singlet Green
function, as is illustrated in
 Fig.~\ref{general-fig}b. The corresponding second
term in
 Eq.~(\ref{3.2}) reads,
 \beqn
 \hat D_2^{(\lambda)}(\vec {r},\vec
{r}^{\,\prime},\alpha;b,z) &=&
 - {1\over2}\,\int\limits_{-\infty}^z
dz_1\,d^2
 {r}^{\prime\prime}\,
 G_{{c\bar c}}^{(1)}(\vec
{r},z_+;{r}^{\,\prime\prime},z)
\\ \nonumber &\times&
 \vec
e^{(\lambda)}\cdot\vec d^{\,\prime\prime}\,
 G_{{c\bar c}}^{(8)}(\vec
{r}^{\,\prime\prime},z;
 \vec {r}^{\,\prime},z_1)\,
 e^{iq_Lz_1}\,
\sigma_{q\bar qG}(\vec
 {r}^{\,\prime},\alpha)\,\rho_A(b,z_1)\ .
\label{3.4}
 \eeqn
 
 The singlet, $G_{{c\bar c}}^{(1)}$, and octet, $G_{{c\bar c}}^{(8)}$,
Green
 functions describe the propagation  of                                    
 color-singlet  and octet ${c\bar c}$, respectively,
in the nuclear medium. They satisfy
 the Schr\"odinger equations,
\beq
 i\,\frac{d}{d\,z}\,
 G_{{c\bar c}}^{(k)}(\vec {r},\vec
{r}^{\,\prime};z,z^\prime) =
 \left[\frac{m_c^2 -
 \Delta_{\vec
{r}}}{2\,E_G\,\alpha\,(1-\alpha)}\,
 + V^{(k)}(\vec {r},\alpha)\right]\,
G_{{c\bar c}}^{(k)}(\vec {r},\vec {r}^{\,\prime};z,z^\prime)\ ,
\label{3.5}
\eeq
 with $k=1,\ 8$ and boundary conditions
 \beq
 G_{{c\bar c}}^{(k)}(\vec
 {r},\vec {r}^{\,\prime};z,z^\prime)
 \Bigl|_{z=z^\prime} =
 \delta(\vec
 {r}-\vec {r}^{\,\prime})\ .
\label{3.6}
\eeq
 
 The imaginary part of the LC potential $V^{(k)}$ is
 responsible for
the attenuation in nuclear matter,
 \beq
 {\rm Im}\,V^{(k)}(\vec {r},\alpha)
=
 - {1\over2}\,\sigma^{(k)}({r},\alpha)\,
 \rho_A(b,z)\ ,
\label{3.7}
\eeq
 where
 \beqn
 \sigma^{(1)}({r},\alpha) &=& \sigma_{{q\bar q}}({r})\ ,
 \nonumber\\
 \sigma^{(8)}({r},\alpha) &=& \sigma_{3}({r},\alpha)\ .
\label{3.8}
\eeqn
 
The real part of the LC potential $V^{(k)}(\vec {r},\alpha)$ describes
the interaction inside the ${c\bar c}$ system. For the singlet state ${\rm
Re}\,V^{(1)}(\vec {r},\alpha)$ should be chosen to reproduce the
charmonium mass spectrum. With a realistic potential (see {\em e.g.} 
Ref.~\cite{HIKT}) one can solve Eq.~(\ref{3.5}) only numerically.  Since we
focus here on the principle problems of understanding of the dynamics of
nuclear shadowing in charmonium production, we chose the oscillator form
of the potential
 \cite{KST},
 \beq
 {\rm Re}\,V^{(1)}(\vec {r},\alpha) =
\frac{a^4(\alpha)\,{r}^2}
 {2\,E_G\,\alpha\,(1-\alpha)}\ ,
\label{3.9}
 \eeq
 where
 \beqn
 a(\alpha)&=&2\,\sqrt{\alpha(1-\alpha)\,
 \mu\,\omega}\ ,
\label{3.10}\\
\mu&=&\frac{m_c}{2}\ ,\ \ \ \ \ \omega = 0.3\,GeV\ .
 \nonumber
 \eeqn
 The LC potential Eq.~(\ref{3.9}) corresponds to a choice of
 a potential,
 \beq
U(\vec R)={1\over2}\,\mu\,\omega\,\vec R^2\ ,
\label{3.11}
 \eeq
 in the nonrelativistic Schr\"odinger equation,
 \beq
\left[-\frac{\Delta}{2\mu} +
 U(\vec R)\right]\,\Psi(\vec R) =
 E\,\Psi(\vec
r)\ ,
\label{3.12}
 \eeq
 which should describe the bound states of a colorless ${c\bar c}$ system.
Of course this is an approximation we are forced to do in order
to solve the evolution equation analytically.
 
To describe color-octet ${c\bar c}$ pairs we fix the corresponding
potential at
 \beq
 {\rm Re}\,V^{(8)}(\vec {r},\alpha)= 0\ ,
\label{3.13}
 \eeq
 in order to reproduce the gluon wave function Eq.~(\ref{1.02}).

To keep calculations simple we use the $r^2$-approximation
Eq.~(\ref{rT2}) for the dipole cross section which is
 reasonable for
small-size heavy quark systems.
 Then, taking into account Eqs.~(\ref{3.7}) -
(\ref{3.9})
 we arrive at the final expressions,
 \beq
 V^{(k)}({r},\alpha)
= {1\over2}\,
 \kappa^{(k)}\,{r}^2\ ,
\label{3.14}
 \eeq
 \beq
 \kappa^{(1)} =
 \frac{a^4(\alpha)}{\alpha(1-\alpha)\,E_G} -
 iC(s)\,\rho_A\ ,
\label{3.15}
 \eeq
 \beq
 \kappa^{(8)} = - iC(s)\,\rho_A\,
 \left\{{9\over8}\,\Bigl[\alpha^2+(1-\alpha)^2\Bigr]-
 {1\over8}\right\}\ .
\label{3.16}
 \eeq

 Making use of this approximation and assuming a constant nuclear
density
 $\rho_A(b,z)=\rho_A$ the Green functions can be obtained in an
analytical
 form,
 \beqn
 G_{{c\bar c}}^{(k)}(\vec {r},\vec
{r}^{\,\prime};z_2,z_1) &=&
 \frac{b^{(k)}}{2\pi\,\sinh(\Omega^{(k)}\,\Delta
z)}
 \nonumber\\ &\times&
 \exp\left\{- \frac{b^{(k)}}{2}\,
\left[\frac{\vec {r}^{\,2} +\vec {r}^{\,\prime\,2}}
{\tanh(\Omega^{(k)}\,\Delta z)} -
 \frac{2\,\vec {r}\cdot\vec
{r}^{\,\prime}}
 {\sinh(\Omega^{(k)}\,\Delta z)}\right]\right\}\ ,
\label{3.17}
 \eeqn
 where
 \beqn
 b^{(k)} &=& \sqrt{\kappa^{(k)}\,
 E_G\,\alpha(1-\alpha)}\ ,\nonumber\\
 \Omega^{(k)} &=&
 \frac{b^{(k)}}{E_G\,\alpha(1-\alpha)}\ ,
 \nonumber\\
 \Delta z &=& z_2-z_1\ .
 \nonumber
 \eeqn
 
\begin{figure}[t] 
\includegraphics{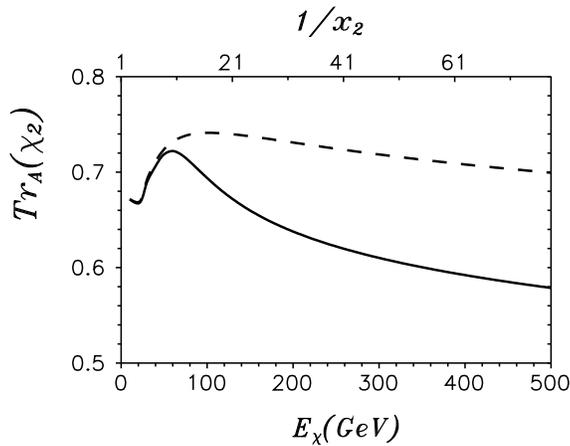}
\begin{center}
\vspace{6.3cm}
 {\caption[Delta]
 {\sl Nuclear transparency for $\chi$ production off lead as function
of energy of the charmonium, or $x_2$ (the upper scale).  The solid curve
includes both effects of coherence and formation, while the dashed curve
corresponds to $l_c=0$. Since transparency scales in $x_2$ according to
(\ref{scaling}), values of $x_2$ are shown on the top axis.}
 \label{e-dep}}
\end{center}
 \end{figure}

 We define the nuclear transparency for $\chi$ production as
 \beq
 Tr_A(\chi)=\frac{\sigma(G\,A \to \chi\,X)}
 {A\,\sigma(G\,N \to
 \chi\,X)}\ .
\label{transparency}
 \eeq
 It depends only on the $\chi$ or projectile gluon energy. We plot our
 predictions for lead in Fig.~\ref{e-dep}.

 Transparency rises at low energy since the formation length
 increases
 and the effective absorption cross section becomes smaller. This
 behavior, assuming $l_c=0$, is shown by dashed curve. However, at higher
 energies the coherence length is switched on and shadowing adds to
 absorption. As a result, transparency
 decreases, as is shown by the solid curve. On top of that, the 
energy dependence of the dipole cross section makes those both curves 
for $Tr_A(E_{\chi})$ fall even faster.
 
Apparently, the nuclear  transparency depends only on the $\chi$ energy, 
rather than the incident energy or $x_1$. It is interesting that this leads 
to $x_2$ scaling. Indeed, the $\chi$ energy
 \beq
E_{\chi}=\frac{M_{\chi}^2}{2\,m_N\,x_2}\ ,
\label{scaling}
 \eeq
depends only on $x_2$. We show the $x_2$ scale in Fig.~\ref{e-dep}
(top) along with energy dependence.

\begin{figure}[tbh]
 \includegraphics{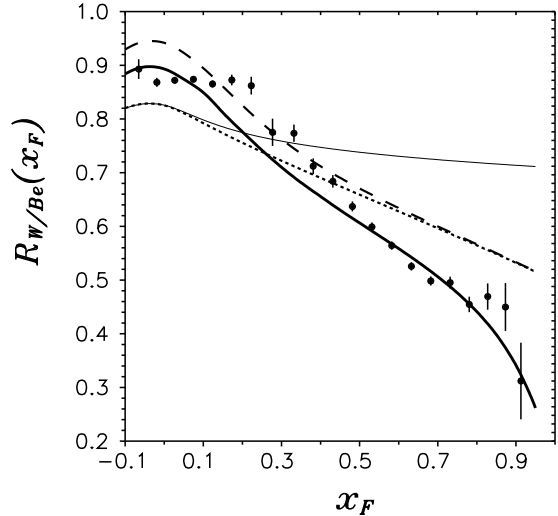}
\begin{center}
\vspace{7.5cm}
 {\caption[Delta]
 {\sl Tungsten to beryllium cross section ratio as function of Feynman
$x_F$ for $J/\Psi$ production at proton energy $800\,GeV$. The thin
solid curve represents contribution of initial state quark shadowing and
final state ${c\bar c}$ attenuation for $\chi$ production. The dotted
curve includes also gluon shadowing. The dashed curve is corrected for
gluon enhancement at large $x_2$ (small $x_F$) using the prescription
from \cite{eks}.  The final solid curve is also corrected for energy
loss and for $\chi\to J/\Psi\gamma$ decay.  Experimental points are from
the E866 experiment \cite{e866}.}
 \label{xf-dep}}
\end{center}
 \end{figure}

We also compare in Fig.~\ref{xf-dep} the contribution of quark shadowing 
and absorption (thin solid curve) with the 
nuclear suppression observed at $800\,GeV$ \cite{e866}.
Since data are  for $W/Be$  ratio, and our constant density approximation 
should not be applied to beryllium, we assume for simplicity that all 
$pA$ cross sections including $pN$ obey the $A^{\alpha(x_F)}$.
We see  that the calculated contribution has quite a different shape
from what is suggested by the data. It also leaves plenty of room for 
complementary mechanisms of suppression at large $x_F$ (see below).

\subsection{Leading twist gluon shadowing}\label{gluons}

Previously we considered only the lowest $|{c\bar c}\ra$ fluctuation
of the gluon, which is apparently an approximation. The higher Fock components
containing gluons should be also included. In fact they are already incorporated
in the phenomenological dipole cross section we use, and give rise to the energy
dependence of $\sigma_{{q\bar q}}$. However, they are still excluded from nuclear
effects. Indeed, although we eikonalize the energy dependent dipole cross section
the higher Fock components do not participate in that procedure, but 
they have to be eikonalized as well. This corrections, as is demonstrated below, 
correspond to suppression of gluon density in nuclei at small $x$.

The gluon density at small $x$ in nuclei is known to be shadowed,
{\em i.e.} reduced compared to a free nucleons. The partonic
 interpretation of this phenomenon looks very different depending on
the reference frame. In the infinite
momentum frame, as was first
 suggested by Kancheli \cite{kancheli}, the
partonic clouds of
 nucleons are squeezed by the Lorentz transformation less
at small than
 at large $x$. Therefore, while these clouds are well separated in
longitudinal
 direction at large $x$, they overlap and can fuse at small $x$,
resulting
 in a diminished parton density \cite{kancheli,glr}.
 
 Different
observables can probe this effect. Nuclear shadowing of the
 DIS inclusive
cross section or Drell-Yan process demonstrate a reduction
 of the sea quark
density at small $x$. Charmonium or open charm
 production is usually
considered as a probe for gluon distribution.
 
 Although observables are
Lorentz invariant, partonic interpretations are
 not, and the mechanism of
shadowing looks quite different in the rest
 frame of the nucleus where it
should be treated as Gribov's inelastic
 shadowing.  This approach seems to go
better along with our intuition,
 besides, the interference or coherence
length effects governing
 shadowing are under a better control.  One can even
calculate
 shadowing in this reference frame in a parameter free way (see 
Refs.~\cite{KRT,kst1,KST}) employing the well developed phenomenology of
 color
dipole representation suggested in Ref.~\cite{ZKL}.  On the other
 hand, within the
parton model one can only calculate the $Q^2$
 evolution of shadowing which is
quite a weak effect.  The main
 contribution to shadowing originates from the
fitted to data input.
 
 In the color dipole representation nuclear shadowing can be
calculated via simple eikonalization of the elastic amplitude for each
Fock component of the projectile light-cone wave function which are
the eigenstates of interaction \cite{ZKL}. Different Fock components
represent shadowing of different species of partons. The $|{q\bar q}\ra$
component in DIS or $|q\gamma^*\ra$ in Drell-Yan reaction should be
used to calculate shadowing of sea quarks. The same components
including also one or more gluons lead to gluon shadowing
\cite{mueller,KST}.

In the color dipole approach one can explicitly see deviations from
QCD factorization, {\em i.e.} dependence of the measured parton
distribution on the process measuring it. For example, the coherence
length and nuclear shadowing in the Drell-Yan process vanish at minimal
$x_2$ (at fixed energy) \cite{eloss}, while the factorization predicts
maximal shadowing. Here we present even more striking deviation from
factorization, namely, gluon shadowing for charmonium production turns
out to be dramatically enhanced compared to DIS.

\subsection{LC dipole representation for the reaction 
\boldmath$G\,N \to \chi\,G\,X$}\label{lc-wf}
 
\begin{figure}[t]
  \centerline{\scalebox{0.5}{\includegraphics{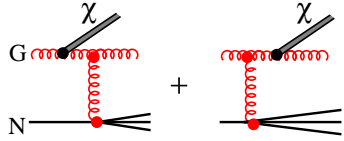}}}
{\caption{\em
      \label{dy}The dominant Feynman diagrams contributing to $\chi$ 
production. 
    }} 
\end{figure}

 In the case of charmonium production, different Fock components of the
projectile gluon, $|({c\bar c})_1nG\ra$ containing a colorless ${c\bar c}$
pair and $n$ gluons ($n=0,1\dots$) build up the cross section of charmonium 
production which steeply rises with energy
(see Ref.~\cite{HIKT}). The cross 
section is expected to factorize in impact parameter representation in
analogy to the DIS and Drell Yan reaction. This representation has the
essential advantage in that nuclear effects can be easily calculated 
\cite{hir,kst1}. Feynman diagrams corresponding $\chi$ production associated
with gluon radiation are shown in Appendix~D of \cite{kth}. 
We treat the interaction of heavy quarks perturbatively in the lowest order
approximation, while the interaction with the nucleon is soft and
expressed in terms of the gluon distribution. The calculations
are substantially simplified if the
radiated gluon takes a vanishing fraction $\alpha_3$ of the total
light-cone momentum and the heavy quarkonium can be treated as a
nonrelativistic system. In this case the amplitude of $\chi\, G$
production has a simple form that corresponds to
the ``Drell-Yan'' mechanism of $\chi$ production illustrated in 
Fig.~\ref{dy}.

 Correspondingly, the cross section of $\chi$ production has the 
familiar factorized form similar to the Drell-Yan reaction 
\cite{hir,bhq,kst1},
 \beq
\alpha_3\,\frac{d\,\sigma(GN\to\chi G X)}
{d\,\alpha_3} = 
\int d^2s\, \left|\Psi_{G\chi}(s,\alpha_3)\right|^2\,
\sigma_{GG}\left[(1-\alpha_3)\vec s,x_2/\alpha_3\right]\ ,
\label{3.2a}
 \eeq
 where $\sigma_{GG}(r,x)=9/4\,\sigma_{{q\bar q}}(r,x)$ 
 is the cross section of interaction of a $G\,G$
dipole with a nucleon. $\Psi(s,\alpha_3)$ is the effective
distribution amplitude for the $\chi-G$ fluctuation of a gluon, which
is the analog to the $\gamma^*\,q$ fluctuation of a quark,
 \beqn
\Psi_{G\chi}(s,\alpha_3) &=& \sum\limits_{\bar\mu\mu}\,
\int d^2r\,d\alpha\, 
\Phi^{\bar\mu\mu}_{\chi}(\vec r,\alpha)\,
\Phi^{\bar\mu\mu}_{G}(\vec r,\alpha) 
\nonumber\\ &\times&
\left[\Phi_{c G}\left(\vec s + 
\frac{\vec r}{2},\frac{\alpha_3}{\alpha}\right)
- \Phi_{c G}\left(\vec s - 
\frac{\vec r}{2},\frac{\alpha_3}
{1-\alpha}\right)\right]\ .  
\label{3.3a}
 \eeqn 
Here,
$\Phi^{\bar\mu\mu}_{\chi}(\vec r,\alpha)$ and $\Phi^{\bar\mu\mu}_{G}(\vec
r,\alpha)$ are the ${q\bar q}$ LC wave functions of the $\chi$ and gluon,
respectively, which depend on transverse separation $\vec r$ and relative
sharing $\alpha$ by the ${q\bar q}$ of the total LC momentum. $\Phi_{c
G}(\vec s,\alpha)$ is the LC wave function of a quark-gluon Fock
component of a quark.

\subsection{Gluon shadowing for \boldmath$\chi$ production off 
nuclei}\label{gluon-shadowing}

The gluon density in nuclei is known to be modified, shadowed at small
Bjorken $x$. Correspondingly, production of $\chi$ treated as gluon-gluon
fusion must be additionally suppressed. In the rest frame
of the nucleus, gluon shadowing appears  as Gribov's inelastic shadowing
\cite{Gribov}, which is 
related to diffractive gluon radiation. The rest frame seems to
be more convenient to calculate gluon shadowing, since techniques are
better developed, and we use it in what follows.
The process $G\,N \to \chi\,X$ considered in the previous section 
includes by default radiation of any number of gluons which give rise to
the energy dependence of the dipole cross section.
 
Extending the analogy between the reactions of $\chi\,G$ production by an
incident gluon and heavy photon radiation by a quark to the case of
nuclear target one can write an expression for the cross section
of reaction $G\,A\to\chi\,G\,X$ in two limiting cases:

{\bf (i)} the production occurs nearly instantaneously over a
longitudinal distance which is much shorter than the mean free path of
the $\chi\,G$ pair in nuclear matter. In this case the cross sections on
a nuclear and nucleon targets differ by a factor $A$ independently of the
dynamics of $\chi\,G$ production.

{\bf (ii)} The lifetime of the $\chi\,G$ fluctuation,
 \beq
t_c = \frac{2\,E_G}{M^2_{\chi G}}\ ,
\label{3.4a}
 \eeq
substantially exceeds the nuclear size. It is straightforward to 
replace the dipole cross section on a nucleon by a nuclear one 
\cite{hir,kst1},
then Eq.~(\ref{3.2a}) is modified to
 \beqn
\frac{d\,\sigma(GA\to\chi G X)}
{d\,(\ln\alpha_3)} &=& 
2\int d^2b\,d^2s\, \left|\Psi_{G\chi}
(\vec s,\alpha_3)\right|^2\nonumber\\
&\times&
\left\{1 - \exp\left[-{1\over2}\,\sigma_{GG}(\vec s,x_2/\alpha_3)\,
T_A(b)\right]\right\}\ .
\label{3.5a}
 \eeqn
 In order to single out the net gluon shadowing we exclude here the size
of the ${c\bar c}$ pair assuming that the cross section responsible for
shadowing depends only on the transverse separation $\vec s$.

{\bf (iii)} A general solution valid for any value of $t_c$ is more complicated
and must interpolate between the above limiting situations.
In this case one can use the methods of the
Landau-Pomeranchuk-Migdal (LPM) theory for photon bremsstrahlung in a
medium generalized for targets of finite thickness in \cite{z,kst1}.
The general expressions for the cross section which reproduces the 
limiting cases $t_c\to0$ ({\bf i}) and $t_c\to\infty$ ({\bf ii})
reads,
 \beqn
&&\frac{d\,\sigma(GA\to\chi G X)}
{d^2b d\,(\ln\alpha_3)} \,
= \left\{\int\limits_{-\infty}^{\infty}
dz\,\rho_A(b,z)\right.\nonumber\\
&\times& \left.\int d^2s\,
\left|\Psi_{G\chi}(s,\alpha_3)\right|^2\right.
\sigma_{GG}\left[(1-\alpha_3)\vec s,x_2/\alpha_3\right]
\nonumber\\ 
&-& \left. {1\over2}\,{\rm Re}
\int\limits_{-\infty}^{\infty}dz_2\,
\rho_A(b,z_2)
\int\limits_{-\infty}^{z_2}dz_1\,
\rho_A(b,z_1)\,
\widetilde\Sigma(z_2,z_1)\,
e^{iq_L(z_2-z_1)}\right\}\ ,
\label{3.6a}
 \eeqn
 where
 \beqn
\widetilde\Sigma(z_2,z_1) &=& 
\int d^2s_1\,d^2s_2\,
\Psi_{G\chi}^*(\vec s_2,\alpha_3)\,
\sigma_{GG}(s_2,x_2/\alpha_3)\,
G(\vec s_2,z_2;\vec s_1,z_1)\nonumber\\
&\times&
\sigma_{GG}(s_1,x_2/\alpha_3)\,
\Psi_{G\chi}(\vec s_1,\alpha_3)\ .
\label{3.7a}
 \eeqn
 
\begin{figure}[tbh]
\includegraphics{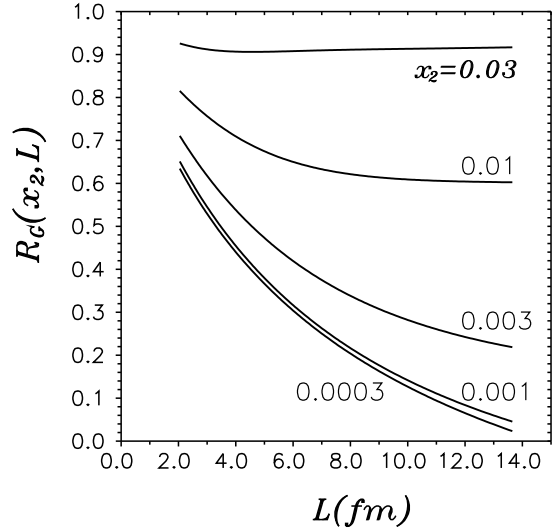}
\begin{center}
\vspace{8cm}
{\caption[Delta]
{\sl Gluon suppression as function of thickness of nuclear matter
with constant density $\rho_A=0.16\,fm^{-3}$.}
\label{glue-shad}}
\end{center}
 \end{figure}

To single out the correction for gluon shadowing one should compare the 
cross section Eq.~(\ref{3.6a}) with the impulse approximation term in 
which absorption is suppressed,
 \beq
R_G(x_2)=
\frac{G_A(x_2)}{A\,G_N(x_2)} = 
1 - \frac{1}{A\,\sigma(GN\to\chi X)}\,
\int\limits_{x_2}^{\alpha_{max}}
d\alpha_3\,\frac{d\sigma(GA\to\chi GX)}
{d\alpha_3}\ .
\label{3.12a}
 \eeq
 For further calculations and many other applications one needs to know
gluon shadowing as function of impact parameter which is calculated as 
follows,
 \beqn\nonumber
R_G(x_2,b)&=&
\frac{G_A(x_2,b)}{T_A(b)\,G_N(x_2)}\\
& =&   
1 - \frac{1}{T_A(b)\,\sigma(GN\to\chi X)}\,
\int\limits_{x_2}^{\alpha_{max}}d\alpha_3\,
\frac{d\sigma(GA\to\chi GX)}                                        
{d^2b\,d\alpha_3}\ .
\label{3.15a}
 \eeqn 
The results of calculations for the 
$b$-dependence of gluon shadowing (\ref{3.15a}) are depicted in
Fig.~\ref{glue-shad} for different values of $x_2$ as function of thickness
of nuclear matter, $L=\sqrt{R_A^2-b^2}$.

 The results confirm the obvious expectation that shadowing increases for
smaller $x_2$ and for longer path in nuclear matter. One can see that for
given thickness shadowing tends to saturate down to small $x_2$, what
might be a result of one gluon approximation. Higher Fock components with
larger number of gluons are switched on at very small $x_2$. At the same
time, shadowing saturates at large lengths what one should have also
expected as a manifestation of gluon saturation. Note that at large
$x_2=0.03$ shadowing is even getting weaker at longer $L$. This is easy
to understand, in the case of weak shadowing one can drop off the
multiple scattering terms higher than two-fold one. Then the shadowing
correction is controlled by the longitudinal formfactor of the nucleus
which decreases with $L$ (it is obvious for the Gaussian shape of the
nuclear density, but is also true for the realistic Woods-Saxon
distribution).

\subsection{Antishadowing of gluons}\label{antishadowing}

Nuclear modification of the gluon distribution is poorly known.
There is still no experimental evidence for that. Nevertheless,
the expectation of gluon shadowing at small $x$ is very solid, and only 
its amount might be disputable. At the same time,
some indications exist that gluons may be enhanced in nuclei 
at medium small $x_2\sim 0.1$. The magnitude of gluon antishadowing
has been estimated in Ref.~\cite{fs} assuming that the total fraction of 
momentum carried by gluons is the same in nuclei and free nucleons
(there is an experimental support for it). Such a momentum conservation
sum rule leads to a gluon enhancement at medium $x$, since gluons
are suppressed in nuclei at small $x$. The effect, up  to $\sim 20\%$
antishadowing in heavy nuclei at $x\approx 0.1$, found in Ref.~\cite{fs} 
is rather large, but it is a result of very strong shadowing which
we believe has been grossly overestimated (see discussion in
Ref.~\cite{KST}).

Fit to DIS data based on evolution equations performed in
Ref.~\cite{eks} also provided an evidence for rather strong antishadowing
effect at $x\approx 0.1$. However, the fit 
employed an ad hoc assumption that gluons 
are shadowed at the low scale $Q_0^2$  
exactly as $F_2(x,Q^2)$ what might be true only by accident. Besides, in 
the $x$ distribution of antishadowing was shaped ad hoc too.

A similar magnitude of antishadowing has been found in the analysis 
\cite{gp} of data on $Q^2$ dependence of nuclear to nucleon ratio of the
structure functions, $F^A_2(x,Q^2)/F^N_2(x,Q^2)$. However it was based on the 
leading order QCD approximation which is not well justified at these values 
of $Q^2$. 

Although neither of  these results seem to be reliable, 
similarity of the scale of the predicted effect
looks convincing, and we included the antishadowing of gluons 
in our calculations. We use the shape of $x_2$ dependence and magnitude of 
gluon enhancement from Ref.~\cite{eks}.
 
\subsection{Comparison with available data and predictions for higher 
energies}\label{results}

The dynamics of $J/\Psi$ suppression at energy $800\,GeV$ 
is rather complicated and includes many effects.
Now we can apply more corrections to the dotted curve in
Fig.~\ref{xf-dep} which involves only quark and gluon shadowing. Namely,
inclusion of the energy loss effect and decay $\chi\to J/\Psi\,\gamma$
leads to a stronger suppression depicted by the dashed curve. Eventually we
correct this curve for gluon enhancement at $x_2\sim 0.1$ (small $x_F$)
and arrive at the final result shown by thick solid curve.

Since our calculation contains no free parameters we think that the 
results agree with the data amazingly well. Some difference in the shape 
of the maximum observed and calculated at small $x_F$ may be a result of 
the used parameterization \cite{eks} for gluon antishadowing. We think that it
gives only the scale of the effect, but neither the ad hoc shape, 
nor the magnitude should be taken literally.
Besides, our calculations are relevant only for those 
$J/\Psi$s which originate from $\chi$ decays which feed only about $40\%$ 
of the observed ones.

\begin{figure}[ht]
\includegraphics{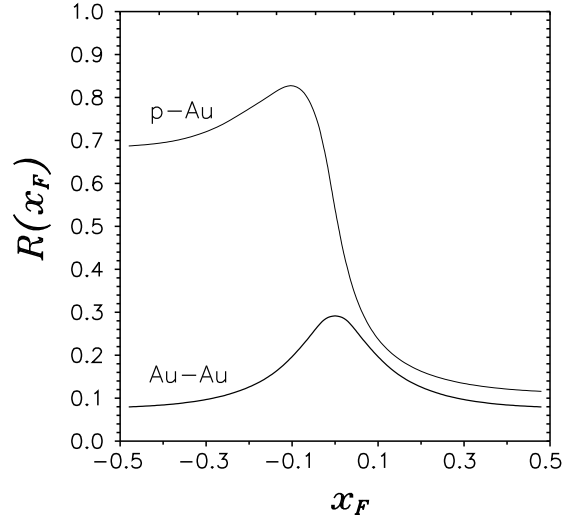}
\begin{center}
\vspace{7.5cm}
{\caption[Delta]
{\sl Nuclear suppression of $J/\Psi$ production in proton-gold 
collisions at $\sqrt{s}=200\,GeV$ as function of $x_F$ (the upper curve)
and in gold-gold collisions (bottom curve). Effects of quark and gluon 
shadowing and gluon antishadowing are included.}
\label{rhic}}
\end{center}
 \end{figure}

At higher energies of RHIC and LHC, the effect of energy loss is 
completely gone and nuclear suppression must expose $x_2$ scaling.
Much smaller $x_2$ can be reached at higher energies. 
Our predictions for proton-gold to proton-proton ratio is depicted in 
Fig.~\ref{rhic}. 
One can see that at $x_F > 0.1$ shadowing suppresses charmonium 
production by nearly an order of magnitude.
We can also estimate the effect of nuclear suppression in heavy ion 
collisions assuming factorization,
 \beq
R_{AB}(x_F) = R_{pB}(x_F)\,R_{pA}(-x_F)\ .
\label{AB}
 \eeq
Our predictions for gold-gold collisions at $\sqrt{s}=200\,\GeV$ are 
shown by the bottom curve in Fig.~\ref{rhic}. Since factorization is 
violated this prediction should be verified.

\section{Summary}
In these lectures, we presented several QCD processes related to 
production of heavy quarks at high energies. We describe all these 
reactions within the same approach, a light-cone color-dipole formalism.
We highlighted the main advantages of this approach in comparison with 
the alternative description based on the standard QCD parton model.
The color-dipole formalism does not involve any uncertain $K$-factors, 
takes care and calculates higher twist corrections, predicts nuclear 
effects. At the same time, this approach is restricted to the small-$x$ 
domain.

We started with the simplest process of diffractive production of
charmonia off protons and nuclei. We made use of the best of our 
knowledge for charmonium wave functions and methods of their boosting to 
the light front. As a result, we arrived at a very good agreement with 
data, achieved without any adjustments. One of the key issues in 
reaching this agreement is inclusion of the Melosh spin rotation 
which substantially changes the production rates of $J/\Psi$ and 
especially $\Psi'$. In particular, it solves the long standing problem of 
understanding photoproduction data for the $\Psi'$ to $J/\Psi$ ratio.
These calculations also provided realistic predictions for 
charmonium-nucleon total cross sections.

We extended the study to charmonium photoproduction on nuclear targets
aiming to study shadowing effects. Surprisingly, we found that higher
twist shadowing which is neglected in parton model calculations, is the
main source of shadowing, at least for foreseen energies. This shadowing
effect is related to a nonzero separation of the produced ${c\bar c}$ and
vanishes in the limit of very heavy quarks. The leading twist shadowing,
so called gluon shadowing, is related to higher Fock components of the
photon containing gluons. It depends only logarithmically on the quark
mass, but its onset is delayed to very high energies.

The next important process is hadroproduction of open heavy flavor. The
color-dipole approach turns out to be quite effective even for a proton
target due to well developed phenomenology for the dipole cross section
fitted to HERA data for $F^p_2(x,Q^2)$. Our predictions for open charm
and beauty production well agrees with available data.

Higher twist quark shadowing for 
open charm hadroproduction off nuclei is a quite weak effect. 
However, we find large higher twist corrections to the leading twist
gluon shadowing, which makes the latter process dependent.
Gluon shadowing in open charm production is found to be stronger than in 
photoproduction of charmonium (and in DIS), but somewhat weaker than 
in the case of hadroproduction of charmonium. Our predictions for RHIC 
will be tested soon.

The last and most complicated case is hadroproduction of charmonia. We
restricted ourselves to the simplest case of production of the $P$-wave
$\chi$ states. Like in photoproduction, we found a substantial higher
twist shadowing, but much stronger leading twist gluon shadowing. For the
first time we explained the steep $x_F$ dependence of nuclear suppression
of charmonia observed in the E772/866 experiments at Fermilab. Gluon
shadowing is dramatically enhanced at the energies of RHIC and LHC and we
predict very strong suppression of charmonia both for $pA$ and $AA$
collisions.

\bigskip {\bf Acknowledgments:} We are grateful to David Blaschke for his
patience in encouraging us to write these lectures. Special thanks go
to our collaborators
J\"org H\"ufner, Yuri Ivanov, Jen-Chieh Peng and Sasha Tarasov
with whom the results presented here have been achieved.
We thank the Institute for Nuclear Theory at the
University of Washington for its hospitality during the workshop {\em The
First Three Years of Heavy-ion Physics at RHIC}, where this work was
completed. B.Z.K.\ is supported by the grant from the Gesellschaft f\"ur
Schwerionenforschung Darmstadt (GSI), grant No.~GSI-OR-SCH, and by the
grant INTAS-97-OPEN-31696. J.R.\ is supported by the U.S.~Department of
Energy at Los Alamos National Laboratory under Contract No.~W-7405-ENG-38.

\end{document}